\documentclass[lettersize,journal]{IEEEtran}
\usepackage{amssymb,amsfonts}
\usepackage{algorithmic}
\usepackage{graphicx}
\usepackage{textcomp}
\usepackage{xcolor}
\usepackage{amsmath}
\usepackage{bm}
\usepackage{empheq}
\usepackage{caption}
\usepackage{subcaption}
\usepackage{xurl}
\usepackage{hyperref}

\usepackage[normalem]{ulem}
\usepackage{cancel}
\usepackage[font=small,labelfont=bf]{caption}
\usepackage{makecell}
\usepackage{amssymb}
\usepackage{cleveref}
\usepackage{cite}

\begin{document}
\title{Water Level Sensing via Communication Signals \\in a Bi-Static System}

\author{Zhongqin Wang,
        J. Andrew Zhang, \IEEEmembership{Senior Member, IEEE}, 		    Kai Wu, \IEEEmembership{Member, IEEE}, 
		Y. Jay Guo, \IEEEmembership{Fellow, IEEE}
\IEEEcompsocitemizethanks{
\IEEEcompsocthanksitem Zhongqin Wang, J. Andrew Zhang (Corresponding Author), and Kai Wu are with the School of Electrical and Data Engineering and the Global Big Data Technologies Centre, University of Technology Sydney, Sydney 2007, Australia. E-mail:\{zhongqin.wang, andrew.zhang, kai.wu\}@uts.edu.au
\IEEEcompsocthanksitem Y. Jay Guo is with the Global Big Data Technologies Centre, University of Technology Sydney, Sydney 2007, Australia. E-mail: jay.guo@uts.edu.au
}
}

\markboth{Journal of \LaTeX\ Class Files,~Vol.~14, No.~8, August~2021}%
{Shell \MakeLowercase{\textit{et al.}}: A Sample Article Using IEEEtran.cls for IEEE Journals}

\IEEEpubid{0000--0000/00\$00.00~\copyright~2021 IEEE}
\maketitle

\begin{abstract}
Accurate water level sensing is essential for flood monitoring, agricultural irrigation, and water resource optimization. Traditional methods require dedicated sensor deployments, leading to high installation costs, vulnerability to interference, and limited resolution. This work proposes PMNs-WaterSense, a novel scheme leveraging Channel State Information (CSI) from existing mobile networks for water level sensing. Our scheme begins with a CSI-power method to eliminate phase offsets caused by clock asynchrony in bi-static systems. We then apply multi-domain filtering across the time (Doppler), frequency (delay), and spatial (Angle-of-Arrival, AoA) domains to extract phase features that finely capture variations in path length over water. To resolve the $2\pi$ phase ambiguity, we introduce a Kalman filter-based unwrapping technique. Additionally, we exploit transceiver geometry to convert path length variations into water level height changes, even with limited antenna configurations. We validate our framework through controlled experiments with 28 GHz mmWave and 3.1 GHz LTE signals in real time, achieving average height estimation errors of 0.025 cm and 0.198 cm, respectively. Moreover, real-world river monitoring with 2.6 GHz LTE signals achieves an average error of 4.8 cm for a 1-meter water level change, demonstrating its effectiveness in practical deployments.
\end{abstract}


\section{Introduction}
\IEEEPARstart{W}{ater} level sensing is a crucial aspect of environmental management, particularly in flood monitoring, agricultural irrigation, and water resource optimization. Accurate water level data supports early warning systems to mitigate flood risks in urban and rural areas, improves irrigation efficiency in agriculture, and ensures the structural safety of reservoirs and dams through proactive monitoring. Traditional water level monitoring systems commonly utilize specialized sensors, such as ultrasonic, pressure, and float-based gauges \cite{mohindru2023development, wu2023review}. These methods have been widely adopted due to their reliability and accuracy across various environments. However, they often require complex installation, incur high maintenance costs, and are susceptible to environmental interference in large-scale deployments. Recent advancements in Perceptive Mobile Networks (PMNs) \cite{zhang2020perceptive, zhang2021overview, zhang2021enabling, liu2022integrated, zhang2023perceptive, lu2024integrated}, which integrate sensing and communication in cellular networks, have opened new opportunities to leverage existing communication infrastructure for water level sensing. By analyzing these signals, particularly Channel State Information (CSI), meaningful environmental features, including water level variations, can be extracted. This approach not only reduces deployment costs but also enables non-invasive, scalable, and real-time monitoring, making it a promising alternative to conventional sensing methods.

Currently, there is limited research on utilizing communication signals for water level sensing. Most existing studies explore 4G Long-Term Evolution (LTE) or 5G signals by analyzing Received Signal Strength Indicators (RSSI), Reference Signal Received Power (RSRP), and Reference Signal Received Quality (RSRQ), and CSI for object localization \cite{hu2024localization, pegoraro2024jump}, rainfall sensing \cite{low2022towards, xu2025smartphone}, vehicle classification \cite{feng2021lte}, and soil moisture \cite{feng2022lte}. RSSI, RSRP, and RSRQ can be accessible via standard mobile device interfaces but suffer from low resolution. In contrast, CSI offers high granularity, making it widely used for precise sensing, though it requires specialized hardware like software-defined radio devices. In a different yet relevant line of research, WiFi sensing has been extensively studied in recent years, leveraging CSI to perform various sensing tasks, such as indoor localization \cite{wang2022single, meneghello2022sharp, wang2024passive}, gait and activity recognition \cite{wang2024rdgait}, pose estimation \cite{wang2024multi}, and vital sign measurement \cite{xie2024robust}. Researchers extract CSI using open-source tools \cite{halperin2011tool, xie2015precise, gringoli2019free}, but most commercial WiFi chipsets restrict raw CSI access, which limits their practicality in Integrated Sensing and Communication (ISAC) applications. In general, WiFi is better for short-range indoor applications, while 4G/5G signals suit large-scale outdoor sensing.

Despite the potential of using these communication signals for water level sensing, several challenges still remain:

\IEEEpubidadjcol
\textit{1) Clock Asynchronization.} In bi-static systems, the lack of a synchronized clock between  the User Equipment (UE) and Base Station (BS) introduces random phase offsets due to Time Offsets (TO) and Carrier Frequency Offsets (CFO), distorting CSI and hindering fine-grained sensing. Existing solutions often rely on multi-antenna techniques like cross-antenna cross-correlation \cite{wang2022single, wang2024passive} or the two-antenna CSI ratio \cite{ni2023uplink}, but they are impractical for single-antenna setups due to hardware constraints or simplified deployments. A unified framework is required to remove the impact of clock asynchronization while avoiding the non-linear transformation of Doppler, delay, and angle features in CSI.

\textit{2) Environmental Interference.} Real-world environments present a significant challenge for accurate water level sensing. Static structures (e.g., buildings, trees) cause persistent multipath reflections, while moving objects like boats, pedestrians, and vehicles induce time-varying interference. Wind, waves, and rainfall add further noise, distorting received signals and masking subtle water level variations. Effective interference suppression is crucial to isolate water surface-reflected signals while mitigating static and dynamic distortions.

\textit{3) Limited Sensitivity to Subtle Variations.} Mobile communication signals face challenges in detecting small water level fluctuations due to bandwidth and spatial resolution limitations. LTE systems, operating within 1.4-20 MHz, have insufficient range resolution to distinguish closely spaced multipath reflections, making it difficult to track precise water level changes, even for a 1-meter height variation. Additionally, typical software define radio-based wireless systems with only 1-4 received antennas also lack fine spatial angular resolution. Thus, detecting small-scale water level variations using mobile communications signals remains challenging.

In this work, we propose \textit{PMNs-WaterSense} (Perceptive Mobile Networks for Water-Level Sensing), a novel framework that utilizes CSI to accurately monitor water levels via existing communication infrastructure. To the best of our knowledge, the framework is the first practical CSI-based water level sensing scheme that operates without dedicated sensors or prior training datasets, enhancing its adaptability for real-world deployment. It can operate with a single-antenna setup for the transmitter and receiver, ensuring a cost-effective design, while also being extendable to multi-antenna setups. It is adaptable across various frequency bands, base station locations, and environmental conditions. To address the above challenges, the framework incorporates several innovative techniques.

First, to mitigate clock asynchrony between the UE and BS, we use a single-antenna self-correlation technique based on CSI power. The technique multiplies an antenna's CSI with its own conjugate. This real-valued CSI power retains phase differences between multipath components, preserving key features for water level sensing. It also maintains the linear relationships among Doppler, delay, and Angle-of-Arrival (AoA) for feature extraction. Additionally, this method eliminates random initial phase offsets from the Phase-Locked Loop (PLL) in a receiver, avoiding the need for complicated antenna calibration in AoA-based spatial filtering.

Second, to suppress environmental noise and extract water level features, we leverage the time domain (Doppler), frequency domain (delay), and spatial domain (AoA) in CSI. A time-domain Fourier transform followed by a frequency-domain Minimum Variance Distortionless Response (MVDR) algorithm generates a Doppler-Delay heatmap. A Constant False Alarm Rate (CFAR) detector is then applied to identify water level changes. If detected, the beamforming weights are used to extract a complex signal feature from the corresponding Doppler and delay bins to represent water level variations. For multi-antenna setups, spatial filtering using a spatial-domain Fourier transform enhances noise suppression.

Third, phase in the extracted complex features exhibits the higher sensing sensitivity compared to amplitude and captures variations in the Non-Line-of-Sight (NLOS) path induced by water level changes. To ensure continuous water level tracking, we use a Kalman filter to resolve the $2\pi$ phase ambiguity. The unwrapped phase is then transformed into actual water level height changes using the transceiver geometry relationship.

Our main contributions are described as follows:

1) We are the first to apply the CSI power for water level sensing, effectively addressing TO and CFO, as well as the phase offsets across Rx antennas. This technique is applicable to single-antenna systems and can extend to multi-antenna setups without additional antenna calibration.

2) We propose a novel water level feature extraction approach using multi-domain filtering. Unlike most existing methods for fast-moving object sensing, our algorithm exploits the slow-changing nature of water levels by performing slow-time Doppler estimation to separate water level variations from other environmental motion patterns. Additionally, frequency- and spatial-domain filtering are employed to suppress interference and enhance water level feature extraction.

3) We develop a water level tracking approach using extracted phase features, offering higher sensitivity to subtle changes. Unlike deep learning-based solutions, this approach does not require pre-labeled data and can be directly deployed in various environments. This approach applies a Kalman filter-based phase unwrapping to resolve the $2\pi$ ambiguity, ensuring continuous tracking. It also leverages transceiver geometry to convert phase variations into water level heights without requiring additional calibration.

4) We develop a real-time prototype using a 5G mmWave and 4G LTE testbeds in a controlled indoor scenario, with a live demo available at \url{https://youtu.be/Mh-VjUEpSuY}. The scheme achieves water level height estimation with the average errors of 0.025 cm and 0.198 cm, respectively. Outdoor experiments collect 2.6 GHz LTE CSI data from a base station at the Parramatta River near Meadowbank in Sydney, Australia. The scheme achieves an average error of 4.8 cm for 1-meter water level change, confirming its real-world applicability.

The remainder of this paper is organized as follows. Section II introduces the CSI model and the water level sensing framework. Section III details the feature extraction process, while Section IV presents water level height tracking. Section V and Section VI evaluate the scheme in controlled and real-world environments, respectively. Section VII reviews related work, and Section VIII concludes the paper.

\section{Signal Models and Scheme Overview}
This section presents the CSI model for water level sensing and provides an overview of PMNs-WaterSense.

\subsection{CSI Model for Water Level Sensing}
\begin{figure}
\centering
\includegraphics[width=0.5\textwidth]{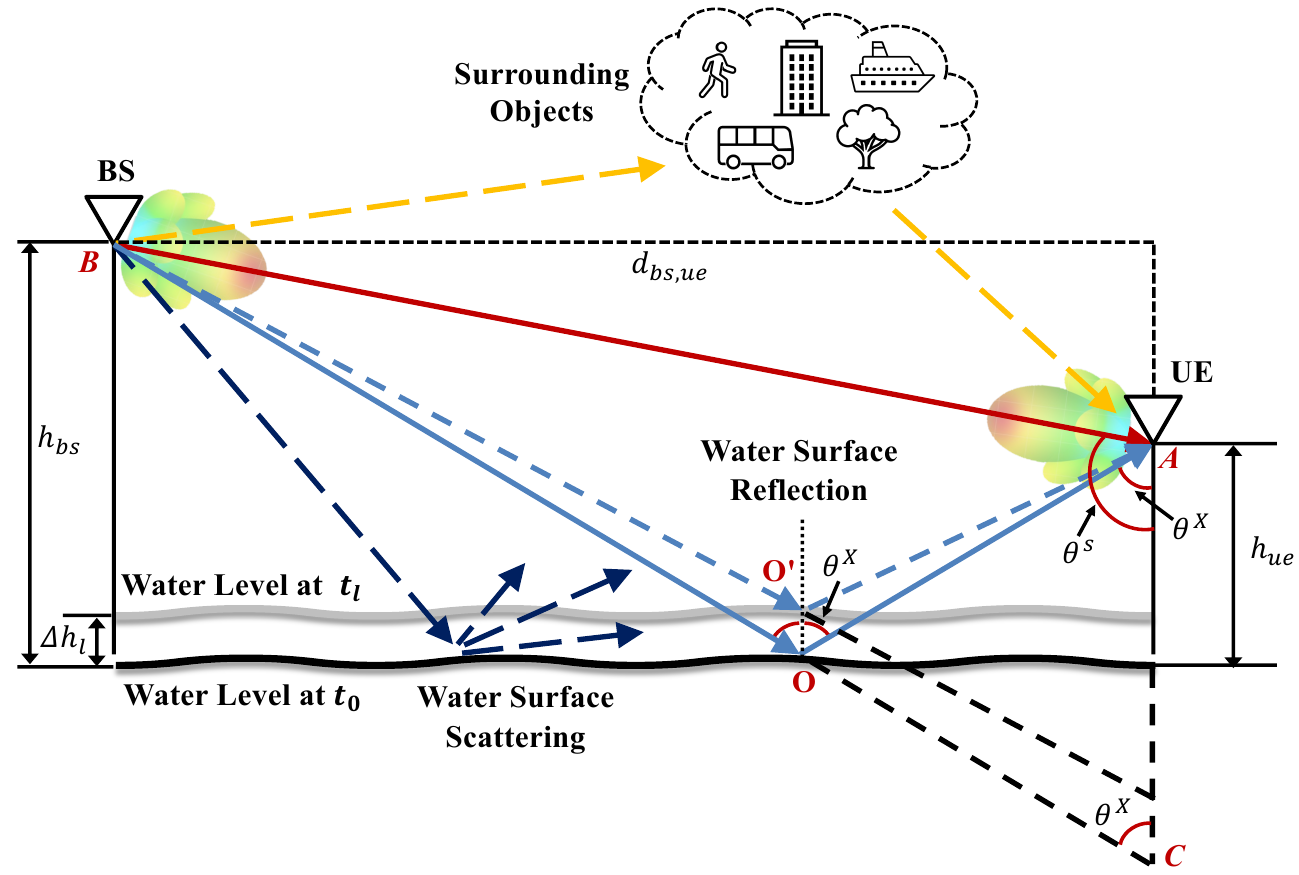}
\caption{Water sensing diagram, where the base station (BS) and the user end (UE) are widely apart as in a bistatic sensing system.}
\label{figure1}
\vspace{-1.5em}
\end{figure}

The communication signal propagation paths are illustrated in Fig. \ref{figure1}. Depending on the deployment scenario, water level sensing can be performed on either the BS or UE side, operating in both uplink and downlink modes. In downlink sensing, the BS transmits signals while the UE receives and processes CSI variations, making it suitable for mobile or low-power devices like Internet of Things (IoT) nodes. In uplink sensing, the UE transmits signals, and the BS, with its advanced processing and multi-antenna capabilities, enhances water level feature extraction. In both cases, signals propagate through multiple paths, including the direct Line-of-Sight (LOS) path (red), NLOS reflections from the water surface (blue), and scattered paths caused by water disturbances such as waves or wind (dark blue). The LOS path is beneficial but not necessary for estimating the water level. Additionally, reflections from surrounding objects (yellow), such as buildings, trees, vehicles, and ships, further contribute to multipath propagation.

In this work, the primary focus is on the NLOS path reflected from the water surface, as changes in water level over time ($t_0$ to $t_l$) result in variations in the reflection path length, with the corresponding water level height change denoted as $\Delta h_l$. In real-world scenarios, where there is a considerable distance between the BS and UE, changes in water level (e.g., rivers, lakes, reservoirs) typically result in small variations in the NLOS path length over tens of minutes. For instance, in our outdoor experiment, real-world data shows an average water level drop of 10 cm per hour. Let the path length (i.e., BS$\rightarrow$O$\rightarrow$UE) be $d^X(t_0)$ at the initial time $t_0$ and $d^X(t_l)$ at the time $t_l$. According to the Taylor series expansion, we have:
\begin{equation}
\begin{aligned}
d^X(t_l) &\approx d^X(t_0) + \frac{\partial d^X(t_0)}{\partial t} \cdot \left(l-1\right) \cdot \Delta t \\
&= d^X(t_0) + v^X \cdot \left(l-1\right) \cdot \Delta t,
\end{aligned}
\label{equation1}
\end{equation}
where $\Delta t$ is the time interval between adjacent samples, and $v^X$ is the slow rate of NLOS path length change. The corresponding frequency shift, known as the \textit{slow-time Doppler Frequency Shift (DFS)}, is denoted as $f^D$ and given by $f^D = {v^X f_c}/{c}$, where $c$ represents the speed of light, and $f_c$ is the carrier frequency of the transmitted signal.

Within a time window, let $\mathit{CSI}_{i,j,k}$ be the CSI from the $i$-th antenna and the $j$-th subcarrier at the time $t_k$ collected at the receiver. Then, we have:
\begin{equation}
\resizebox{\linewidth}{!}{$
\begin{aligned}
&\mathit{CSI}_{i,j,k} = \rho_k^{\text{tx}} \rho_k^{\text{agc}} H_i^{e} H_{j,k}^{h} \left(H_{i,j}^{S} + H_{i,j,k}^{X}\right) \\
&= \underbrace{\gamma_k}_{\text{Power}} \underbrace{e^{-J \left(\phi_{j,k}^{\text{TO}} + \phi_k^{\text{CFO}}\right)}}_{\text{Phase Offsets}} \underbrace{e^{-J\phi_i^{\text{h}}}}_{\text{Hardware}} \\
&\quad \Bigg( \underbrace{\sum_{l_1} \rho_{i,j}^{S}[l_1]  \underbrace{e^{-J2\pi f_j \tau_{l_1}^S}}_{\text{Delay}}   \underbrace{e^{-J\pi(i-1) \sin\theta^S[l_1]}}_{\text{AoA}}}_{\text{Static Components}} \\
&\quad + \underbrace{\sum_{l_2} \rho_{i,j,k}^{X}[l_2] \underbrace{e^{-J2\pi f_j \tau_{l_2}^X}}_{\text{Delay}} \underbrace{e^{-J2\pi f^D[l_2](k-1)\Delta t}}_{\text{Slow-Time Doppler}} \underbrace{e^{-J\pi(i-1) \sin\theta^X[l_2]}}_{\text{AoA}}}_{\text{Dynamic Components}}\Bigg),
\end{aligned}
$}
\label{equation2}
\end{equation}
where $J$ is the imaginary unit, and $f_j$ represents the subcarrier frequency. Other parameters are defined as follows:

\begin{figure*}
\centering
\includegraphics[width=\textwidth]{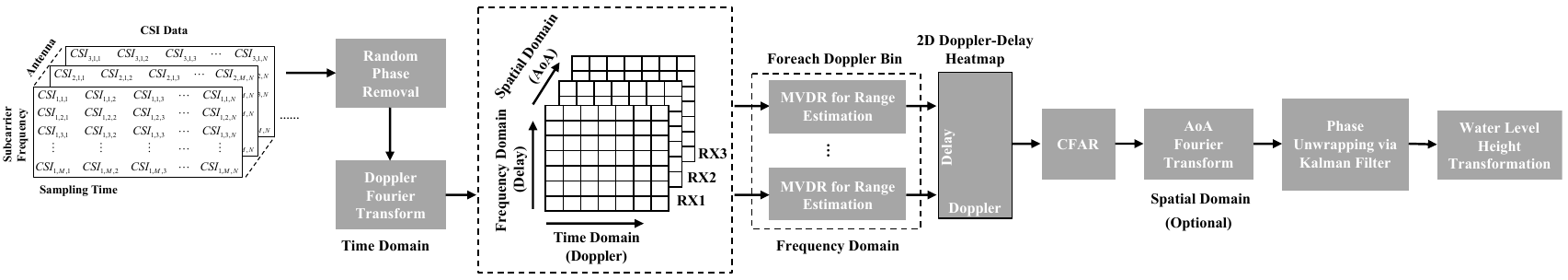}
\caption{Signal processing workflow.}
\label{figure2}
\vspace{-1.5em}
\end{figure*}

1) $\gamma_k = \rho_k^{\text{tx}} \gamma_k^{\text{agc}}$ represents the time-varying power factors at the transmitter and receiver. Transmission power control adjusts signal strength for reliable communication, while the automatic gain control (AGC) fine-tunes receiver gain to stabilize signal levels. Both introduce unintended CSI variations due to dynamic signal conditions. However, in environments with water bodies such as rivers, lakes, and reservoirs, the open surroundings and gradual water level changes may result in relatively smooth power variations, allowing us to assume power fluctuations remain stable within the time window.

2) $H_{j,k}^e = e^{-J(\phi_{j,k}^{\text{TO}} + \phi_k^{\text{CFO}})}$ is the random phase offsets caused by TO and CFO due to the transceiver asynchronization.

3) $H_i^h = e^{-J\phi_i^h}$ represents the phase offset of an antenna channel, caused by the antenna cable and other hardware components. Notably, since the receiver's voltage-controlled oscillator (VCO) starts with a random initial phase at each power-up, the PLL component cannot fully correct it \cite{tadayon2019decimeter}. Most existing systems require calibration before each operation to ensure accurate AoA estimation.

4) $H_{i,j}^s$ denotes static components from surrounding reflections off the ground, buildings, bridges, etc. In its expanded form, $l_1$ is the $l_1$-th static path, $\rho_{i,j}^s[l_1]$ is its amplitude, $\tau^S[l_1]$ is the propagation delay, and $\theta^S[l_1]$ is the AoA at the receiver.  

5) $H_{i,j,k}^X$ represents dynamic components from water level variations and moving objects like cars, ferries, and pedestrians. $l_2$ denotes the $l_2$-th dynamic path, $\rho_{i,j,k}^X[l_2]$ is its amplitude, $\tau^X[l_2]$ is the initial NLOS propagation delay, and $\theta^X[l_2]$ is the dynamic NLOS AoA. These components capture environmental variations crucial for sensing. 

For accurate water level sensing, the impact of power variations, TO, CFO, and hardware imbalance should be eliminated. Moreover, it is essential to isolate water surface reflections from static and irrelevant dynamic components. 

\subsection{Overview of the Proposed Water Sensing Scheme}
Our scheme, PMNs-WaterSense, leverages CSI measurements from communication signals for water level monitoring without modifying existing infrastructure or requiring prior training datasets. It supports various frequency bands, center frequencies, single- or multi-antenna receivers, base station locations, and environmental conditions, ensuring broad adaptability. Since water level changes lead to gradual length variations of the signal reflection path, the scheme can suppress unrelated environment interference and isolate the dominant signal from the water surface for accurate sensing. Fig. \ref{figure2} shows the main sensing modules as summarized below.

\textit{1) Water Level Feature Extraction.} In Section III, we first eliminate the impact of random phase while preserving the linear relationships of CSI signal features. The signal features are then extracted across time, frequency, and spatial domains. The process includes Doppler estimation (time-domain Fourier transform), delay estimation (frequency-domain MVDR), CFAR detection, and spatial filtering (spatial-domain Fourier transform, skipped in single-antenna setups). 

\textit{2) Water Level Height Tracking.} In Section IV, this step extracts phase variations from the NLOS signal reflected off the water surface, applies phase unwrapping to resolve the $2\pi$ ambiguity, and converts the results into water level heights.

\section{Water Level Feature Extraction}
In this section, we first remove random offsets and then suppress noise across the time, frequency, and spatial domains to extract features associated with water level changes.

\subsection{Exploitation of Information-Bearing CSI Power}
Most existing methods \cite{wang2022single, wang2024passive, zeng2019farsense} rely on multiple receiver antennas to mitigate random CSI phase offsets caused by asynchronous clocks between the transmitter and receiver. However, they are limited in single-antenna setups and require complex antenna calibration. This includes compensating for hardware variations and addressing random initial phase shifts from the PLL, complicating spatial AoA filtering. To provide a simpler, universal solution, we propose to use CSI power for sensing by multiplying the CSI signal with its conjugate,
\begin{equation}
\resizebox{\linewidth}{!}{$
\begin{aligned}
& \text{P}_{i,j,k}=CSI_{i,j,k} \overline{CSI}_{i,j,k} \\
& = \gamma_k^2 H^e_{j,k} H^{h}_i \left( H^S_{i,j} + H^X_{i,j,k} \right) 
\overline{H}^e_{j,k} \overline{H}^{h}_i \left( \overline{H}^S_{i,j} + \overline{H}^X_{i,j,k} \right) \\
&= \gamma_k^2 \left[ \left\| H^S_{i,j} \right\|^2 + \left\| H^X_{i,j,k} \right\|^2 + 2 \left\| H^S_{i,j} \overline{H}^X_{i,j,k} \right\| \cos \left( \angle H^S_{i,j} \overline{H}^X_{i,j,k} \right)  \right]
\end{aligned}
$}
\label{equation3}
\end{equation}
where $\left\| \cdot \right\|$ denotes the magnitude, and $\left\| H^e_{j,k} H^h_i \right\| = 1$.

According to Eq. (\ref{equation2}), we have
\begin{equation}
\begin{aligned}
& \left\| H_{i,j}^S H_{i,j,k}^X \right\| \cos \left( \angle H_{i,j}^S H_{i,j,k}^X \right) = \\
& \quad \sum_{l_1} \sum_{l_2} \rho_{i,j}^S[l_1] \rho_{i,j,k}^X[l_2] \cos \big( 
    \varphi^{\text{Doppler}}_{k}[l_2]+ \varphi^{\text{Delay}}[l_1,l_2] + \\
& \quad\quad\quad\quad\quad\quad \quad\quad\quad\quad\quad\quad\quad  \varphi^{\text{AoA}}_{i,j}[l_1,l_2] + \varphi[l_1,l_2] 
\big),
\end{aligned}
\label{equation4}
\end{equation}
where the intermediate variables are given by 
\begin{equation}
\left\{ 
\begin{aligned}
&\varphi^{\text{Doppler}}_{k}[l_2] = 2 \pi f^D[l_2] \left(k - 1\right) \Delta t, \\
&\varphi^{\text{Delay}}_j{[l_1,l_2]} = 2 \pi \Delta f \left(j - 1\right)\left( \tau^X[l_2] - \tau^S[l_1] \right), \\
&\varphi^{\text{AoA}}_{i,j}[l_1,l_2] = \pi \left(i - 1\right) \left( \sin \theta^X[l_2] - \sin \theta^S[l_1] \right), \\
&\varphi[l_1,l_2] = 2 \pi f_c \left( \tau^X[l_2] - \tau^S[l_1] \right).
\end{aligned}
\right.
\label{equation5}
\end{equation}
Here, $\Delta f$ represents the subcarrier frequency spacing.

Jointly inspecting Eq.(\ref{equation3})--Eq.(\ref{equation5}), we can conclude that: (1) $\gamma_k$ varies slowly over time as it reflects gradual adjustments in transmission power and AGC. In the following feature extraction, a window function is applied to reduce its impact. (2) $\scriptstyle \left\| H^S_{i,j} \right\|$ represents static reflections from stationary objects (e.g., ground, buildings). Within the time window, these static components can be assumed to remain constant. (3) When a LOS path exists between the BS and UE, its power dominates over other reflections. In the absence of an LOS path, the signal propagation between the BS and UE generally undergoes fewer bounces than reflections from the water surface. As a result, $\scriptstyle{ \left\| H^S_{i,j} \overline{H}^X_{i,j,k} \right\|}$ often remains stronger than $\scriptstyle \left\| H^X_{i,j,k} \right\|^2$, allowing the latter to be treated as noise in subsequent processing. (4) While CSI power contains only the real part of the signal, it retains essential phase information and the inherent linear dependencies among features involving Doppler, delay, and AoA. This ensures that the key signal characteristics for water level sensing can be accurately extracted.

\subsection{Time-domain Doppler Fourier Transform}
In real-world environments, water level changes occur much more slowly than the motion of dynamic objects such as vehicles, pedestrians, or wind-induced disturbances. This difference allows us to differentiate DFS components related to water level variations from those of faster-moving objects in the frequency domain. Within a time window of tens of minutes, for the $i$-th Rx antenna and the $j$-th subcarrier, we first subtract the mean of CSI over time to suppress static components and highlight dynamic variations:
\begin{equation}
\mathcal{P}_{i,j,k} =  \text{P}_{i,j,k} - \frac{1}{L} \sum_{k=0}^{L-1}  \text{P}_{i,j,k},
\label{equation6}
\end{equation}
where $L$ is the total number of samples in the time window.

Due to power variations over time, static components cannot be entirely removed, introducing residual low-frequency components. To further mitigate this effect, we apply a window function $\mathsf{w}_k$ (Hamming window in this work) when performing the Doppler Fast Fourier Transform (FFT) along the CSI sampling time. The FFT result at a Doppler bin $f^D$ is:
\begin{equation}
\begin{aligned}
X_{i,j}\left(f^D\right) &=\mathcal{F}\left\{ \mathsf{w}_k \cdot \mathcal{P}_{i,j,k} \right\} \\
&=
\begin{cases}
w H^S_{i,j}\sum \limits_{l} \rho_{i,j}^X[l] e^{J \varphi_{i,j}^X[l] } + \mathcal{N}_{i,j}, & f^D > 0 \\ 
0, & f^D = 0 \\ 
w \overline{H}^S_{i,j} \sum \limits_{l} \rho_{i,j}^X[l] e^{-J \varphi_{i,j}^X[l] } + \overline{\mathcal{N}}_{i,j}, & f^D < 0,
\end{cases}
\end{aligned}
\label{equation7}
\end{equation}
where
\begin{equation}
\varphi_{i,j}^X = 2 \pi \Delta f \left(j - 1\right)\tau^X + \pi \left(i - 1\right) \sin \theta^X + 2 \pi f_c \tau^X,
\label{equation8}
\end{equation}
and $w$ denotes the window function term after FFT.

Here are some considerations: (1) Multiple targets at different delays and directions can exhibit the same Doppler velocity in a given bin; (2) Since CSI power retains only the magnitude (real value), Doppler ambiguity prevents distinguishing whether the water level is rising or falling; (3) CSI sampling can be uneven over time, depending on actual hardware platforms. The solutions to the first two issues will be provided next. Here, we further note that the third issue can be tackled by applying the non-uniform fast Fourier transform (NUFFT) \cite{fessler2003nonuniform}. Any windowing function applied needs to match the non-uniform sampling timestamps.

\subsection{Frequency-domain Delay MVDR}
To resolve the Doppler ambiguity arising from the conjugate symmetry of Doppler FFT results for $+f^D$ and $-f^D$, which have the same amplitude but opposite phases, we resort to the delay information in the frequency domain. For each Doppler bin, we apply the MVDR algorithm to estimate the power spectrum, achieving a more concentrated power distribution across various delays, as follows:
\begin{equation}
P_{\text{mvdr}} \left( \Delta \tau\right) = \frac{1}{\left\| \bm{\mathsf{a}}^\mathsf{H}\left( \Delta \tau\right) \bm{R}^{-1} \bm{\mathsf{a}}\left( \Delta \tau\right) \right\|},
\label{equation9}
\end{equation}
where $\bm{\mathsf{a}}$ is the steering vector for $M$ subcarriers:
\begin{equation}
\bm{\mathsf{a}}\left( \Delta \tau\right) = \left[ 1 \quad e^{-J2\pi \Delta f \Delta \tau} \quad \cdots \quad e^{-J2\pi \Delta f  \left(M-1\right) \Delta \tau} \right]^\mathsf{T},
\label{equation10}
\end{equation}
and $\Delta \tau = \tau^X - \tau^S > 0$ because the NLOS path reflected from the water surface is expected to be longer than either the direct path between the BS and UE or other dominant static NLOS paths. Additionally, $\bm{R} = E\left( \bm{\Lambda} \bm{\Lambda}^\mathsf{H} \right)$ is the covariance matrix of the observation matrix $\bm{\Lambda}$, given by:
\begin{equation}
\bm{\Lambda}=
\begin{bmatrix}
X_{1,1}[f^D] & X_{2,1}(f^D) & \hdots & X_{N,1}(f^D) \\
X_{1,2}[f^D] & X_{2,2}(f^D) & \hdots & X_{N,2}(f^D) \\
\vdots & \vdots & \vdots & \vdots \\
X_{1,M}[f^D] & X_{2,M}(f^D) & \hdots & X_{N,M}(f^D)
\end{bmatrix}_{M \times N},
\label{equation11}
\end{equation}
where $N$ is the number of antennas, forming the $M \times N$ observation matrix. We apply the forward-backward smoothing \cite{shan1985spatial} to enhance the accuracy of the covariance matrix estimation. When only a single antenna is used, the covariance matrix reduces to an outer product of the Doppler FFT results. This process is repeated for each Doppler bin to generate a Doppler-Delay heatmap, where, to enhance clarity, we express delay in terms of range. The range denotes the difference between the water surface reflection path and the strongest static path (e.g., the LOS path between the BS and UE).

\begin{figure}
\centering
\begin{subfigure}{0.24\textwidth}
	\centering
	\includegraphics[width=\textwidth]{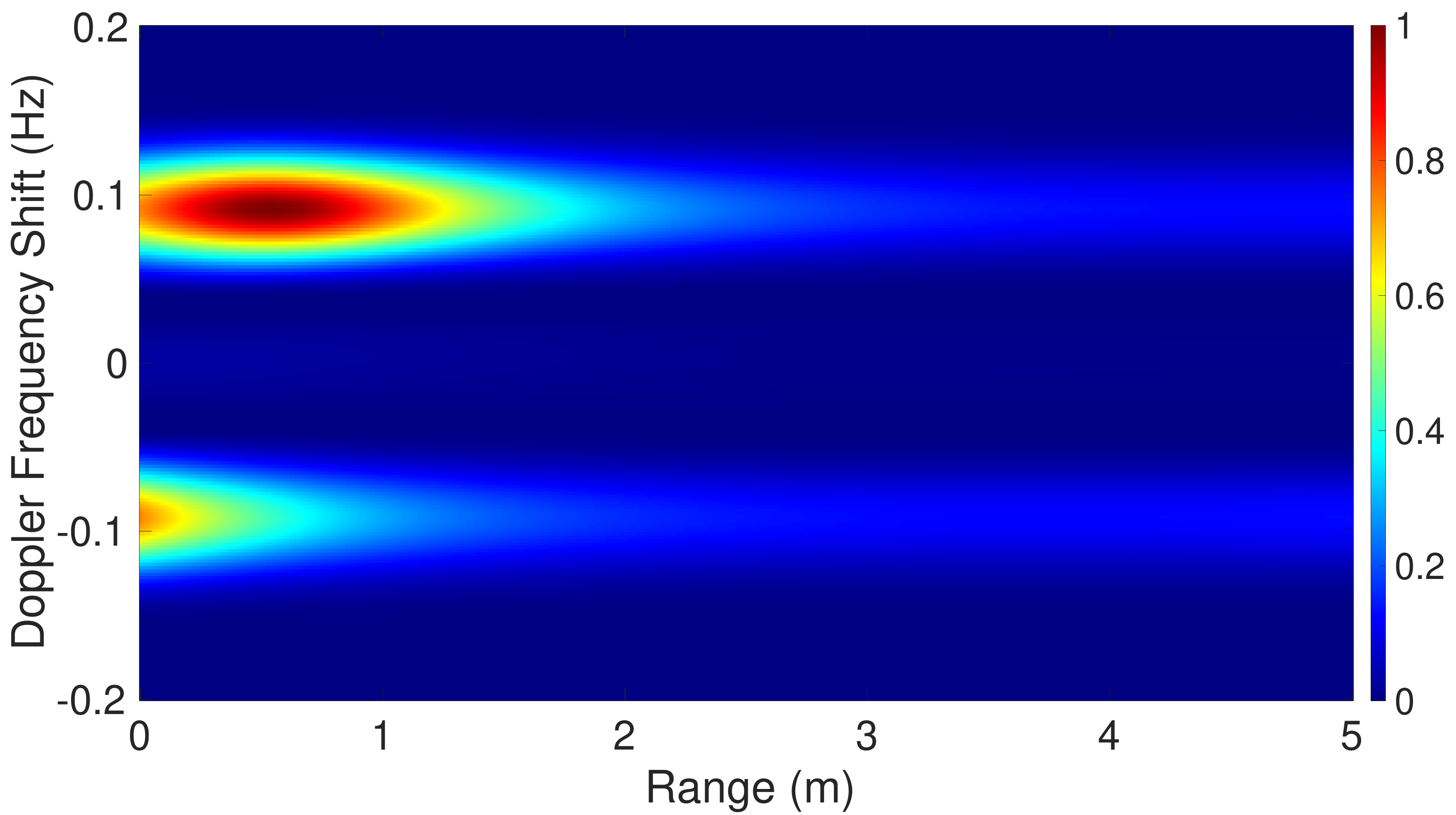}
	\subcaption{28 GHz mmWave signal}
	\label{Fig3b}
\end{subfigure}
\begin{subfigure}{0.24\textwidth}
	\centering
	\includegraphics[width=\textwidth]{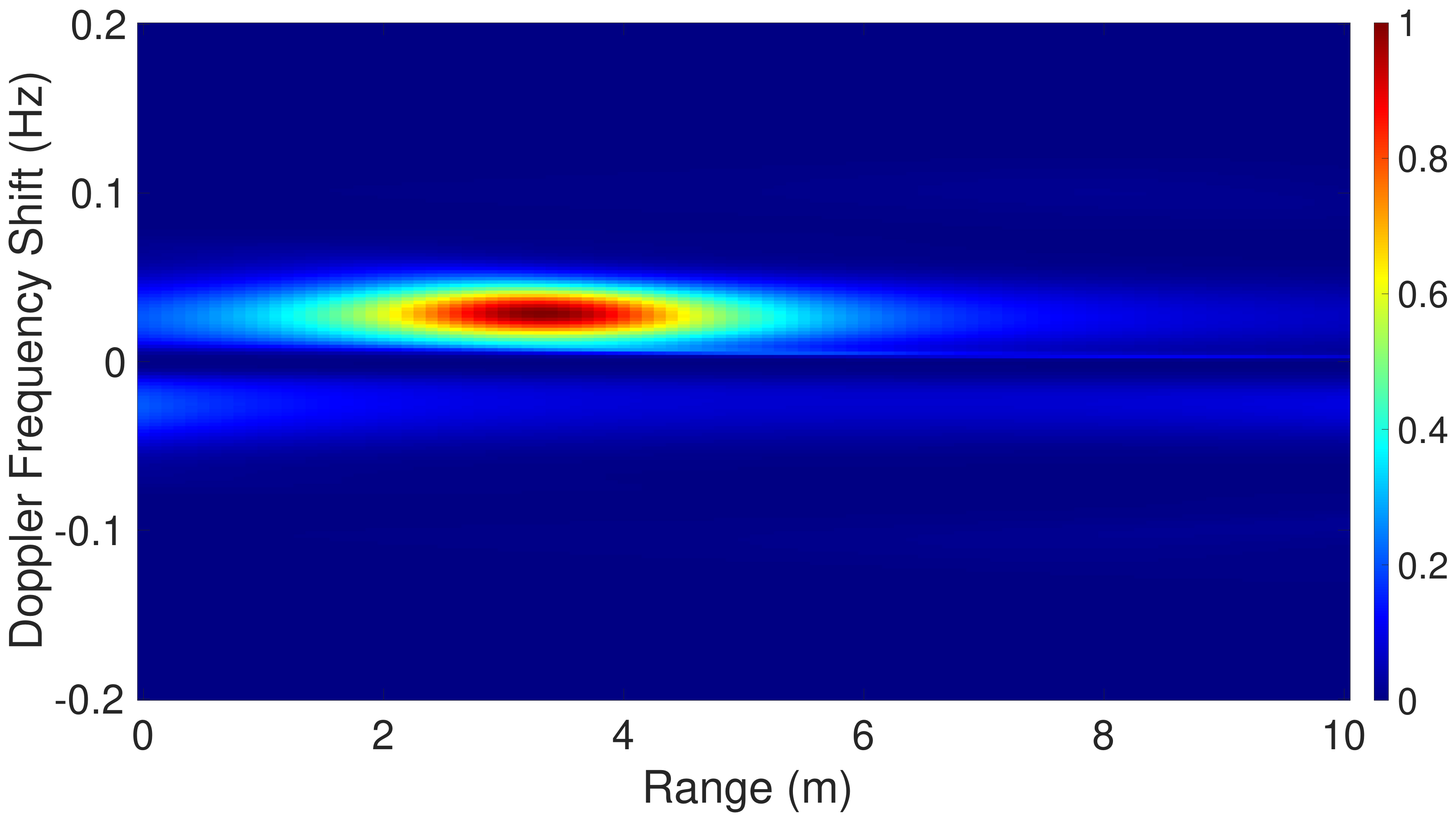}
	\subcaption{3.1 GHz LTE signal}
	\label{Fig3a}
\end{subfigure}
\caption{Doppler-Range heatmap.}
\label{Fig3}
\vspace{-1.5em}
\end{figure}

Here we employ two testbeds to generate a 28 GHz mmWave signal and a 3.1 GHz LTE signal, respectively, for water level sensing with different sampling frequency (see Section V), where the mmWave setup consists of a single antenna, while the LTE setup utilizes three antennas. Fig. \ref{Fig3} shows the Doppler-Range heatmaps based on the proposed method. By leveraging subcarrier information, Doppler ambiguity is effectively suppressed. The red peaks represent high-power reflections, highlighting potential water surface reflections. Additionally, mmWave provides higher accuracy in range estimation due to its shorter wavelength.

\subsection{CFAR for Water Level Variation Detection}  
Accurately detecting water level variations is essential, as a static water surface primarily contributes to background noise, potentially causing false detections. We first convert the 2D Doppler-Range heatmap into a 1D Doppler profile by averaging power across the range dimension based on the finer Doppler resolution. A 1D average CFAR algorithm is then applied to detect water level changes by identifying Doppler peaks. The algorithm dynamically adjusts the detection threshold for improving robustness against noise.

\subsection{Feature Extraction for Water Level Variation}
Once the CFAR detects changes, we identify peaks at specific Doppler and range bins in the Doppler-Range heatmap. The scheme assumes the bins with the highest power correspond to water surface reflections and uses them to compute MVDR beamformer weights for precise feature extraction:
\begin{equation}
\mathbf{w}_{\text{delay}} = \frac{\mathsf{R}^{-1} \bm{\mathsf{a}}}{\bm{\mathsf{a}}^\mathsf{H} \mathsf{R}^{-1} \bm{\mathsf{a}}}.
\label{equation12}
\end{equation}
When only a single antenna is available, the limited number of snapshots in the observation matrix introduces high noise into the beamforming weights. In this case, employing the delay-and-sum beamformer, where the weights are defined as $\mathbf{w}_{\text{delay}} = \bm{\mathsf{a}}(\Delta \tau)$, provides a more stable alternative. Additionally, since water level changes occur gradually, the Doppler and range estimates remain stable within adjacent windows. To ensure accurate water level tracking, an outlier removal can be applied to identify the reliable Doppler and range bins based on the measurements in a short-time duration. 

For each antenna, these weights are applied to combine CSI measurements across all subcarriers, yielding the extracted dynamic components:
\begin{equation}
\begin{aligned}
Y_i(f^D, \Delta \tau) &= \sum_{j=1}^{M} \mathbf{w}_{\text{delay}}^\mathsf{H} (\Delta \tau) X_{i,j}(f^D) \\
&\approx \rho_{i} e^{J  \left[ 2 \pi f_c(\tau^X - \tau^S) + \pi (i - 1) \left( \sin \theta^X - \sin \theta^S \right) \right]}.
\end{aligned}
\label{equation13}
\end{equation}

When the water level changes, the complex signal feature $Y_i(f^D, \Delta \tau)$ vary accordingly: (1) Phase. When the water level rises, the reflection path from the water surface shortens, resulting in a decrease in the phase; conversely, when the water level drops, the reflection path increase, leading to an increase in the phase. Due to $2\pi$ phase periodicity, precise phase unwrapping is required for continuous tracking of water level changes over time (see Section IV). (2) Amplitude. The amplitude $\rho_{i}$ theoretically varies inversely with the path length. The amplitude does not suffer from the ambiguity issue, but it offers lower resolution and accuracy for detecting water level changes compared to phase information.

\subsection{Extension to Multi-Antenna Setups for Spatial Filtering}
With multiple Rx antennas, spatial filtering can further suppress noise in extracting the NLOS path reflected from the water surface. Let the extracted signals at a specific Doppler and range bin across all antennas be:
\begin{equation}
\bm{Y} = \left[ Y_{1}(f^D, \Delta \tau),Y_{2}(f^D, \Delta \tau),\cdots, Y_{N}(f^D, \Delta \tau) \right],
\label{equation14}
\end{equation}
For the $1 \times N$ vector $\bm{Y}$, an FFT is performed along the antenna dimension. The spatial AoA FFT result $Z(f^D, \Delta \tau, \theta)$ with the highest amplitude is considered the refined feature from the water surface reflection:
\begin{equation}
Z(f^D, \Delta \tau, \theta) \approx  \rho e^{J 2 \pi f_c(\tau^X - \tau^S)}.
\label{equation15}
\end{equation}
The beamformed result $Z(f^D, \Delta \tau, \theta)$ theoretically achieves a higher SNR due to spatial combining gain. The proposed time-frequency-domain processing methods effectively suppress interference within the identified Doppler and range bins for water sensing. Additionally, joint time-frequency-space-domain processing could further improve water reflection detection and enhance water level sensing accuracy. As this work aims to establish a unified sensing framework, we shall leave more comprehensive development to future work.


\section{Water Level Height Tracking}  
Phase generally provides higher sensitivity to subtle variations compared to amplitude, making it more suitable for precise water level tracking. Thus, we leverage the phase to achieve continuous monitoring of water level height changes.

\subsection{Phase Unwrapping with Kalman Filtering}  
The extracted phase $\Delta \phi$ is within $[-\pi, \pi]$, introducing a $2\pi$ ambiguity in estimating the NLOS path length variations. To resolve this, we propose to use the Kalman filter for phase unwrapping in continuous water sensing.

\textit{1) State Initialization.} 
The Kalman filter's initial state vector is defined as $\mathbf{x}_k = [\Delta \phi_k]$, representing the raw phase. The state transition matrix $\mathbf{F} = [1]$ assumes that the phase changes smoothly over time with minimal abrupt variations.

\textit{2) State Prediction.} The predicted state $\mathbf{\hat{x}}_{k+1}$ is derived as:
\begin{equation}
\mathbf{\hat{x}}_{k+1} = \mathbf{F} \mathbf{x}_k.
\label{equation16}
\end{equation}
The predicted covariance matrix $\mathbf{P}_{k+1}$ is updated as:
\begin{equation}
\mathbf{P}_{k+1} = \mathbf{F} \mathbf{P}_k \mathbf{F}^\mathsf{T} + \mathbf{Q},
\label{equation17}
\end{equation}
where $\mathbf{Q}$ denotes the process noise covariance, accounting for uncertainties in the state transition model. In our implementation, $\mathbf{Q}$ is set to $0.01$.

\textit{3) Phase Unwrapping and Residual Correction.} 
Residual computation is critical for addressing phase discontinuities caused by the $2\pi$ ambiguity. In each iteration, the residual is calculated by comparing the predicted unwrapped phase $\Delta \hat{\phi}_k$ with the measured wrapped phase $\Delta \phi_k$. To maintain phase continuity, the residual is first adjusted to remain within the valid range of $[-\pi, \pi]$ using a modulo operation:
\begin{equation}
\Delta \hat{\phi}_k = \text{mod}(\Delta \hat{\phi}_k + \pi, 2\pi) - \pi.
\label{equation18}
\end{equation}
Then, the initial residual is computed as:
\begin{equation}
\Delta \phi_\text{residual} = \Delta \phi_k - \Delta \hat{\phi}_k.
\label{equation19}
\end{equation}
If the residual exceeds $\pi$ or falls below $-\pi$, a $2\pi$ correction is applied to bring it back into the valid range. The correction factor $\Delta \mathbb{K}_k$ is determined as follows:
\begin{equation}
\Delta \mathbb{K}_k = 
\begin{cases}
-1, & \Delta \phi_\text{residual} > \pi, \\
0, & -\pi \leq \Delta \phi_\text{residual} \leq \pi, \\
1, & \Delta \phi_\text{residual} < -\pi.
\end{cases}
\label{equation20}
\end{equation}

Thus, the corrected residual is expressed as:
\begin{equation}
y_k = \Delta \phi_\text{residual} + 2\pi \Delta \mathbb{K}_k.
\label{equation21}
\end{equation}

\textit{4) Kalman Gain Calculation.} The Kalman gain, $\mathbf{K}_{k+1}$, determines the balance between the predicted state and the new measurement, which is computed as:
\begin{equation}
\mathbf{K}_{k+1} = \mathbf{P}_{k+1} \mathbf{H}^\mathsf{T} \left(\mathbf{H} \mathbf{P}_{k+1} \mathbf{H}^\mathsf{T} + \mathbf{R}\right)^{-1},
\label{equation22}
\end{equation}
where $\mathbf{H} = \mathbf{I}$ is the identity matrix, reflecting a direct measurement model, and $\mathbf{R}$ is the measurement noise covariance and is set to $0.25$ in our implementation.

\textit{5) State Update.} The state is updated using the Kalman gain and the corrected residual:
\begin{equation}
\mathbf{x}_{k+1} = \mathbf{\hat{x}}_{k+1} + \mathbf{K}_{k+1} y_k,
\label{equation23}
\end{equation}
and the updated covariance matrix is:
\begin{equation}
\mathbf{P}_{k+1} = (\mathbf{I} - \mathbf{K}_{k+1} \mathbf{H}) \mathbf{P}_{k+1}.
\label{equation24}
\end{equation}

The Kalman filter factors in the localized changing trend of the water level and fuses the measurements with prediction to reduce noise, which can effectively solve the 2$\pi$ phase jumping issue. This ensures continuous and accurate tracking of path variations for unambiguous water level sensing.

\begin{figure*}
\centering
\includegraphics[width=0.95\textwidth]{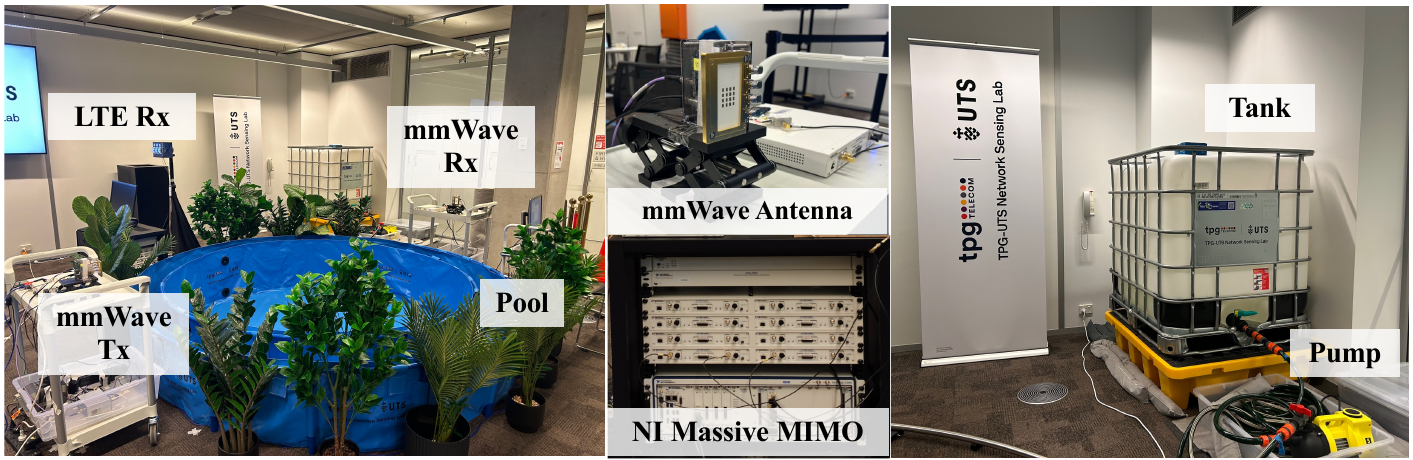}
\caption{Experiment setup in a controlled lab environment for water level sensing using mmWave and LTE signals.}
\label{figure4}
\vspace{-1.5em}
\end{figure*}

\subsection{Water Level Height Transformation} 
Once the unwrapped phase is obtained, it can be converted into actual water level height changes using the angle $\theta^X$. 

\textit{1) With Precise AoA Estimation.} If the reflection angle $\theta^X$ can be accurately estimated using multi-antenna arrays (e.g., in uplink sensing), we can directly use it to compute the water level change. In the scenarios where water level variations within a minute-level time window are minimal, as is typical for rivers and reservoirs, $\theta^X$ can be approximated as constant. Under this assumption, based on the geometric relationship in Fig.~\ref{figure1}, the water level change $\Delta h$ between two adjacent time windows, $l$ and $l+1$, is estimated as:
\begin{equation}
\Delta h \approx \frac{\lambda}{4\pi}  \frac{\Delta \phi_{l+1} -\Delta \phi_{l} }{\sin \theta^{X}},
\label{equation25}
\end{equation}
where $\lambda$ is the wavelength, $\Delta \phi_{l+1}$ is the unwrapped phase, and $\theta^{X}$ should be updated for a long-term tracking.

\textit{2) Without Precise AoA Estimation.} In downlink sensing, most UE devices are equipped with a single or only a few antennas, making precise AoA estimation challenging. Instead, the angle $\theta^X$ can be inferred from the geometric relationship between the signal path and transceiver configuration. As shown in Fig.~\ref{figure1}, let the BS height be $h_{bs}$, UE height be $h_{ue}$, and the horizontal distance between them be $d_{bs,ue}$\footnote{BS locations and heights are publicly available via the Australian Communications and Media Authority (ACMA) at \url{https://web.acma.gov.au}.}. The NLOS path is assumed to originate from the strongest reflection off the water surface, satisfying $OC = OA$ and $AC = 2h_{ue}$. The reflection angle $\theta^X$ is approximated as:
\begin{equation}
\theta^X = \arctan \left(\frac{d_{bs,ue}}{h_{bs} + h_{ue}}\right).
\label{equation26}
\end{equation}
Substituting the estimated angle into Eq. (\ref{equation25}) leads to the water level height estimation.

\section{Experiments in a Controlled Lab Environment}
This section presents experimental results in a lab environment, where comprehensive performance can be tested using controllable configurations.

\begin{figure*}
     \centering
		  \begin{minipage}[b]{0.245\linewidth}  
			\centering
    		\begin{subfigure}{\textwidth}
        		\centering
        	\includegraphics[width=1.01\textwidth]{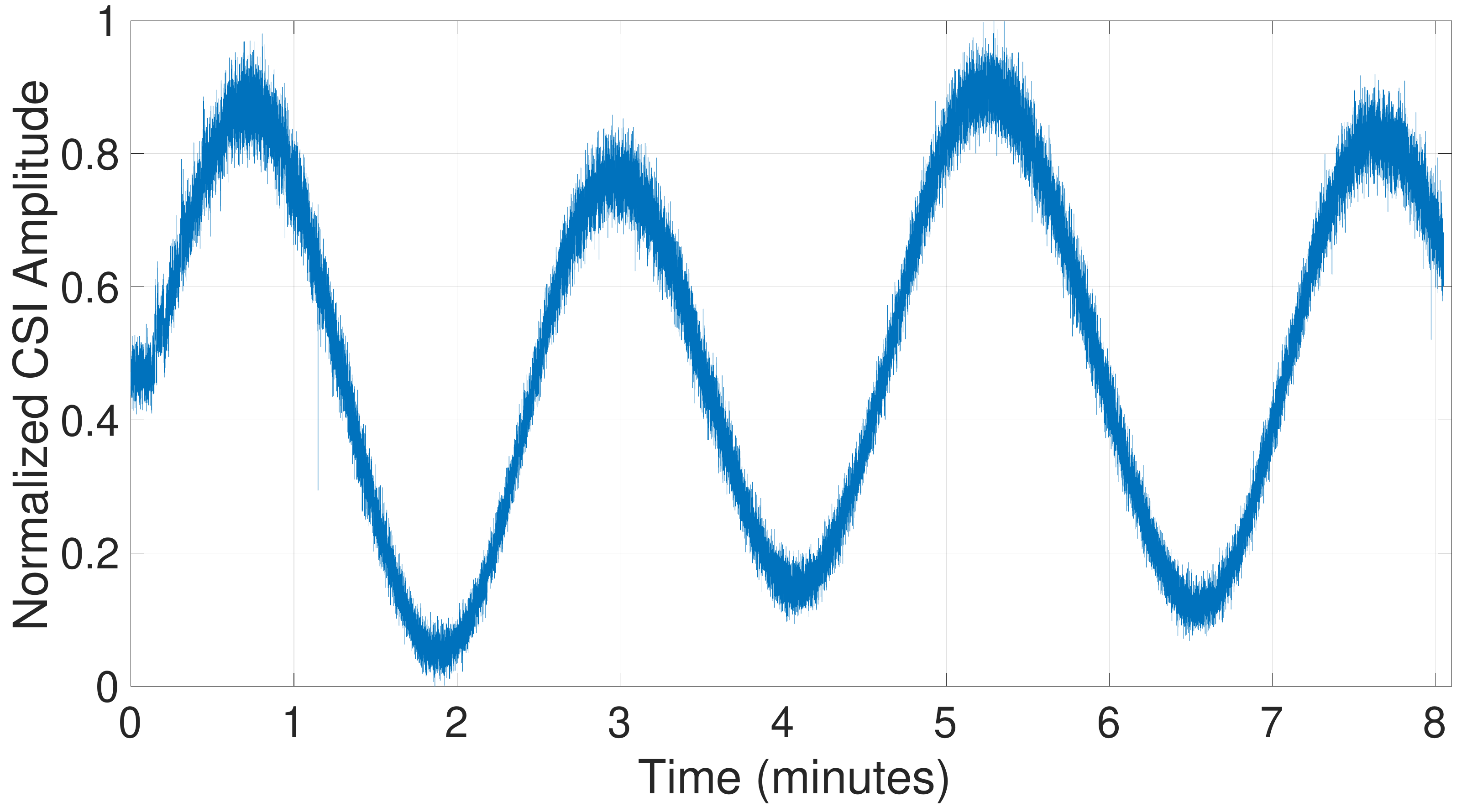}
        		\subcaption{Raw mmWave CSI}
        		\label{Fig5a}
    		\end{subfigure}\\
\begin{subfigure}{\textwidth}
        		\centering
        	\includegraphics[width=1.01\textwidth]{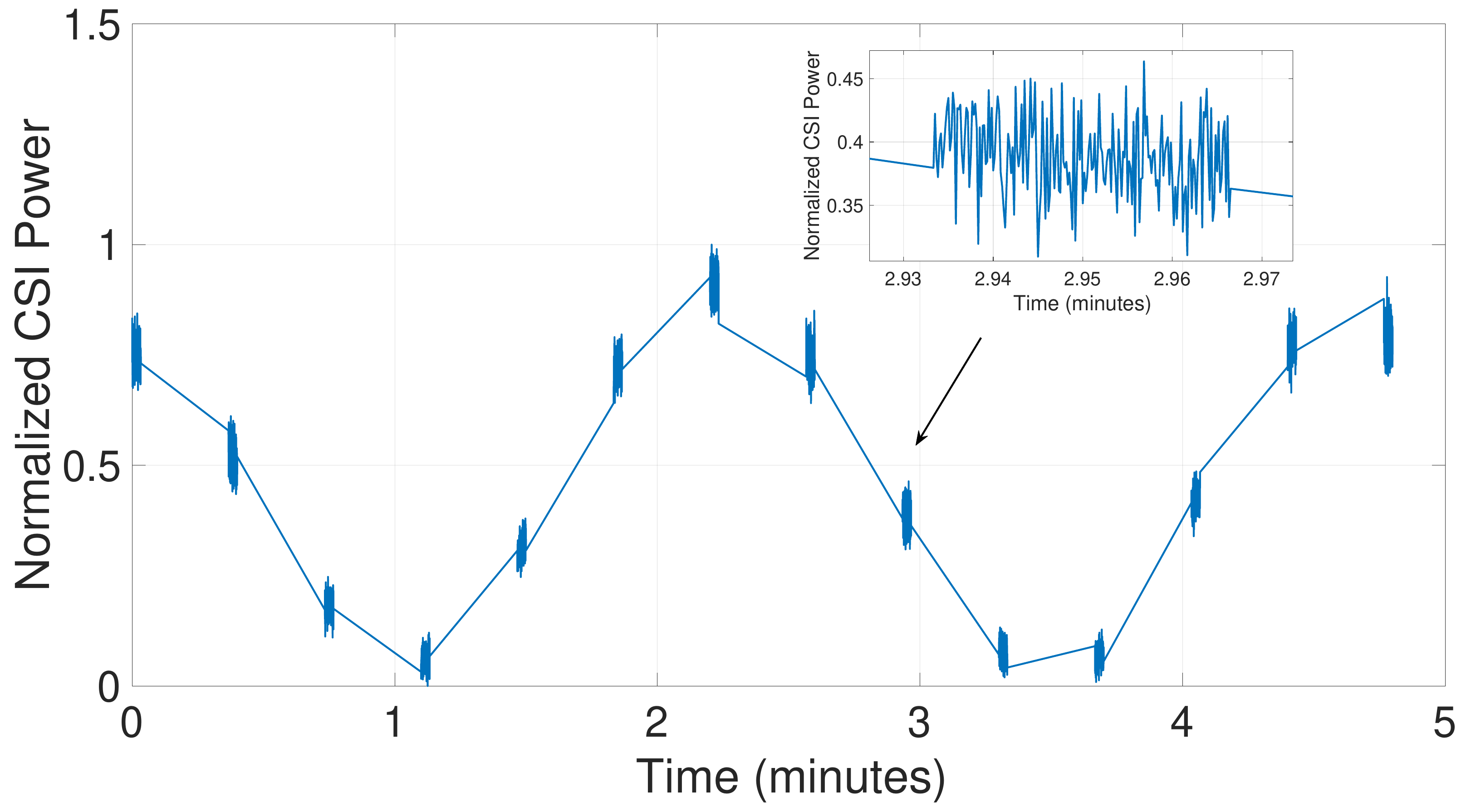}
        		\subcaption{Downsampled mmWave CSI \\ \centering in a time window}
        		\label{Fig5b}
    		\end{subfigure}\\
\begin{subfigure}{\textwidth}
        		\centering
        	\includegraphics[width=1.01\textwidth]{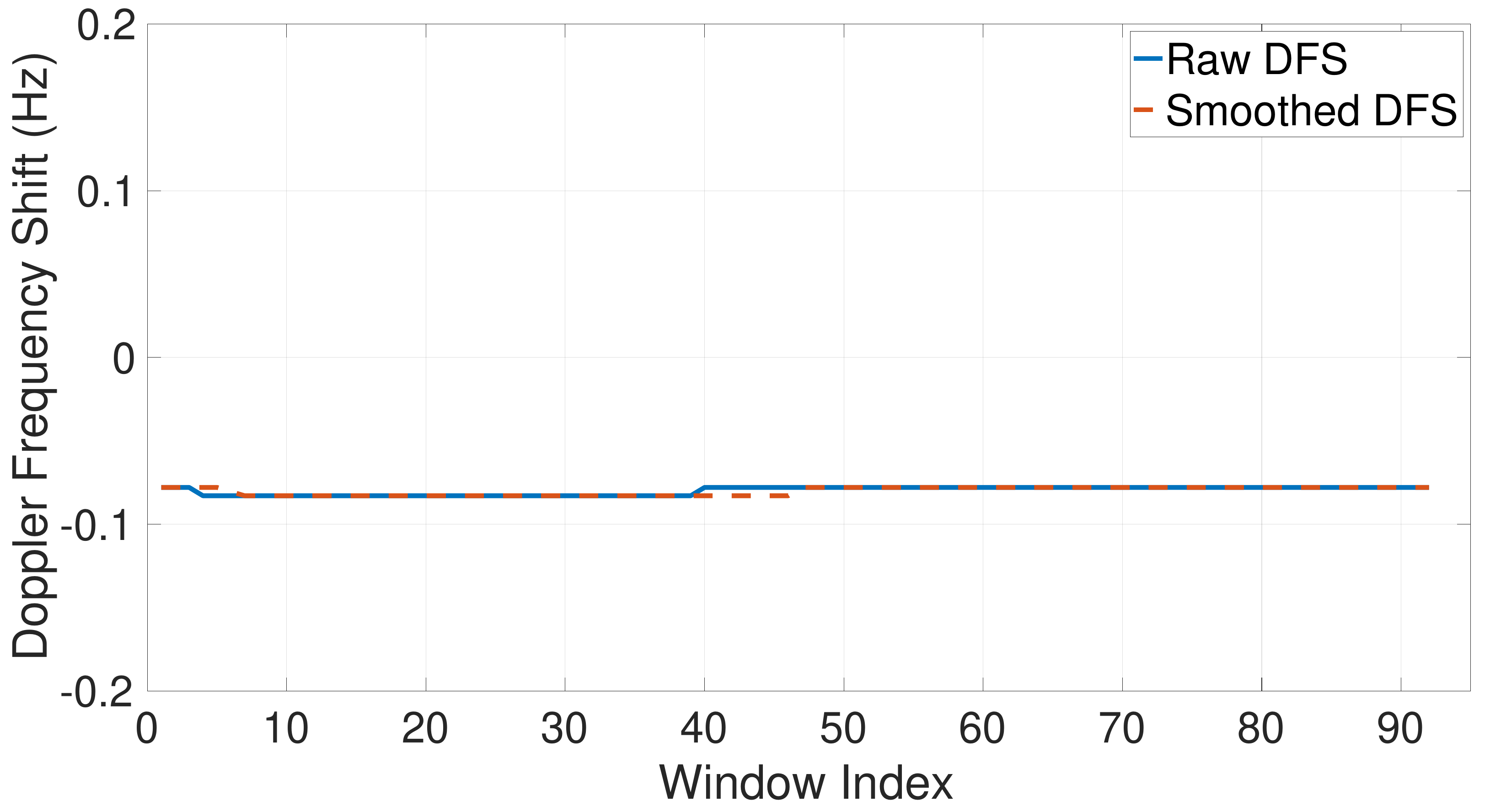}
        		\subcaption{Doppler profile}
        		\label{Fig5c}
    		\end{subfigure}\\
			\begin{subfigure}{\textwidth}
        		\centering
        		\includegraphics[width=\textwidth]{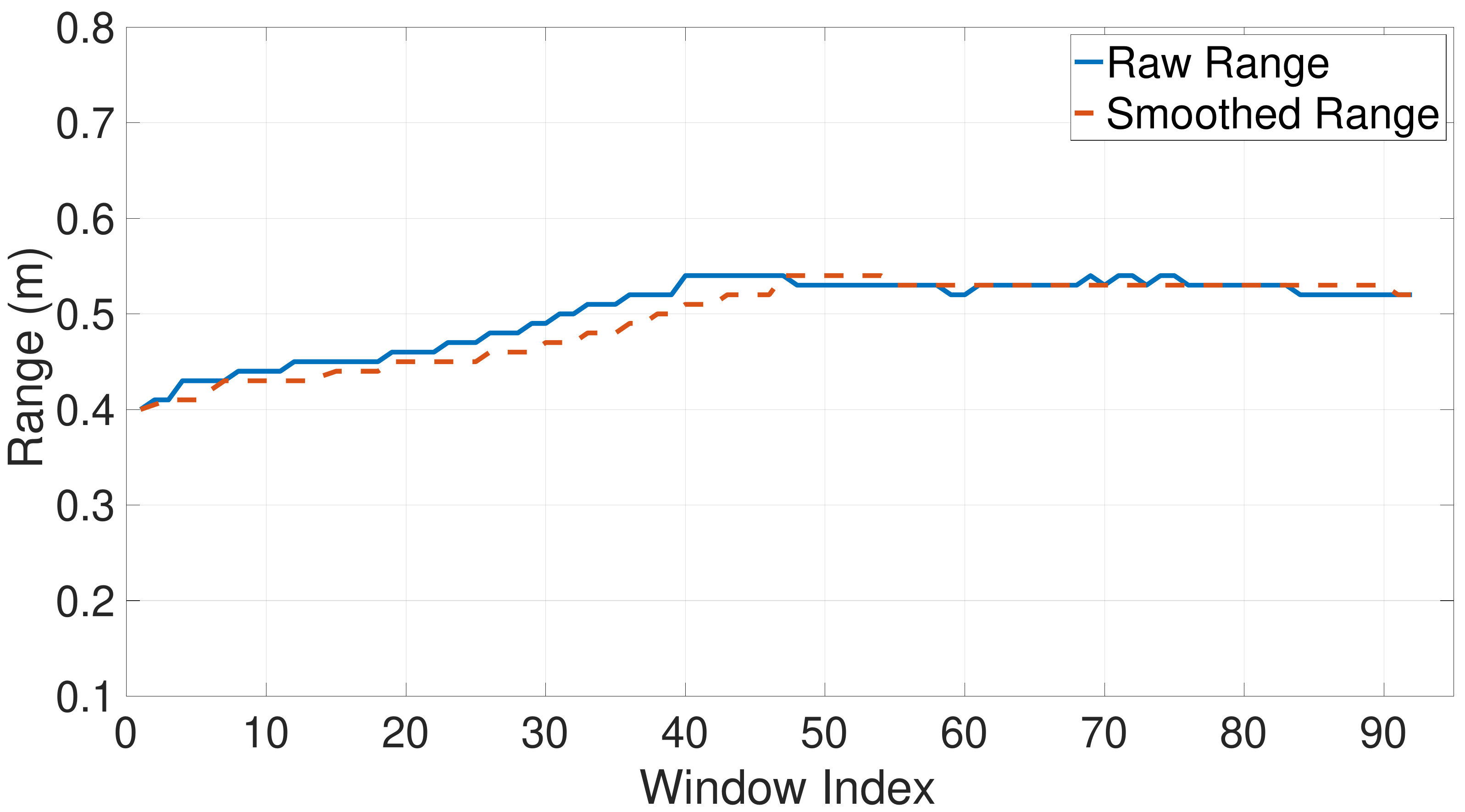}
        		\subcaption{Range profile}
        		\label{Fig5d}
			\end{subfigure}\\
			\begin{subfigure}{\textwidth}
        		\centering
        		\includegraphics[width=1.01\textwidth]{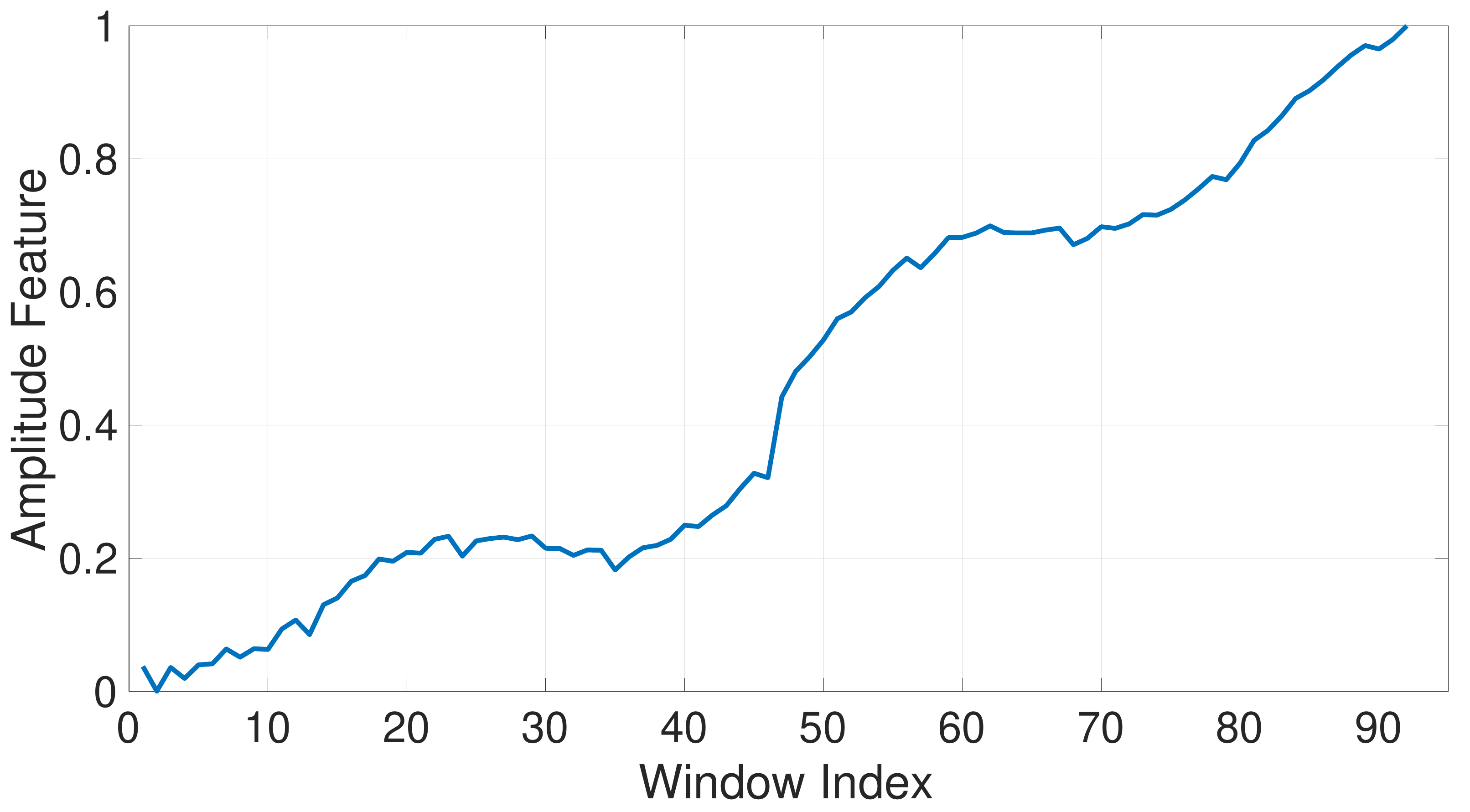}
        		\subcaption{Amplitude feature}
        		\label{Fig5e}
    		\end{subfigure}\\
			\begin{subfigure}{\textwidth}
        		\centering
        		\includegraphics[width=\textwidth]{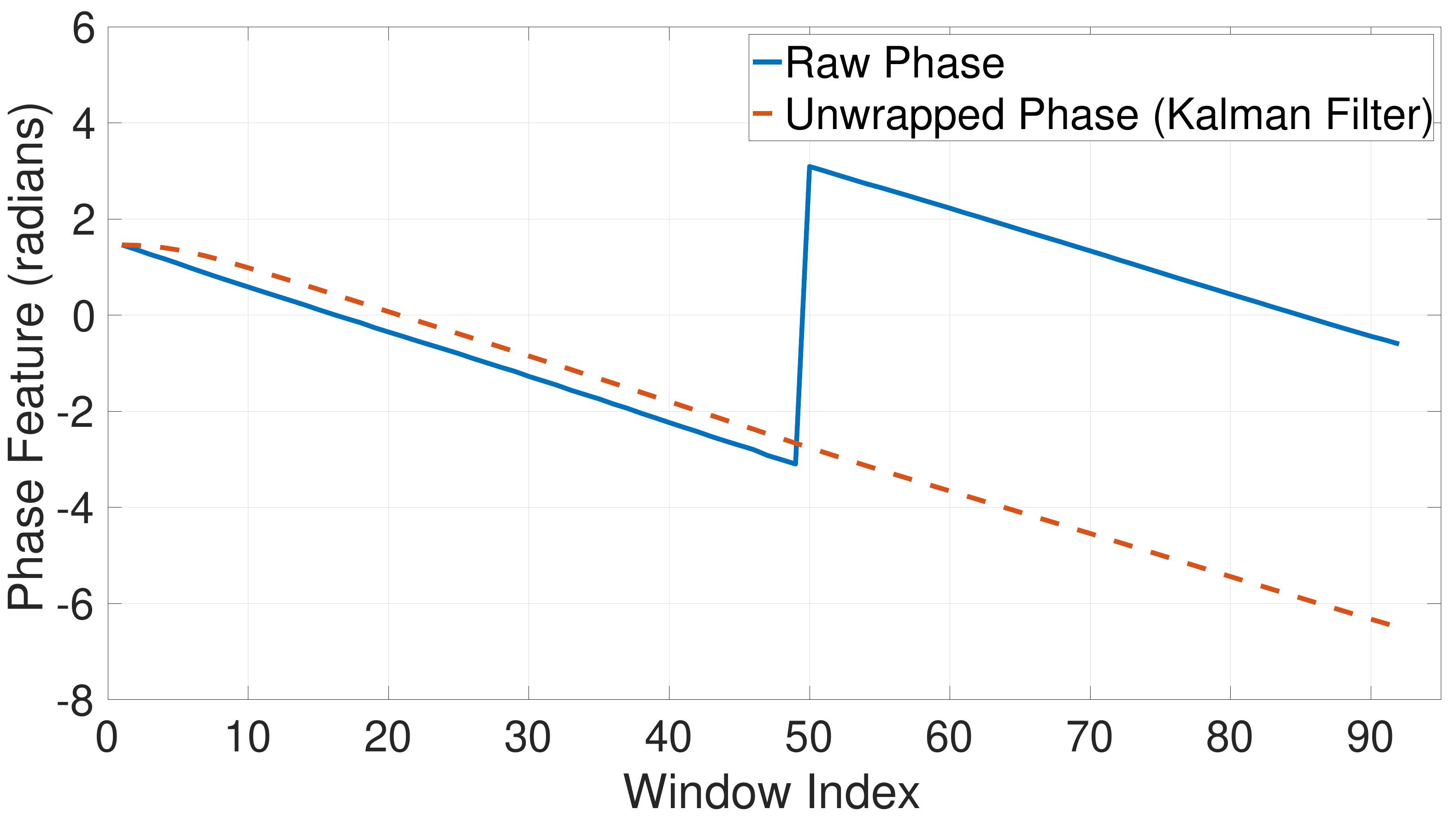}
        		\subcaption{Phase feature}
        		\label{Fig5f}
			\end{subfigure}
\begin{subfigure}{\textwidth}
        		\centering
        		\includegraphics[width=\textwidth]{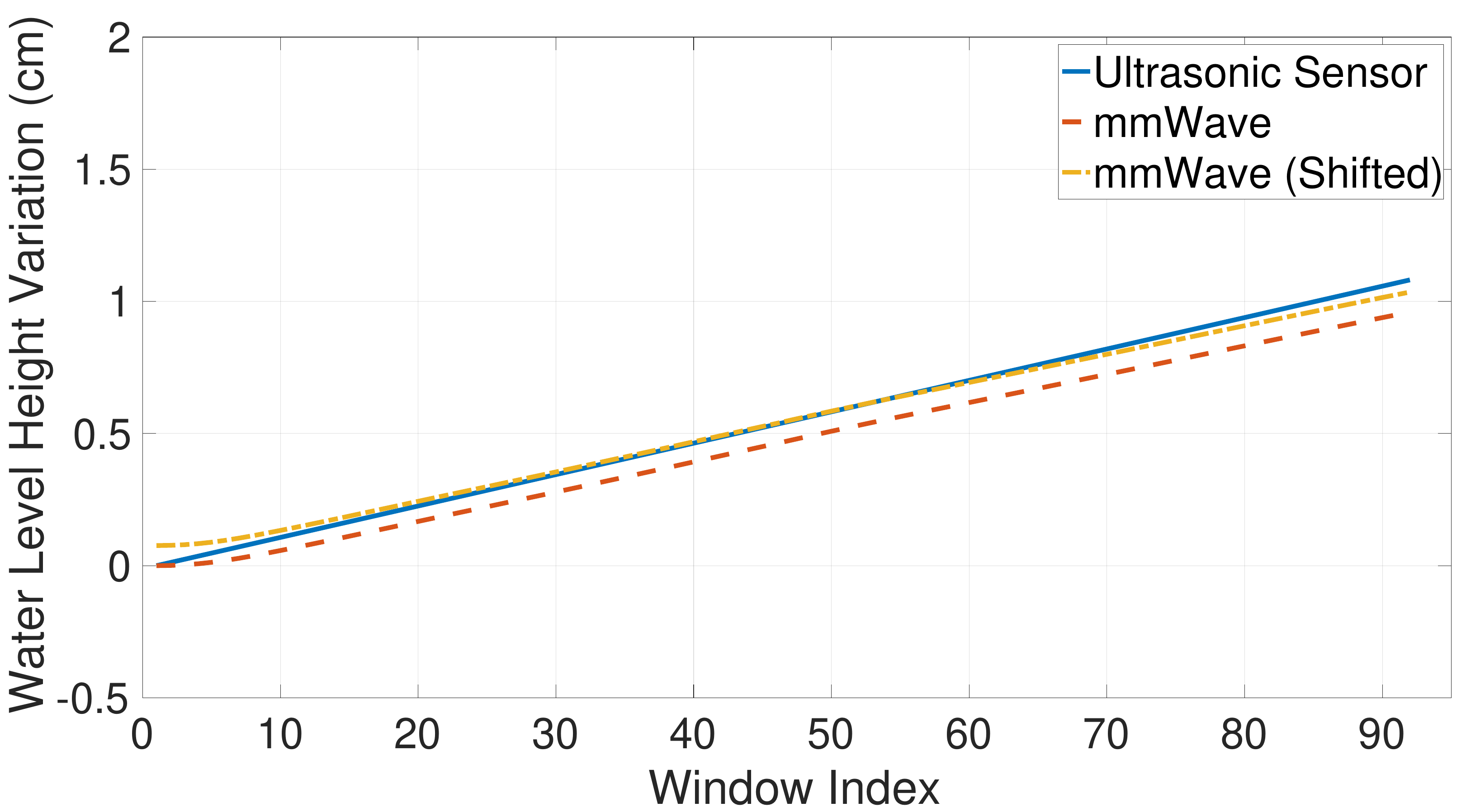}
        		\subcaption{Height variation}
        		\label{Fig5g}
			\end{subfigure}
		\caption{Tank$\rightarrow$Pool (mmWave).}
		\label{Fig5}
		\end{minipage} 
\begin{minipage}[b]{0.245\linewidth}    
			\centering
    		\begin{subfigure}{\textwidth}
        		\centering
        	\includegraphics[width=\textwidth]
{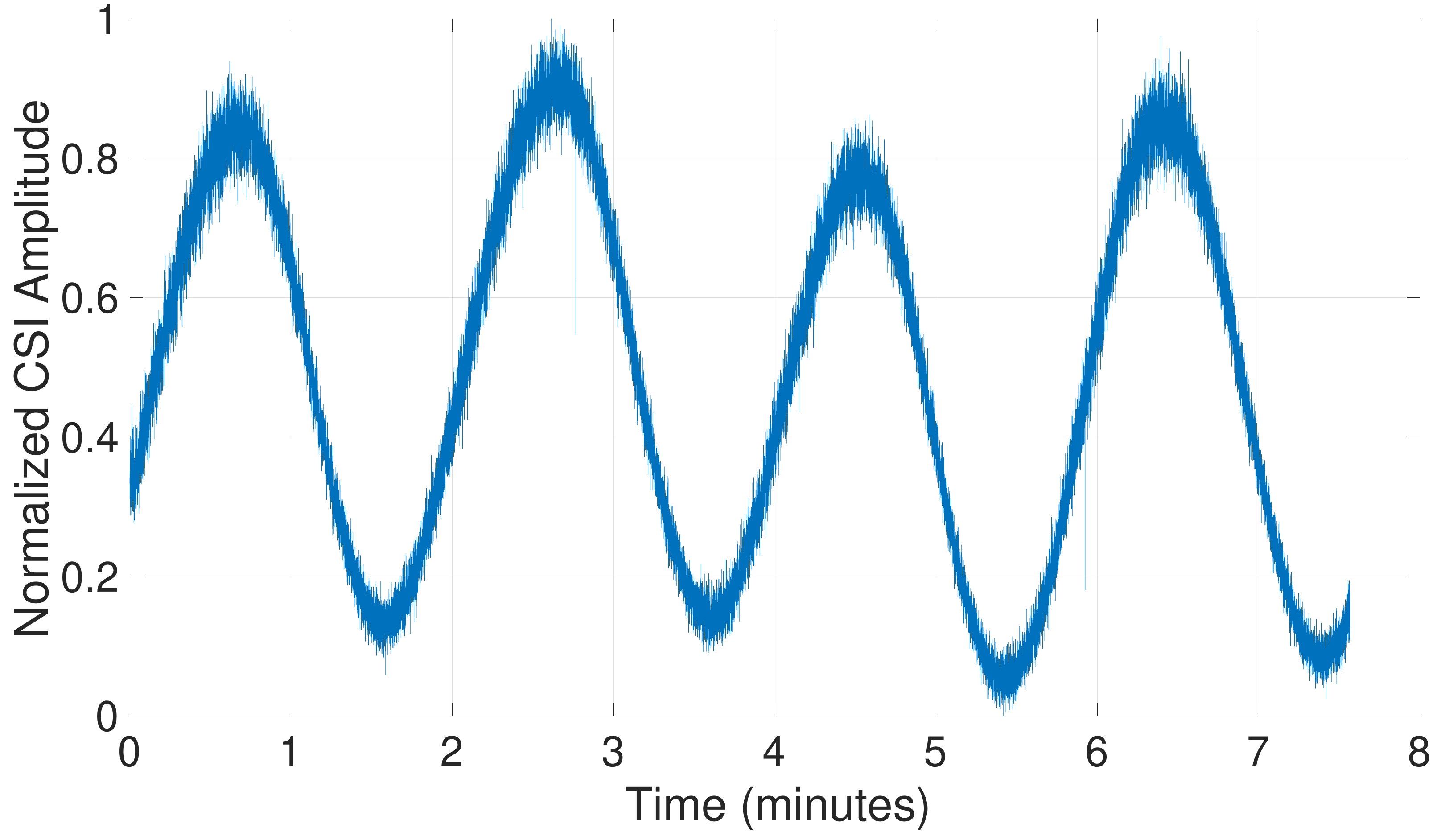}
        		\subcaption{Raw mmWave CSI}
        		\label{Fig6a}
    		\end{subfigure}\\
\begin{subfigure}{\textwidth}
        		\centering
        	\includegraphics[width=\textwidth]{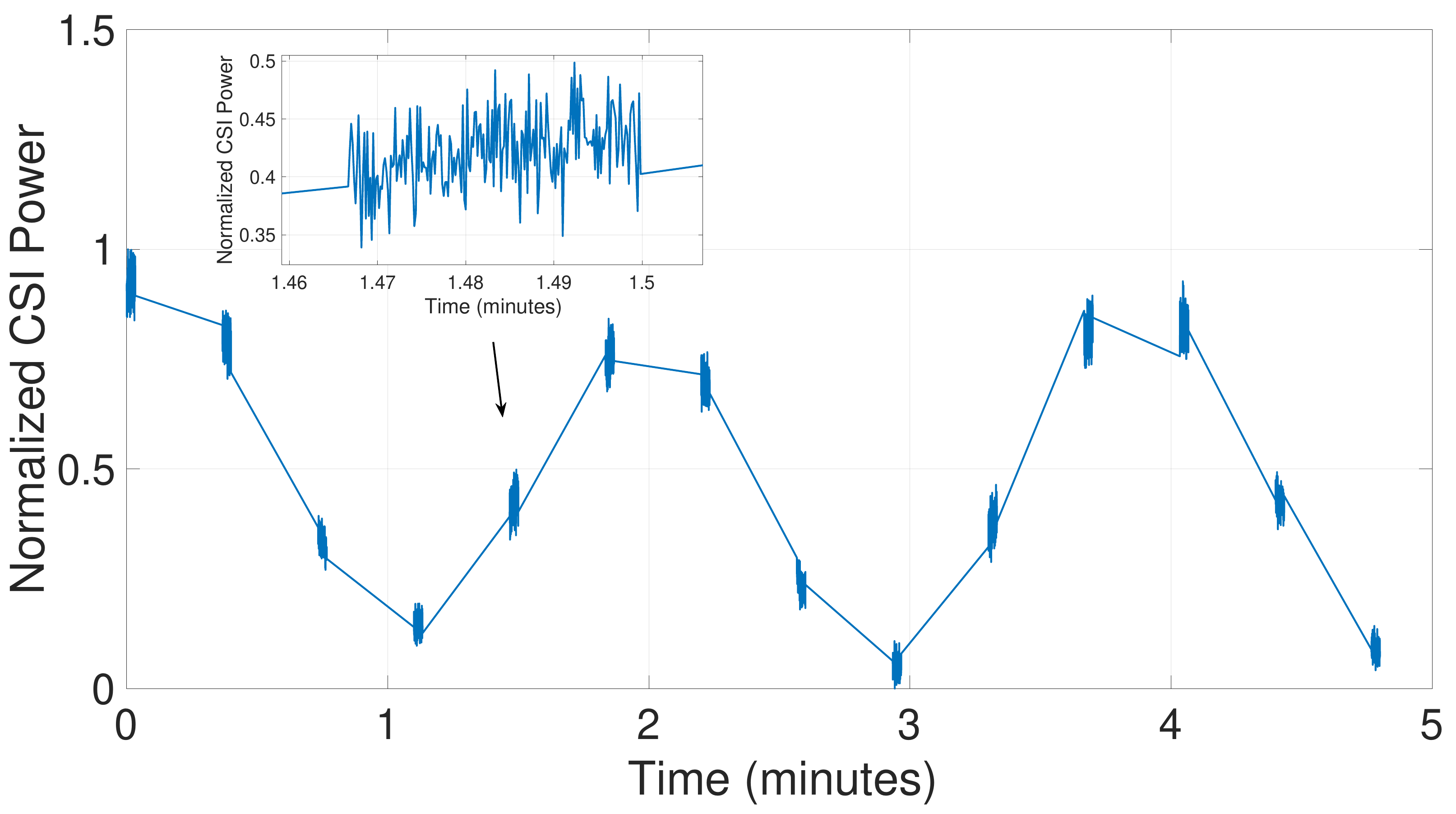}
        		\subcaption{Downsampled mmWave CSI \\ \centering in a time window}
        		\label{Fig6b}
    		\end{subfigure}\\
\begin{subfigure}{\textwidth}
        		\centering
        	\includegraphics[width=\textwidth]
{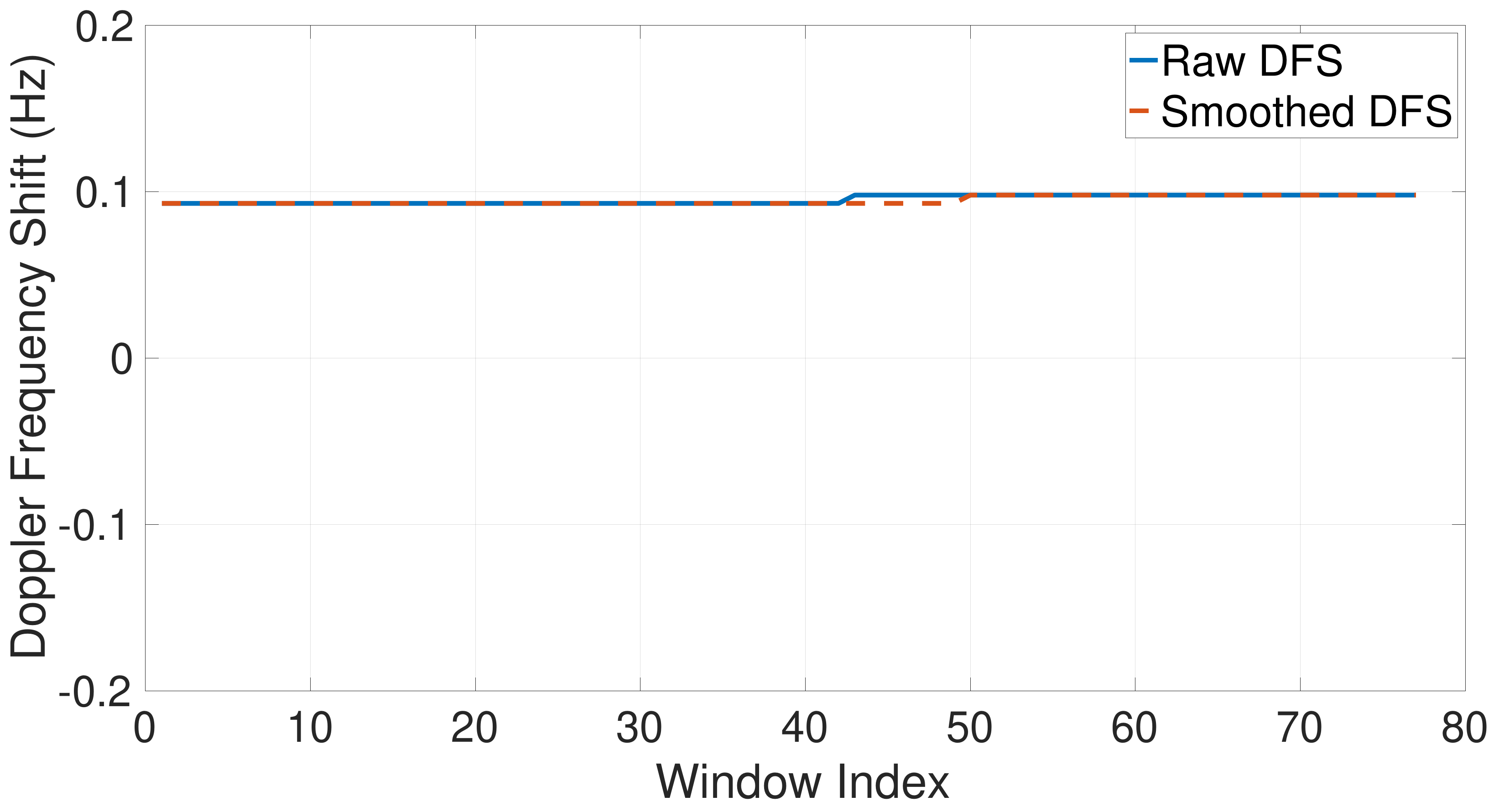}
        		\subcaption{Doppler profile}
        		\label{Fig6c}
    		\end{subfigure}\\
			\begin{subfigure}{\textwidth}
        		\centering
        		\includegraphics[width=\textwidth]{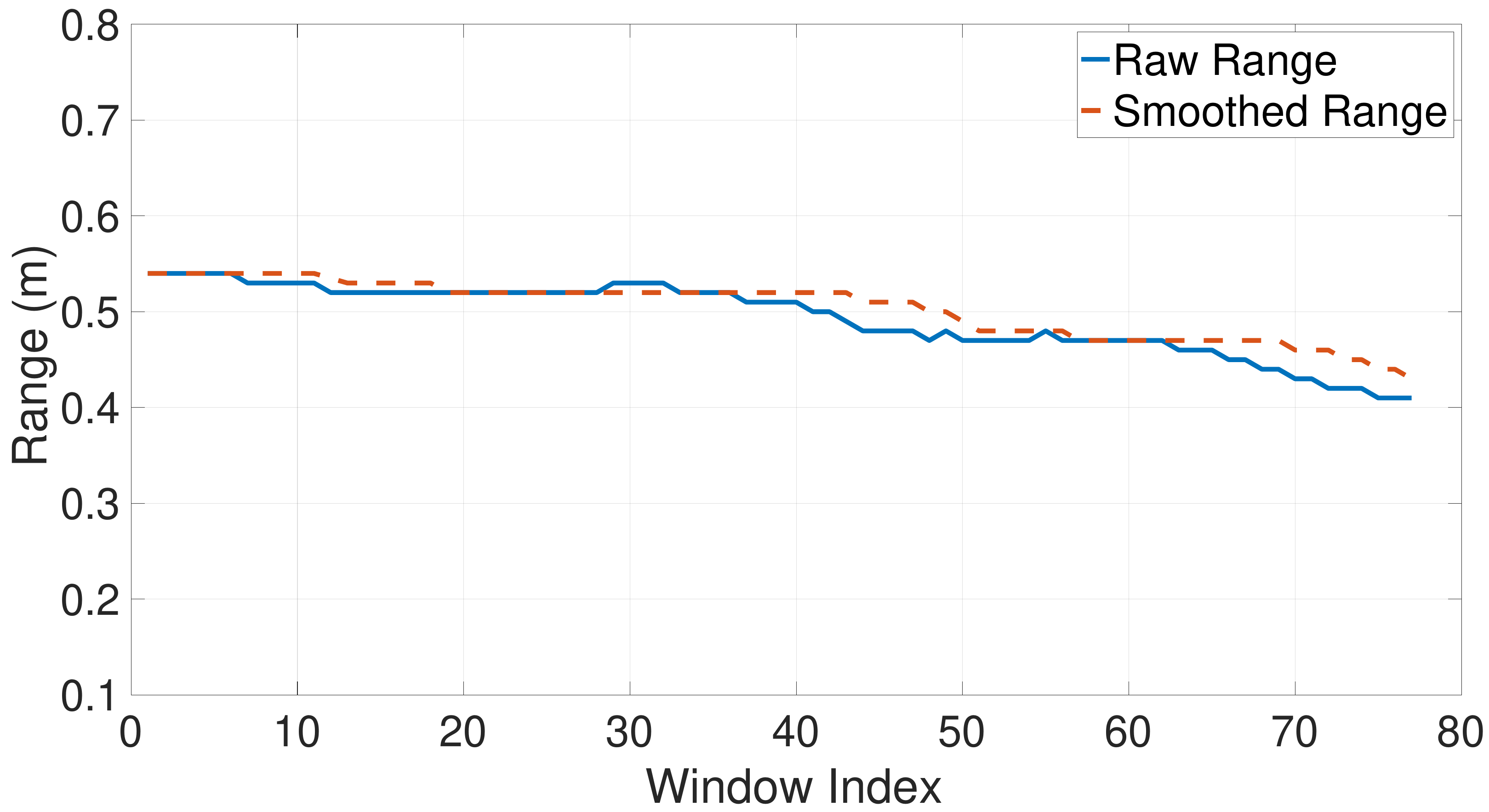}
        		\subcaption{Range profile}
        		\label{Fig6d}
			\end{subfigure}\\
			\begin{subfigure}{\textwidth}
        		\centering
        		\includegraphics[width=\textwidth]{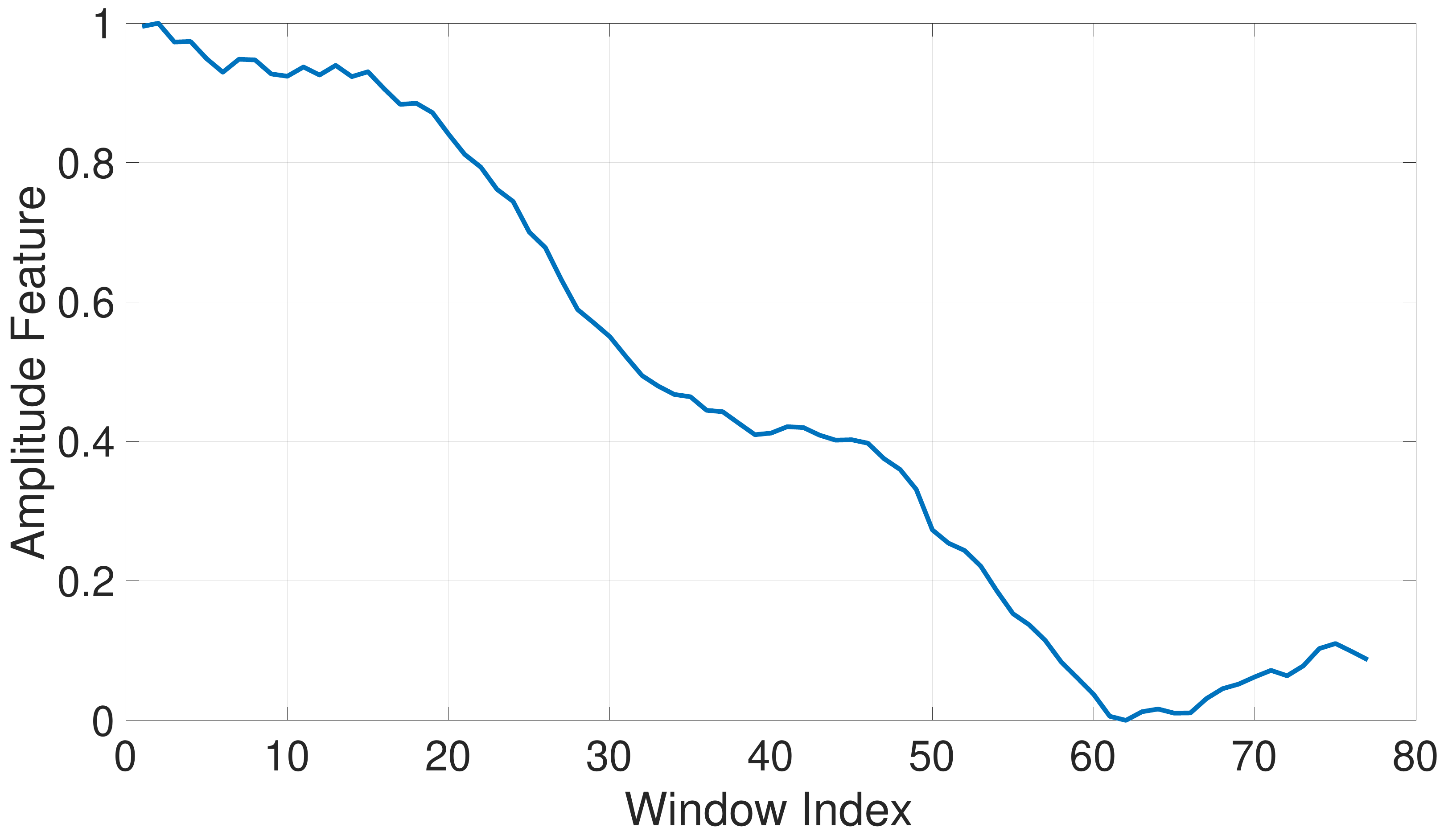}
        		\subcaption{Amplitude feature}
        		\label{Fig6e}
    		\end{subfigure}\\
			\begin{subfigure}{\textwidth}
        		\centering
        		\includegraphics[width=\textwidth]{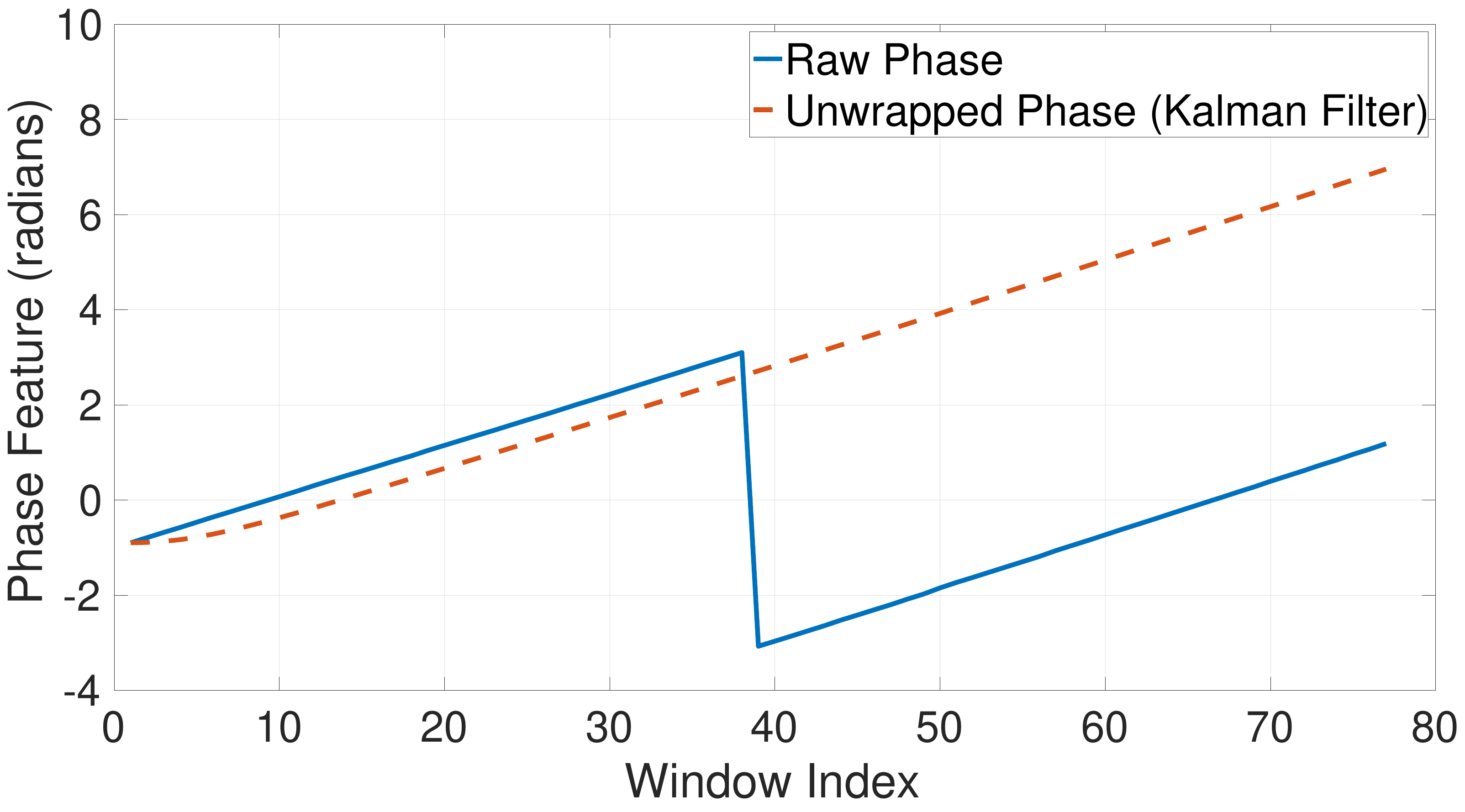}
        		\subcaption{Phase feature}
        		\label{Fig6f}
			\end{subfigure}
\begin{subfigure}{\textwidth}
        		\centering
        		\includegraphics[width=\textwidth]{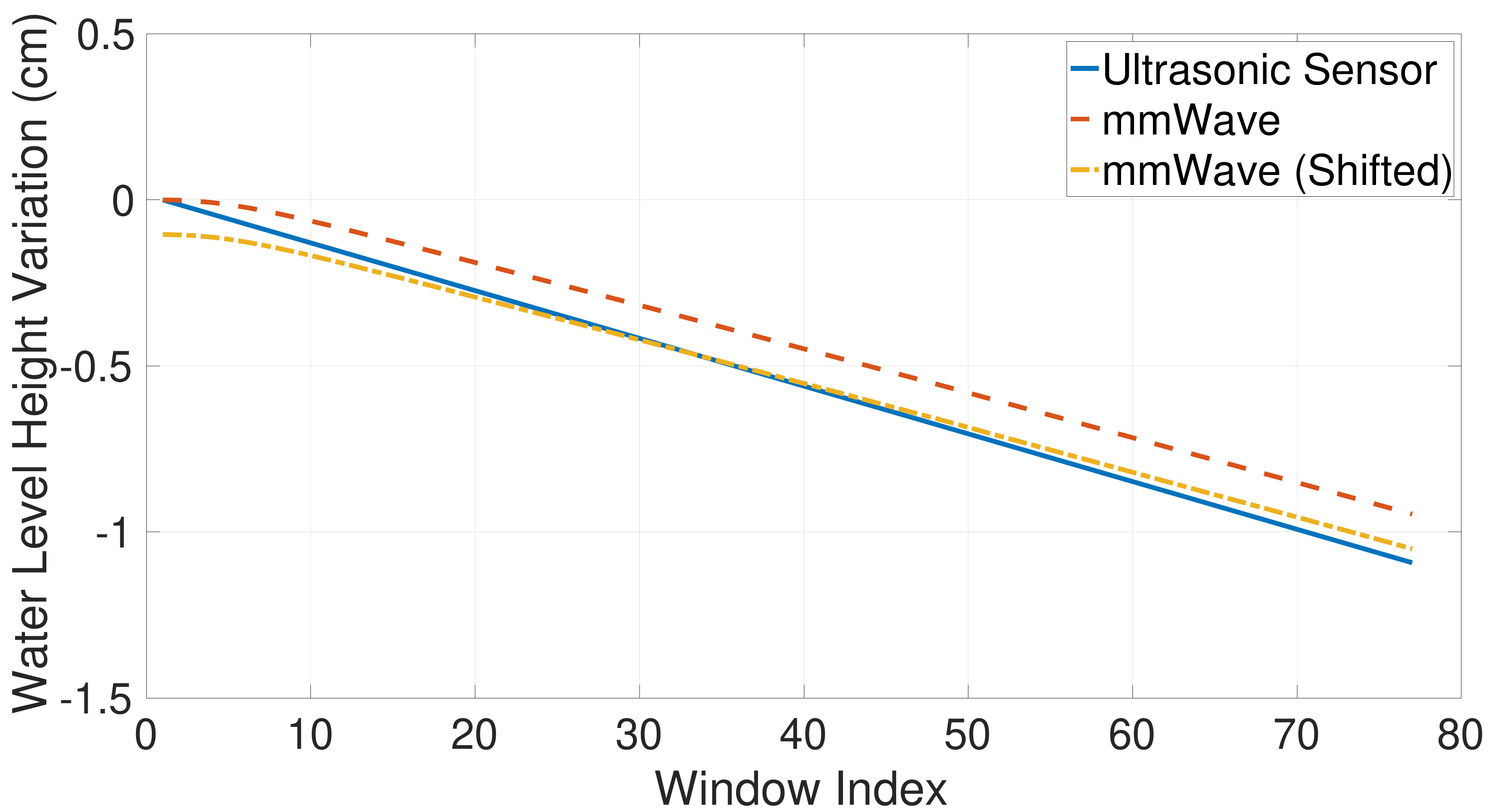}
        		\subcaption{Height variation}
        		\label{Fig6f}
			\end{subfigure}
		\caption{Pool$\rightarrow$Tank (mmWave).}
		\label{Fig6}
		\end{minipage}
\begin{minipage}[b]{0.245\linewidth} 
			\centering
    		\begin{subfigure}{\textwidth}
        		\centering
        	\includegraphics[width=\textwidth]{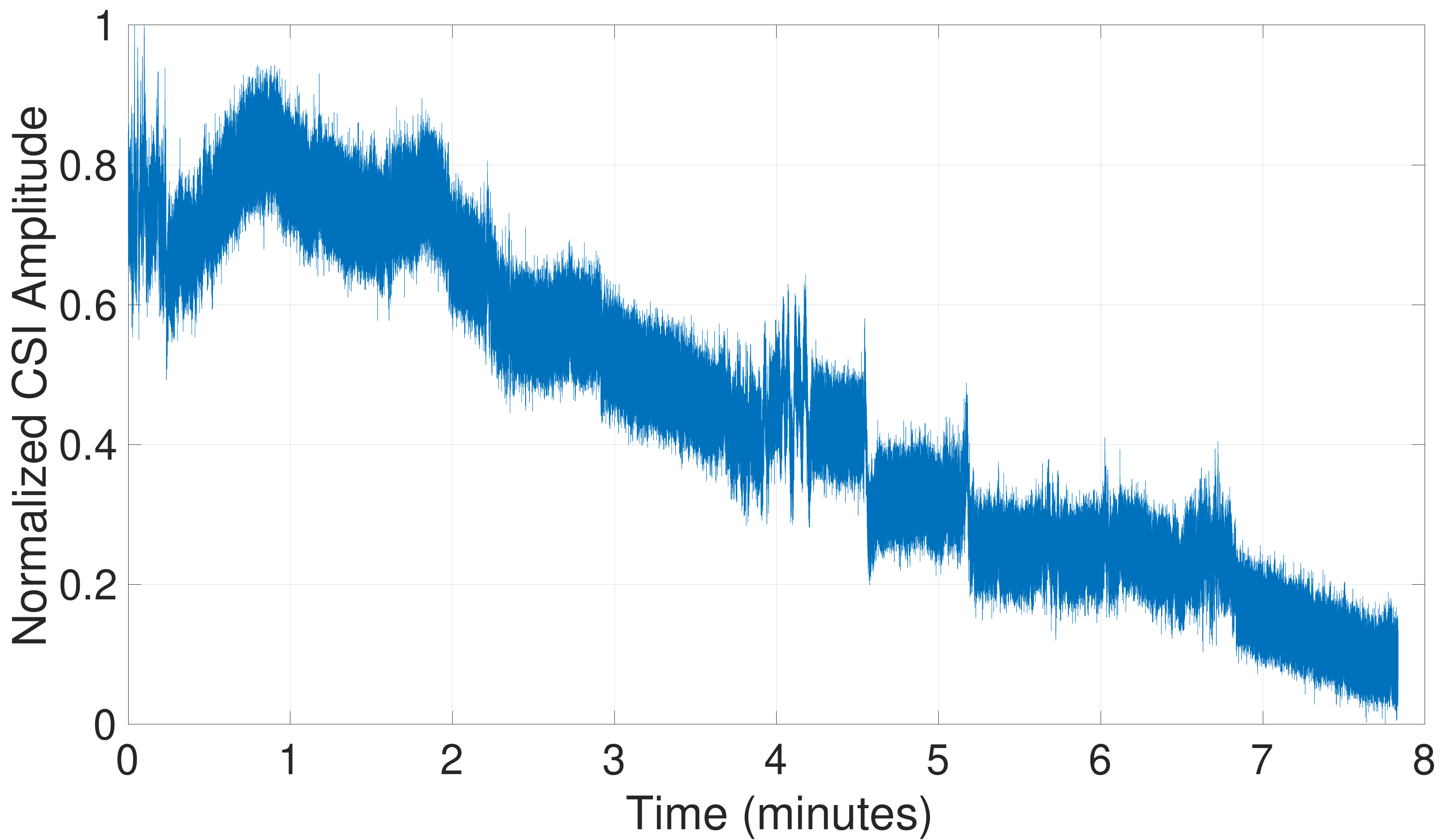}
        		\subcaption{Raw LTE CSI}
        		\label{Fig7a}
    		\end{subfigure}\\
\begin{subfigure}{\textwidth}
        		\centering
        	\includegraphics[width=\textwidth]{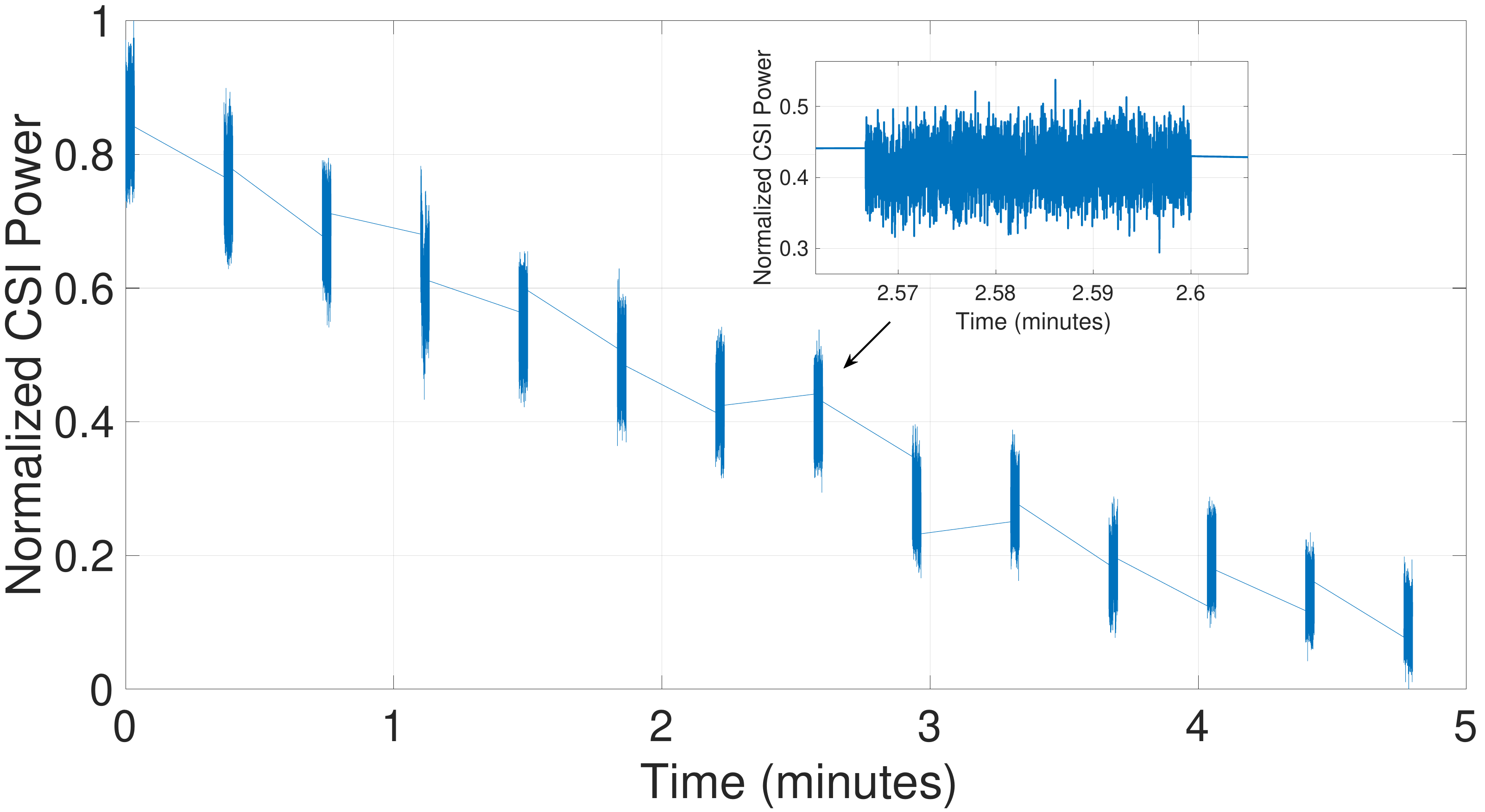}
        		\subcaption{Downsampled LTE CSI in \\ \centering a time window}
        		\label{Fig7b}
    		\end{subfigure}\\
\begin{subfigure}{\textwidth}
        		\centering
        	\includegraphics[width=\textwidth]{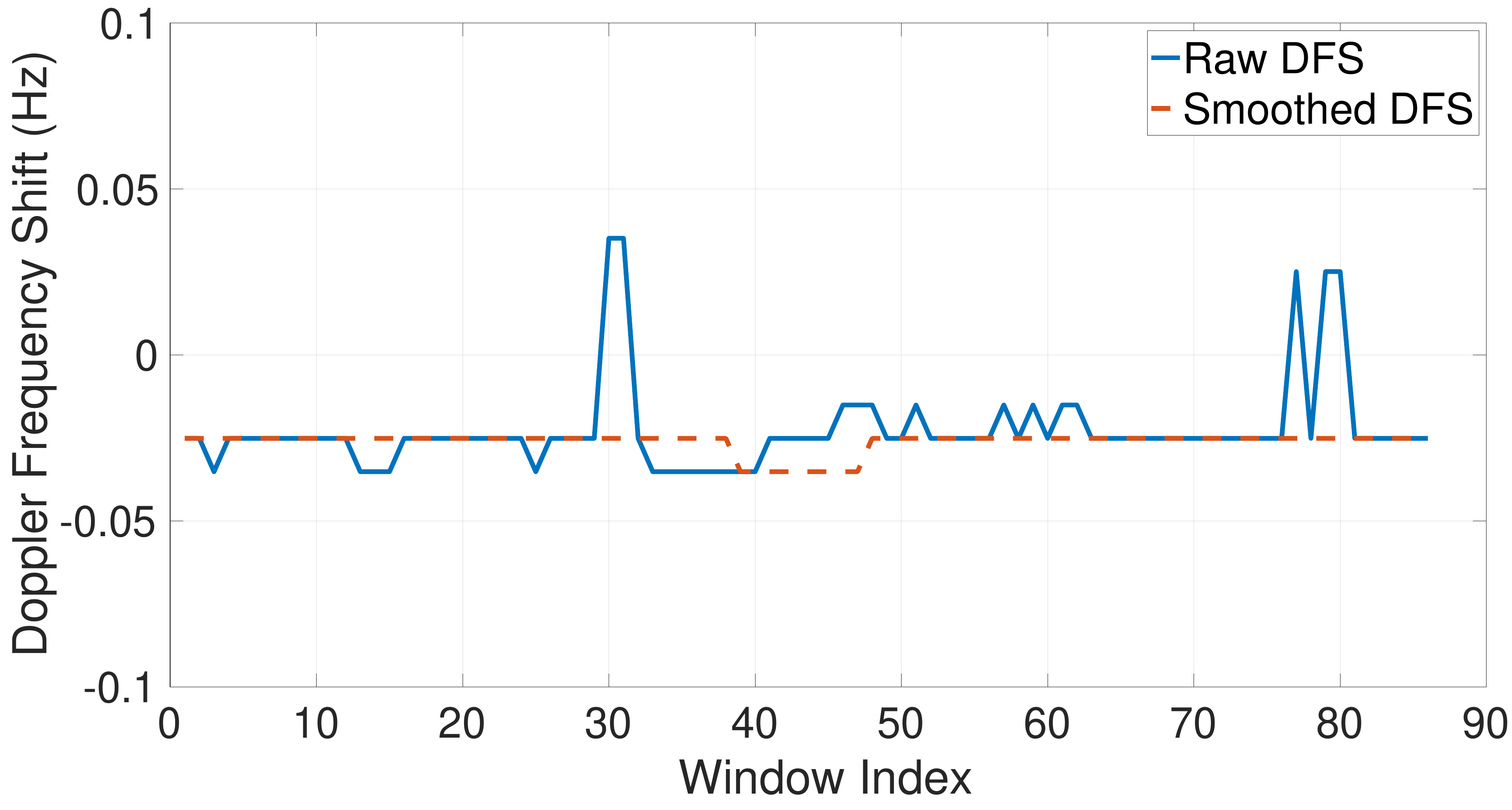}
        		\subcaption{Doppler profile}
        		\label{Fig8c}
    		\end{subfigure}\\
			\begin{subfigure}{\textwidth}
        		\centering
        		\includegraphics[width=0.98\textwidth]{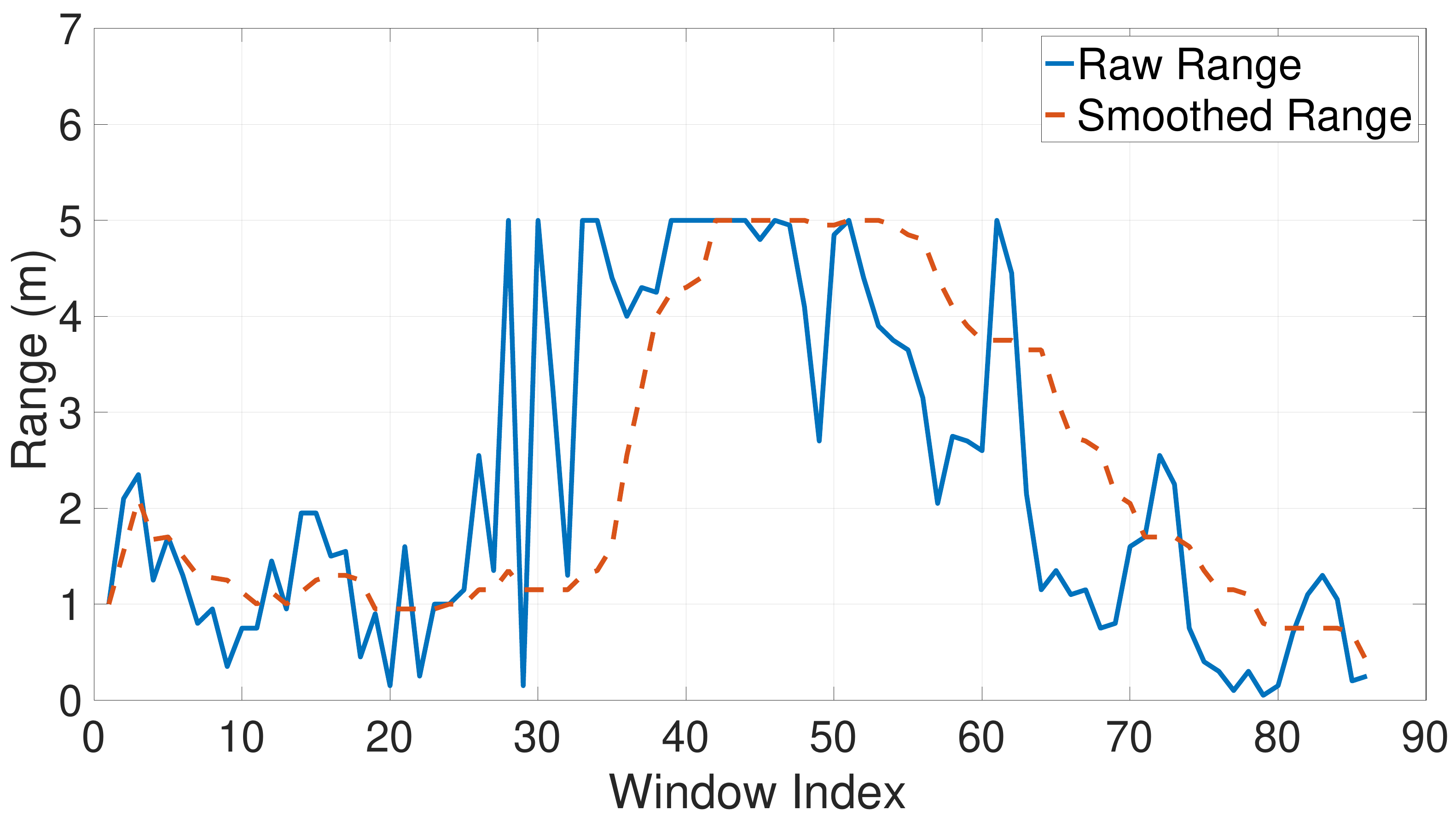}
        		\subcaption{Range profile}
        		\label{Fig7d}
			\end{subfigure}\\
			\begin{subfigure}{\textwidth}
        		\centering
\includegraphics[width=1.01\textwidth]
{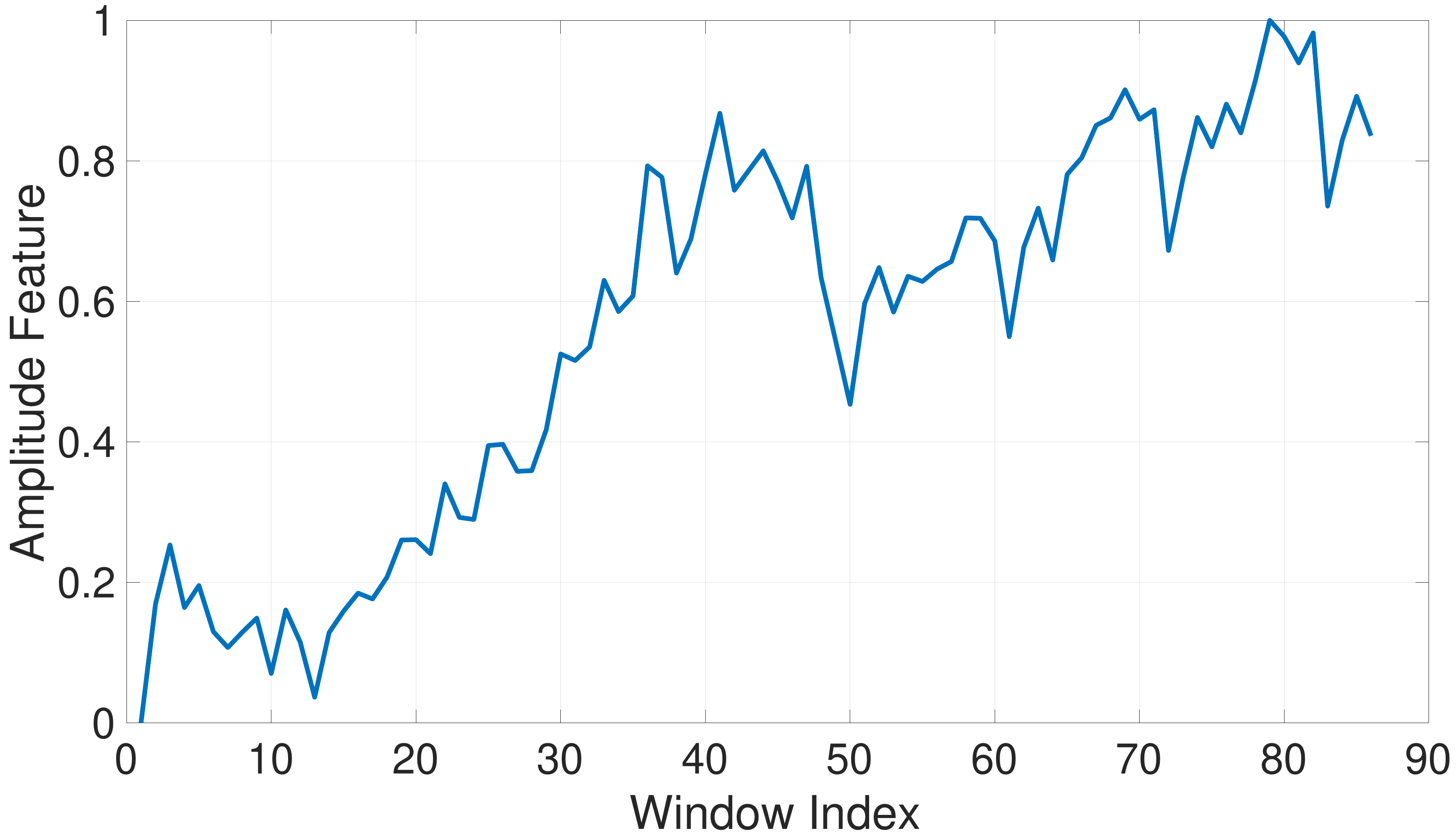}
        		\subcaption{Amplitude feature}
        		\label{Fig7e}
    		\end{subfigure}\\
			\begin{subfigure}{\textwidth}
        		\centering
        		\includegraphics[width=\textwidth]{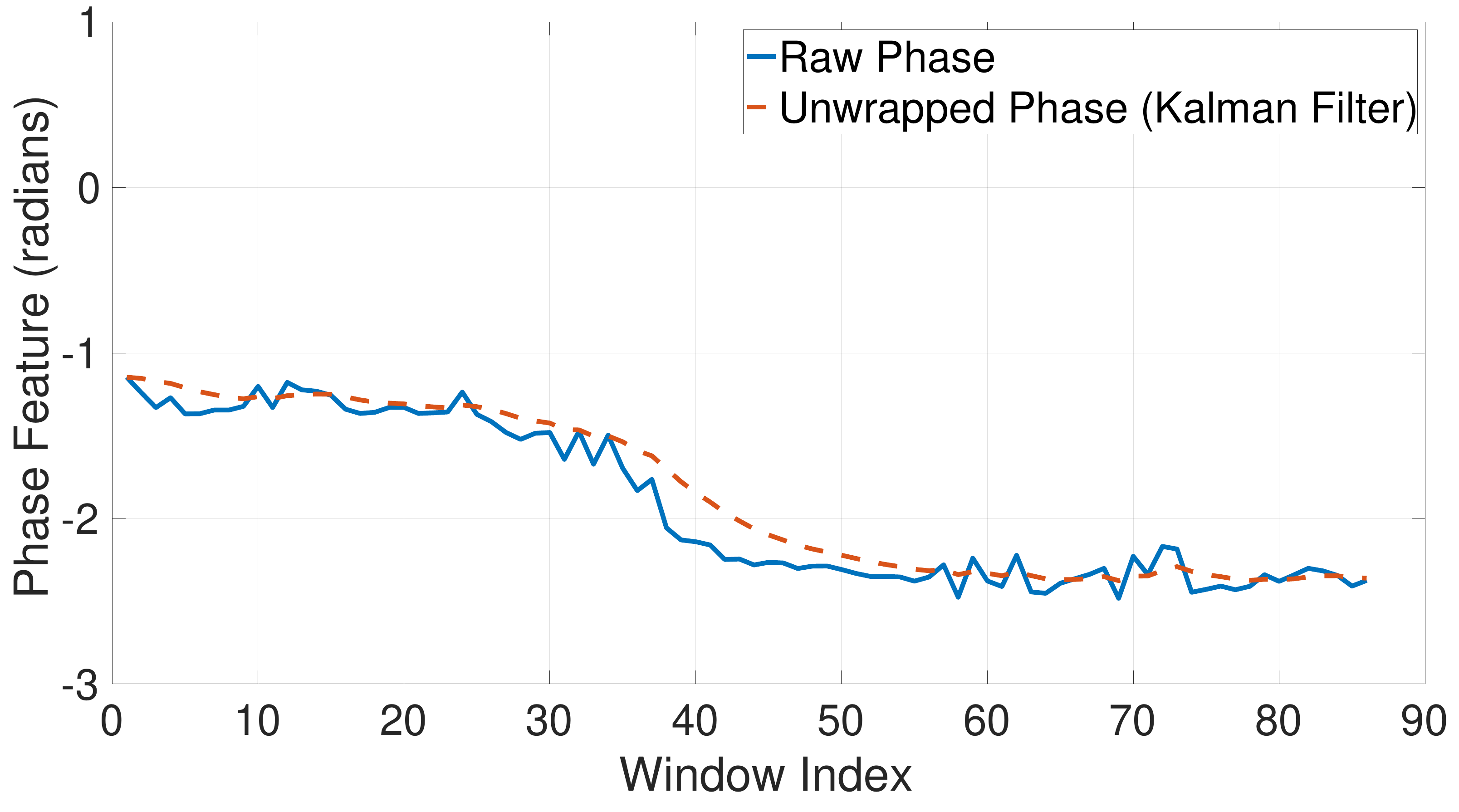}
        		\subcaption{Phase feature}
        		\label{Fig7f}
			\end{subfigure}
\begin{subfigure}{\textwidth}
        		\centering
        		\includegraphics[width=0.98\textwidth]{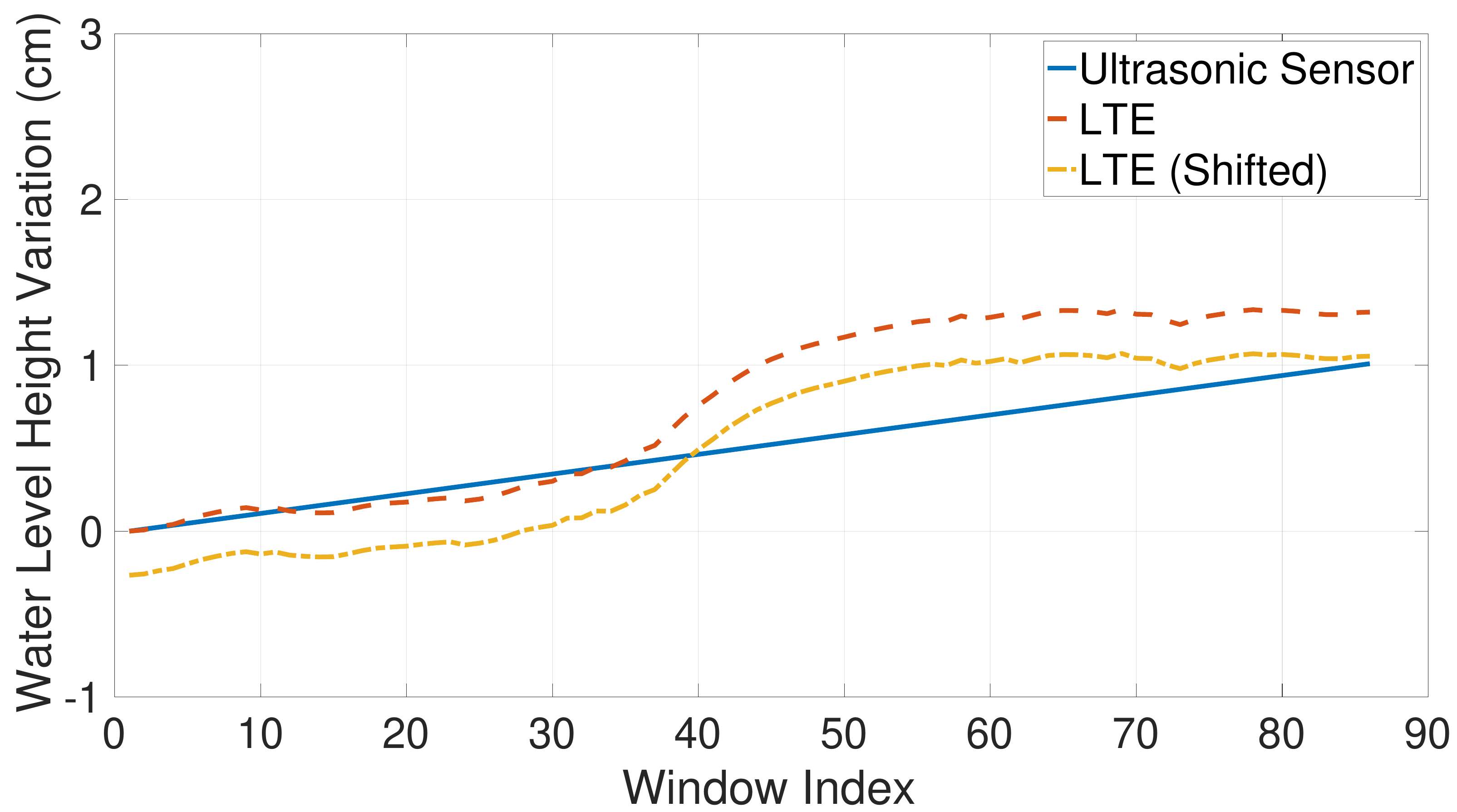}
        		\subcaption{ Height variation}
        		\label{Fig7g}
			\end{subfigure}
		\caption{Tank$\rightarrow$Pool (LTE).}
		\label{Fig7}
		\end{minipage} 
		\begin{minipage}[b]{0.245\linewidth}    
			\centering
    		\begin{subfigure}{\textwidth}
        		\centering
        	\includegraphics[width=0.98\textwidth]
{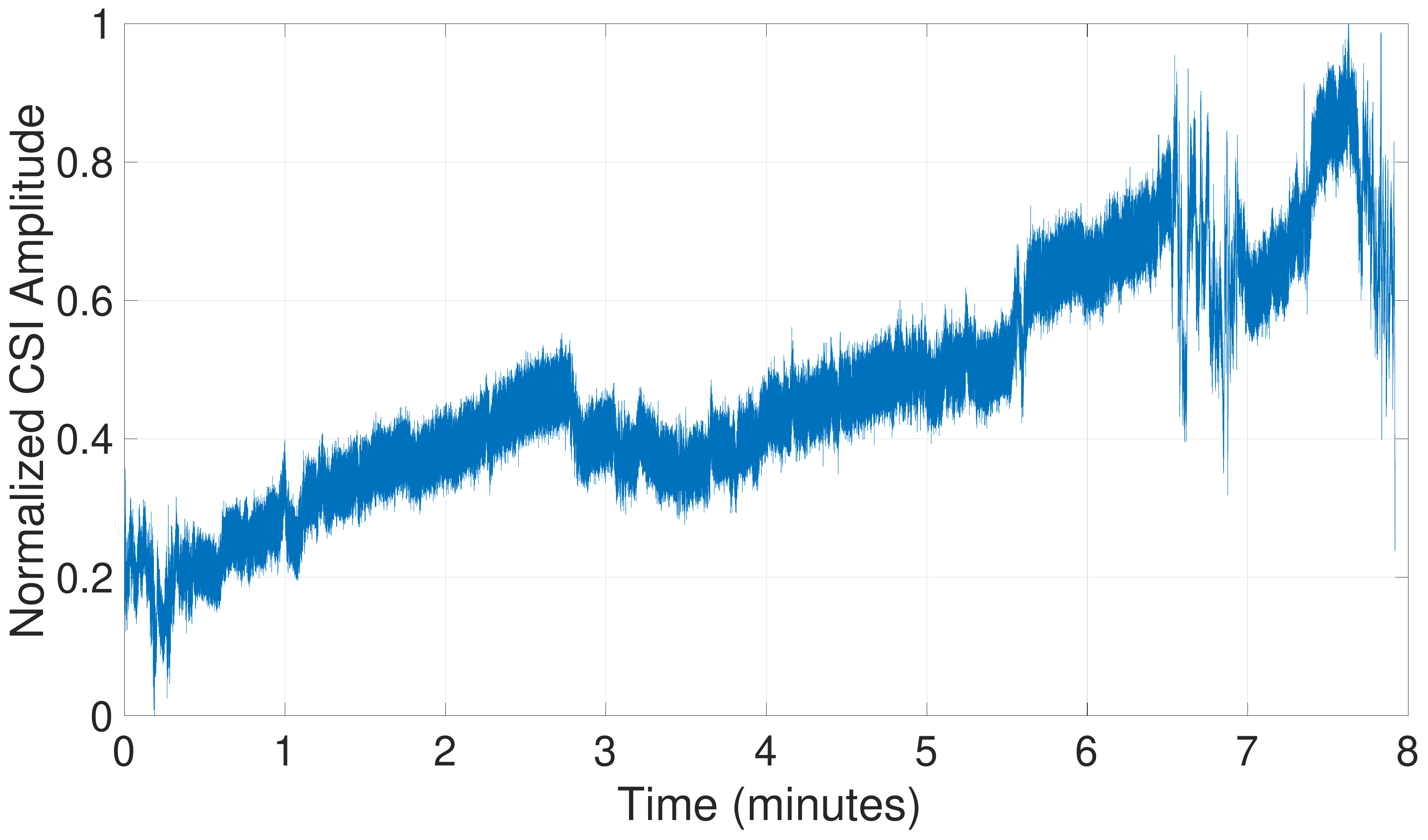}
        		\subcaption{Raw LTE CSI}
        		\label{Fig8a}
    		\end{subfigure}\\
\begin{subfigure}{\textwidth}
        		\centering
        	\includegraphics[width=\textwidth]{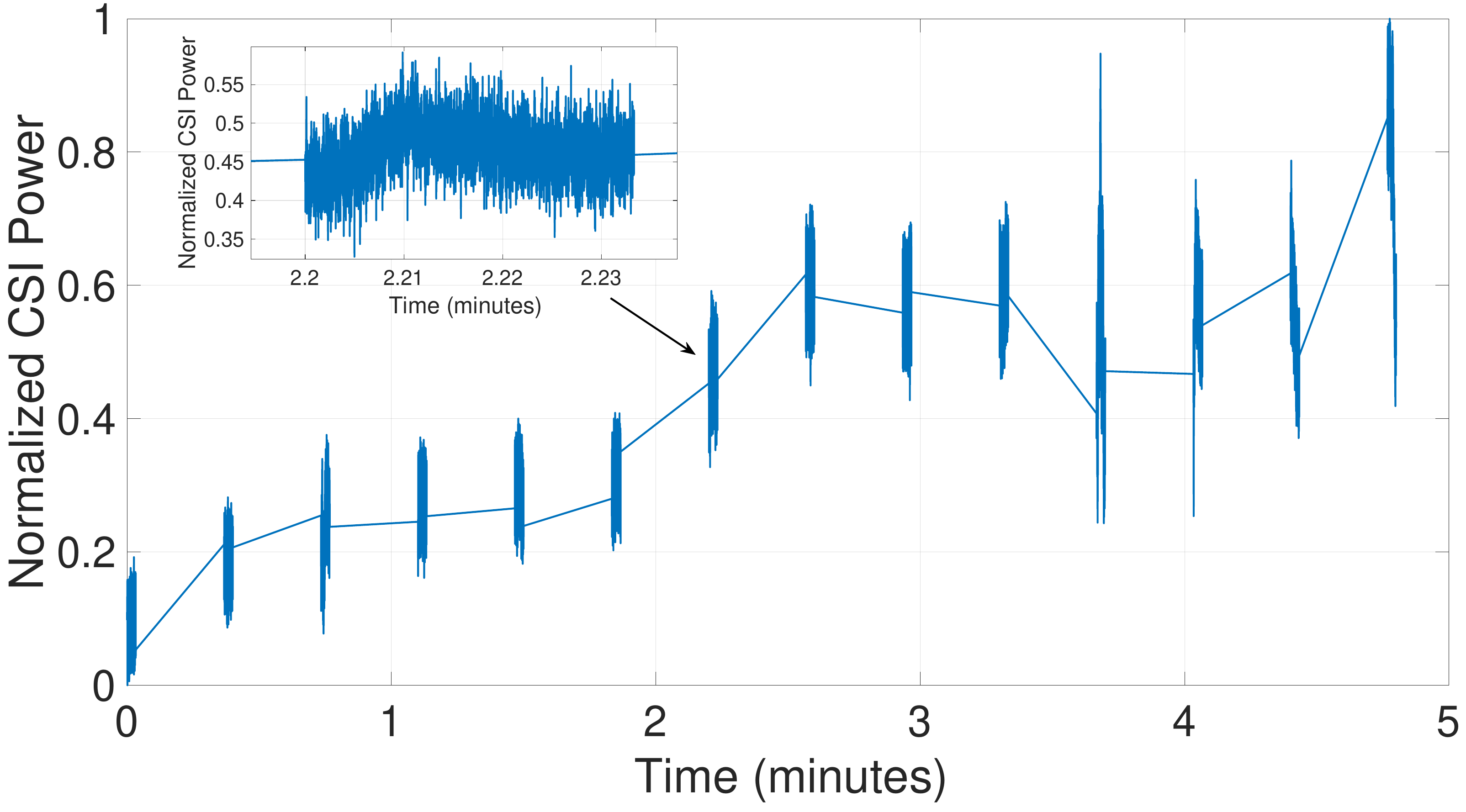}
        		\subcaption{Downsampled LTE CSI in\\ \centering a time window}
        		\label{Fig8b}
    		\end{subfigure}\\
\begin{subfigure}{\textwidth}
        		\centering
        	\includegraphics[width=\textwidth]{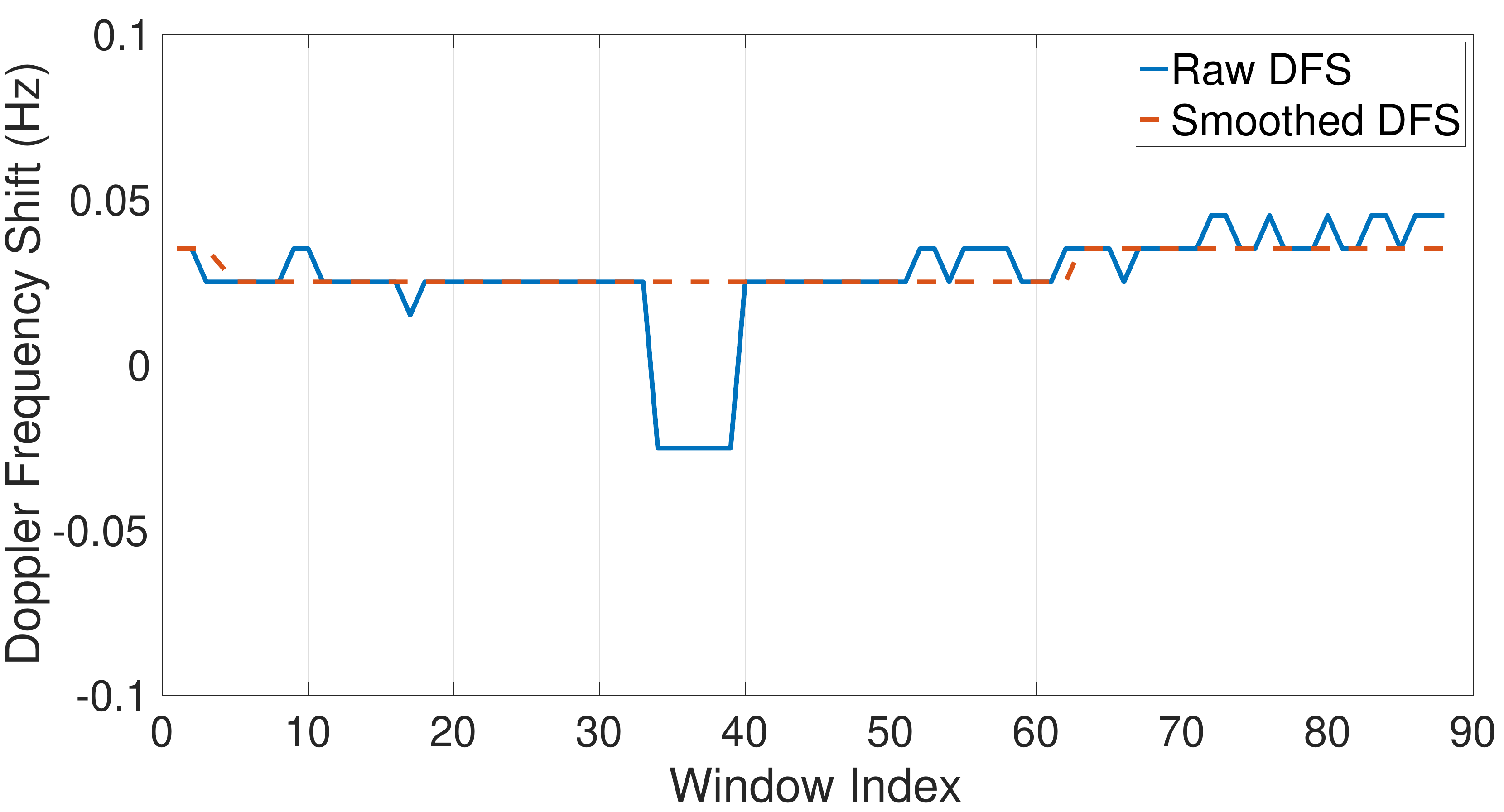}
        		\subcaption{Doppler profile}
        		\label{Fig8c}
    		\end{subfigure}\\
			\begin{subfigure}{\textwidth}
        		\centering
        	\includegraphics[width=0.98\textwidth]{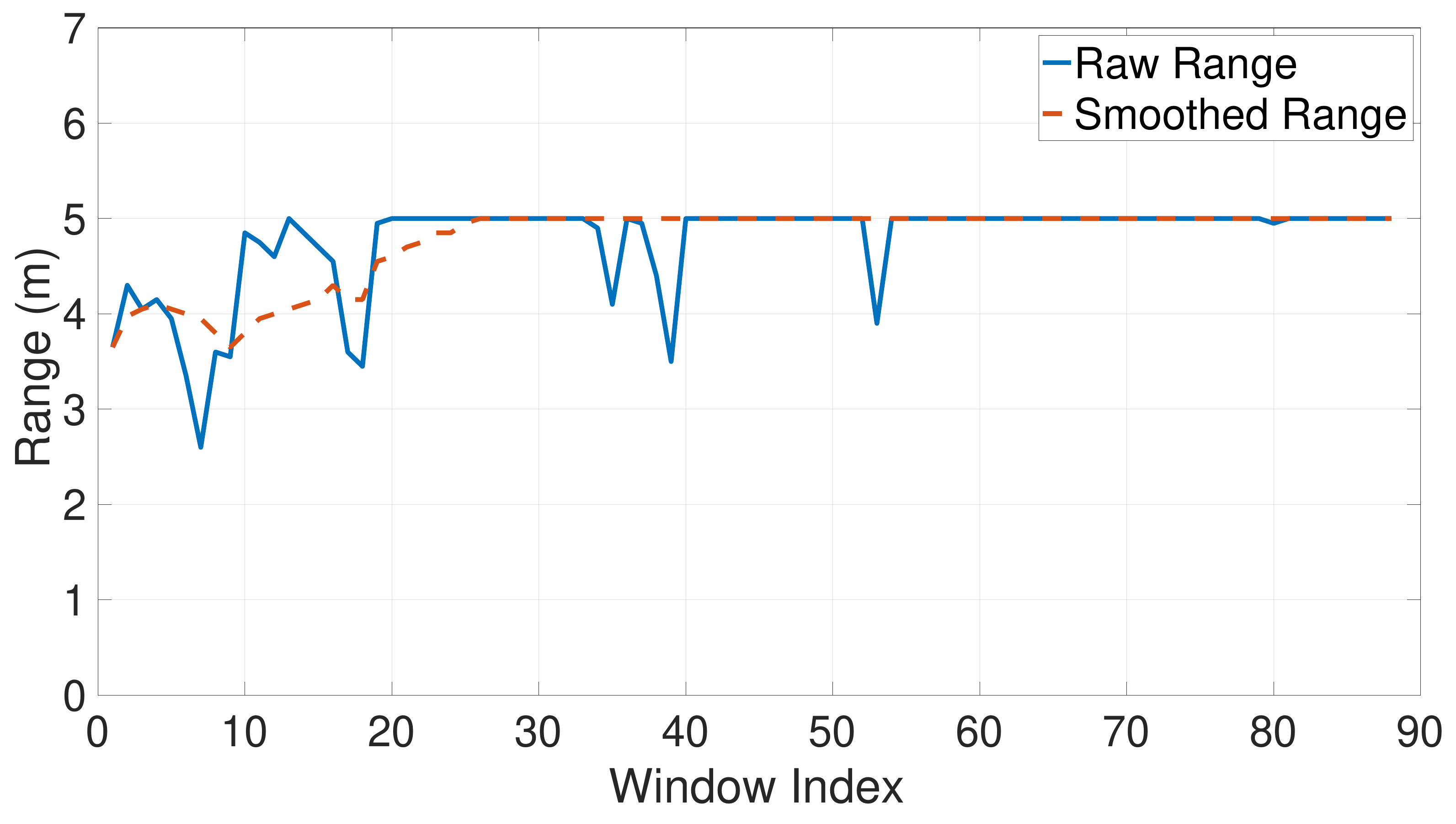}
        		\subcaption{Range profile}
        		\label{Fig8d}
			\end{subfigure}\\
			\begin{subfigure}{\textwidth}
        		\centering
        		\includegraphics[width=\textwidth]{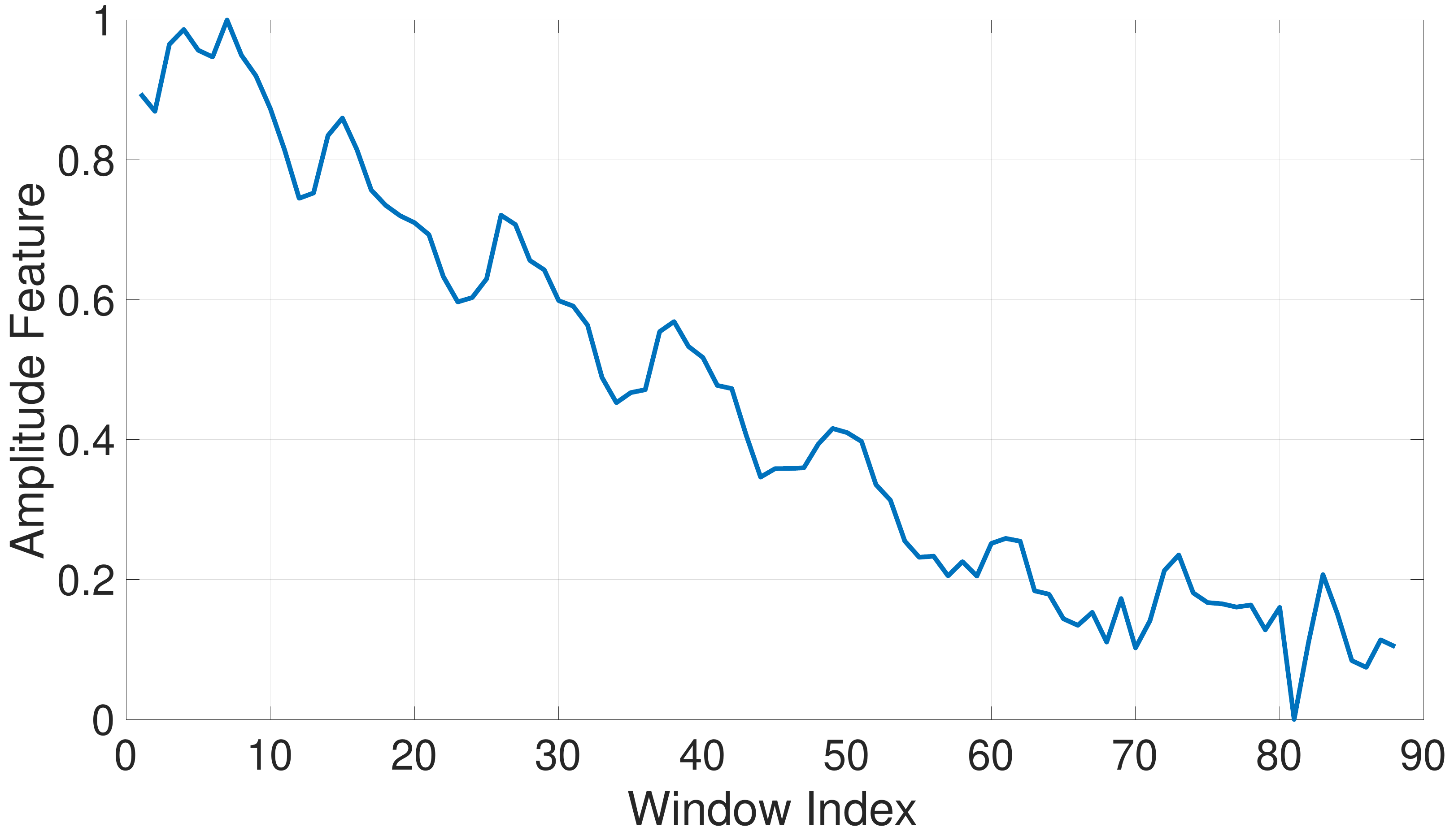}
        		\subcaption{Amplitude feature}
        		\label{Fig8e}
    		\end{subfigure}\\
			\begin{subfigure}{\textwidth}
        		\centering
        		\includegraphics[width=\textwidth]{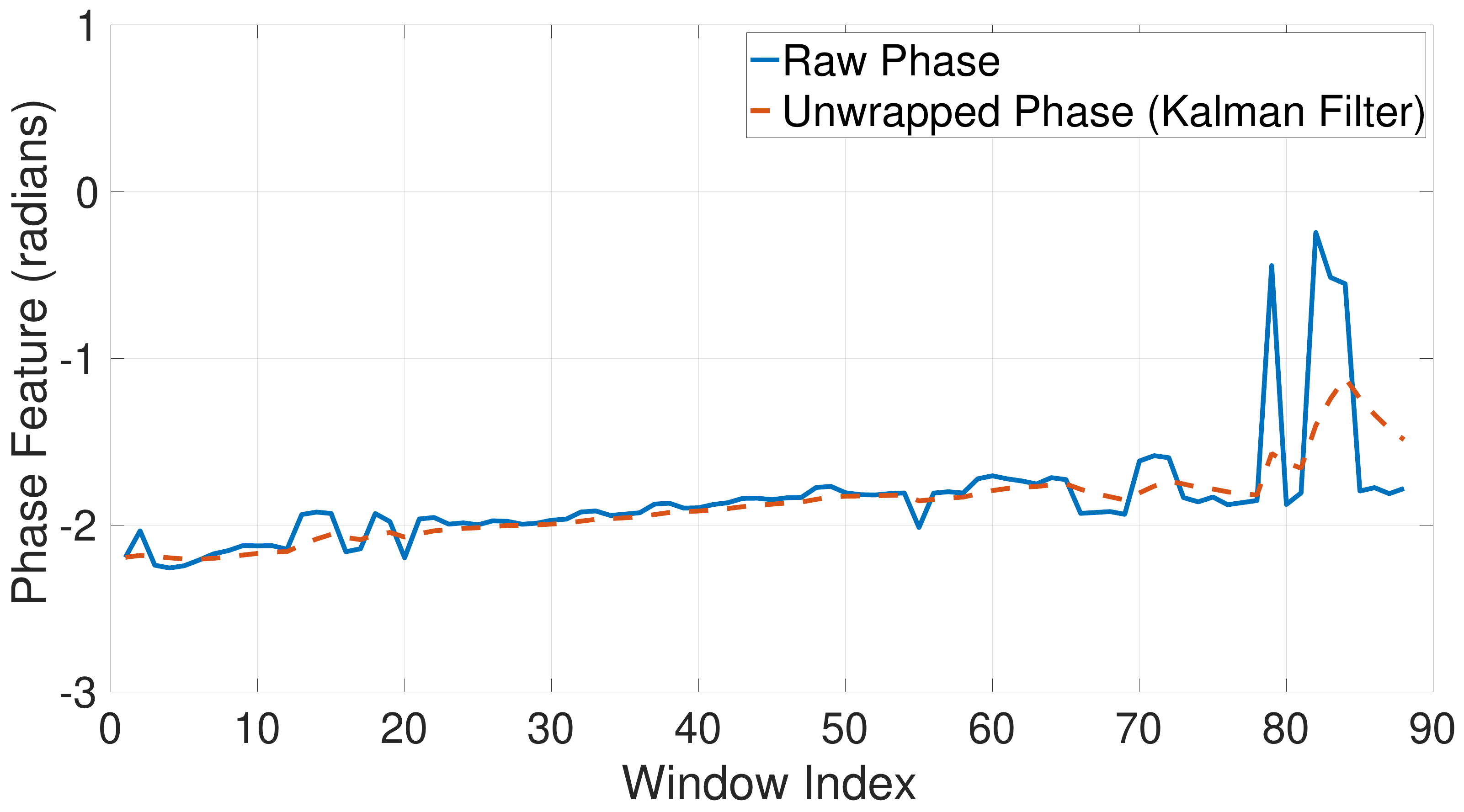}
        		\subcaption{Phase feature}
        		\label{Fig8f}
			\end{subfigure}
\begin{subfigure}{\textwidth}
        		\centering
        		\includegraphics[width=0.98\textwidth]{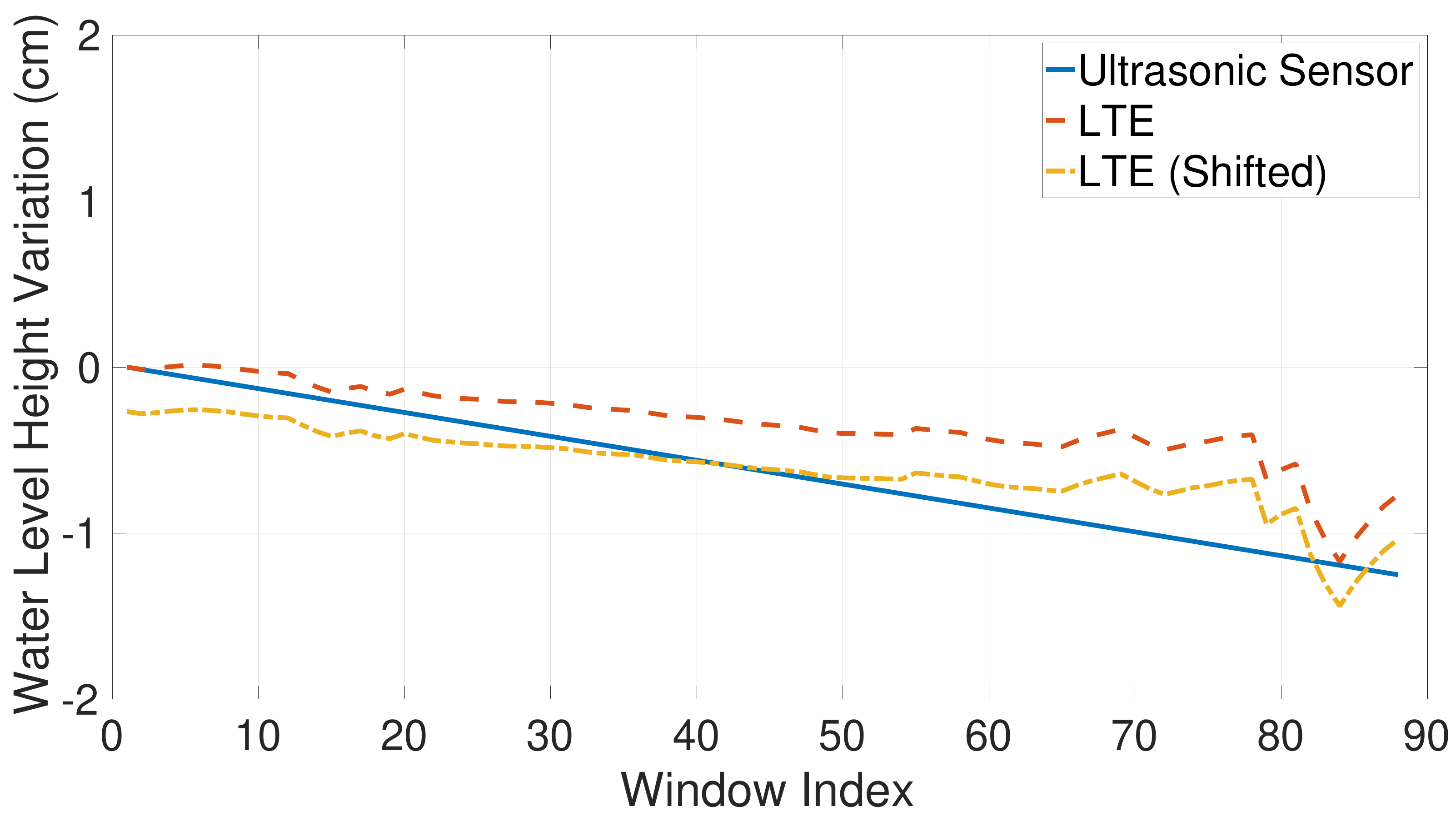}
        		\subcaption{Height variation}
        		\label{Fig8g}
			\end{subfigure}
		\caption{Pool$\rightarrow$Tank (LTE).}
		\label{Fig8}
		\end{minipage} 
\end{figure*}

\subsection{Implementation}
\textit{Experimental Scenario.} In Fig. \ref{figure4}, an indoor environment is constructed to simulate water level changes. The setup includes a circular pool, with water levels controlled via a tank and an electronic pump. The pump alternates between filling and draining, inducing water level variations of 3.5 cm over 8 minutes. To introduce real-world multipath effects, artificial plants are placed around the pool, while air conditioning vents create dynamic airflow, adding environmental variability.

\textit{Signal Generation.} During water level changes, CSI data is simultaneously collected using two systems operating at 28 GHz (mmWave) and 3.1 GHz (LTE). The mmWave signals provide higher resolution due to their shorter wavelength, whereas the LTE signals offer longer-range coverage but lower sensitivity. \textit{(1) mmWave System.} The mmWave system is built using the TMYTEK BBox One 5G Development Tool, including a baseband module and a phased-array antenna module. Both the transmitter and receiver are equipped with phased-array antennas for analog beamforming, producing a single-channel signal. We configure it to operate with a 70 MHz bandwidth, 46 subcarriers and a CSI sampling rate of 100 Hz. The transmitter and receiver are positioned 5 m apart, with antennas placed at a height of 1 m above the water surface. \textit{(2) LTE System.} The LTE system consists of an NI Massive MIMO testbed as the receiver and a USRP device as the transmitter. It operates with a 20 MHz bandwidth, 100 subcarriers and a CSI sampling rate of 2 KHz. The transmitter has a single omnidirectional antenna, while the receiver has a three-element directional antenna array. They are placed 8.5 m apart, with antennas at 1.5 m above the water surface.

\textit{Non-Uniform Sampling Simulation.} In real-world applications, since water level changes occur gradually, continuous high-frequency CSI extraction is often unnecessary. Periodic sampling can capture water level variations while reducing computational load, energy consumption, and storage requirements. While our mmWave and LTE systems support uniform sampling, we emulate practical conditions using interval-based sampling in our experiments. By default, the CSI is collected for 2 seconds per session, followed by a gap of several seconds before the next round. A time window is set to 5 minutes, with a 20-second interval between sessions. Each window consists of 14 sessions, each lasting 22 seconds, including the 2-second CSI collection period and the 20-second gap.

\textit{Ground Truth.} An ultrasonic sensor is installed at the edge of the pool, aligned with the water surface, providing millimeter-level accuracy for water level measurement. Since our scheme focuses on detecting water level variations, it needs to collect CSI data over a time window at the initial stage, which introduces a temporal delay relative to the actual water level variation. In our experiment, we first normalize the estimated water levels by subtracting the mean of all estimates across time windows. Next, we set the initial value as a reference point, shifting all values to ensure the height starts from zero. The same process is applied to the ultrasonic sensor data. Finally, we interpolate the estimated values over time to align them with the ground truth for height error estimation.

\begin{figure*}
\centering
\begin{minipage}[t]{0.49\linewidth}
\vspace{0pt}  
\centering
\begin{subfigure}{0.49\textwidth}
	\centering
	\includegraphics[width=\textwidth]{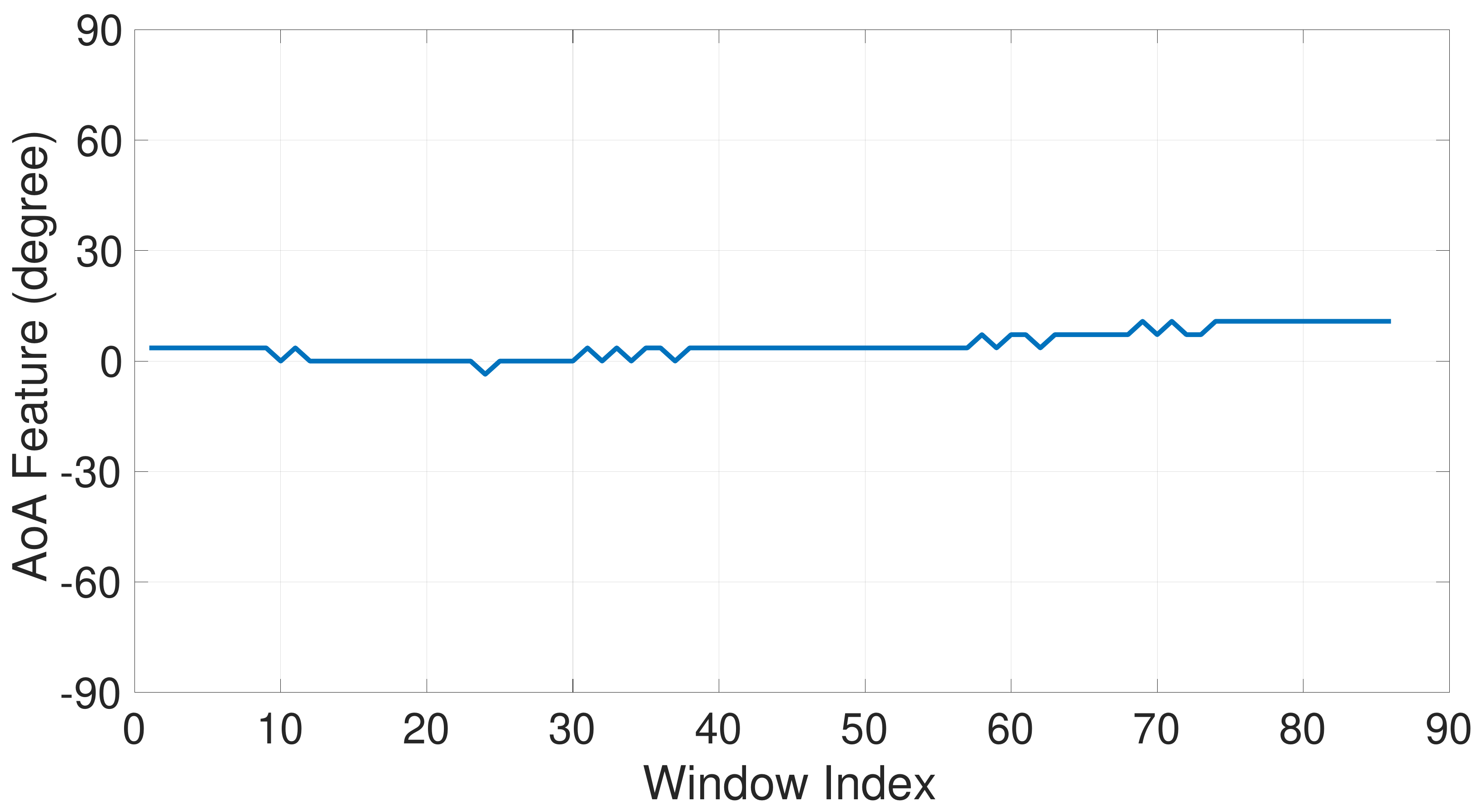}
	\subcaption{Tank$\rightarrow$Pool (LTE)}
	\label{Fig9a}
\end{subfigure}
\begin{subfigure}{0.49\textwidth}
	\centering
	\includegraphics[width=\textwidth]{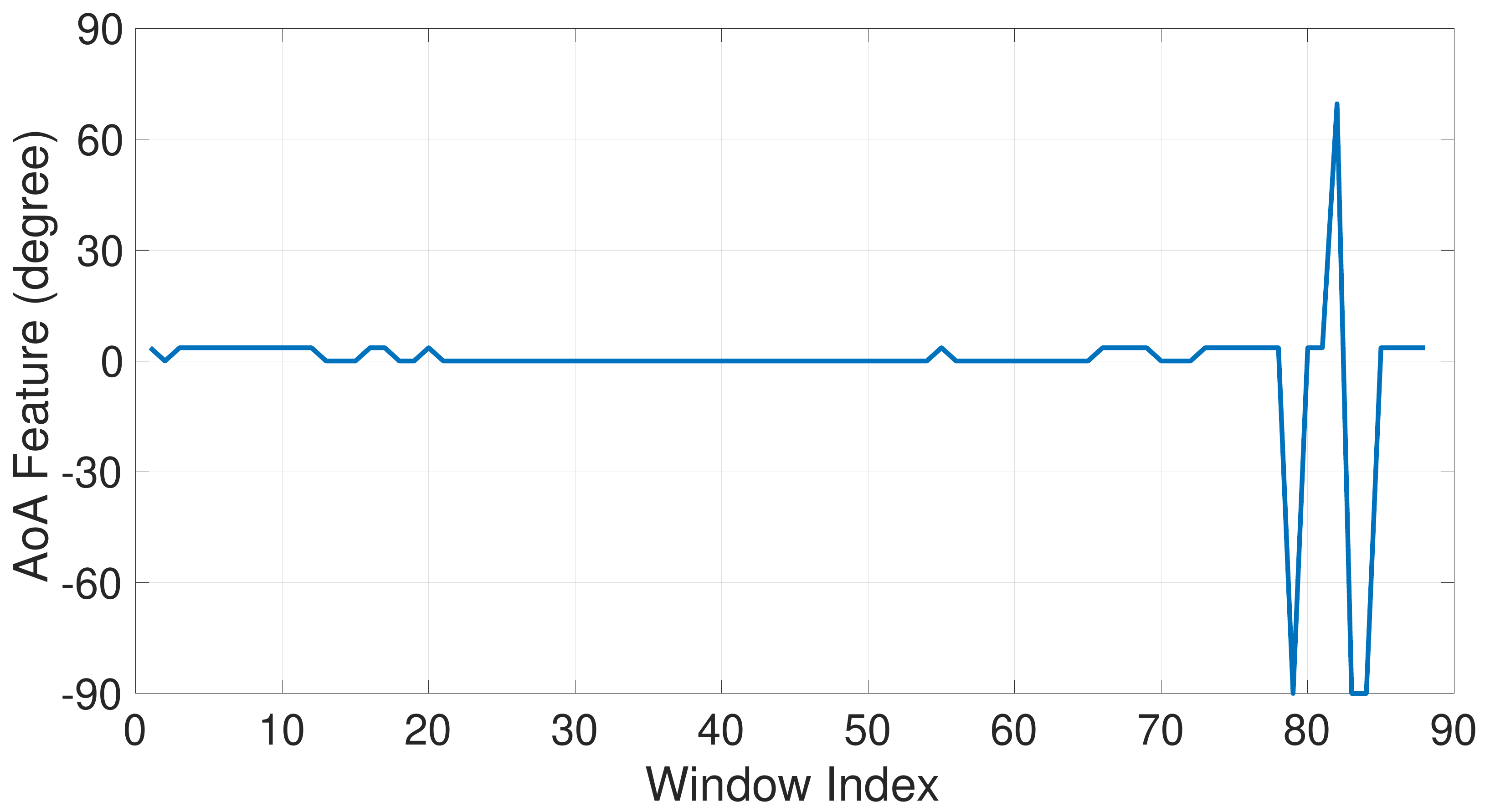}
	\subcaption{Pool$\rightarrow$Tank (LTE)}
	\label{Fig9b}
\end{subfigure}
\caption{AoA feature.}
\label{Fig9}
\end{minipage}
\begin{minipage}[t]{0.49\linewidth}
\vspace{0pt} 
\centering
\begin{subfigure}[t]{0.49\textwidth}
	\centering
	\includegraphics[width=\textwidth]{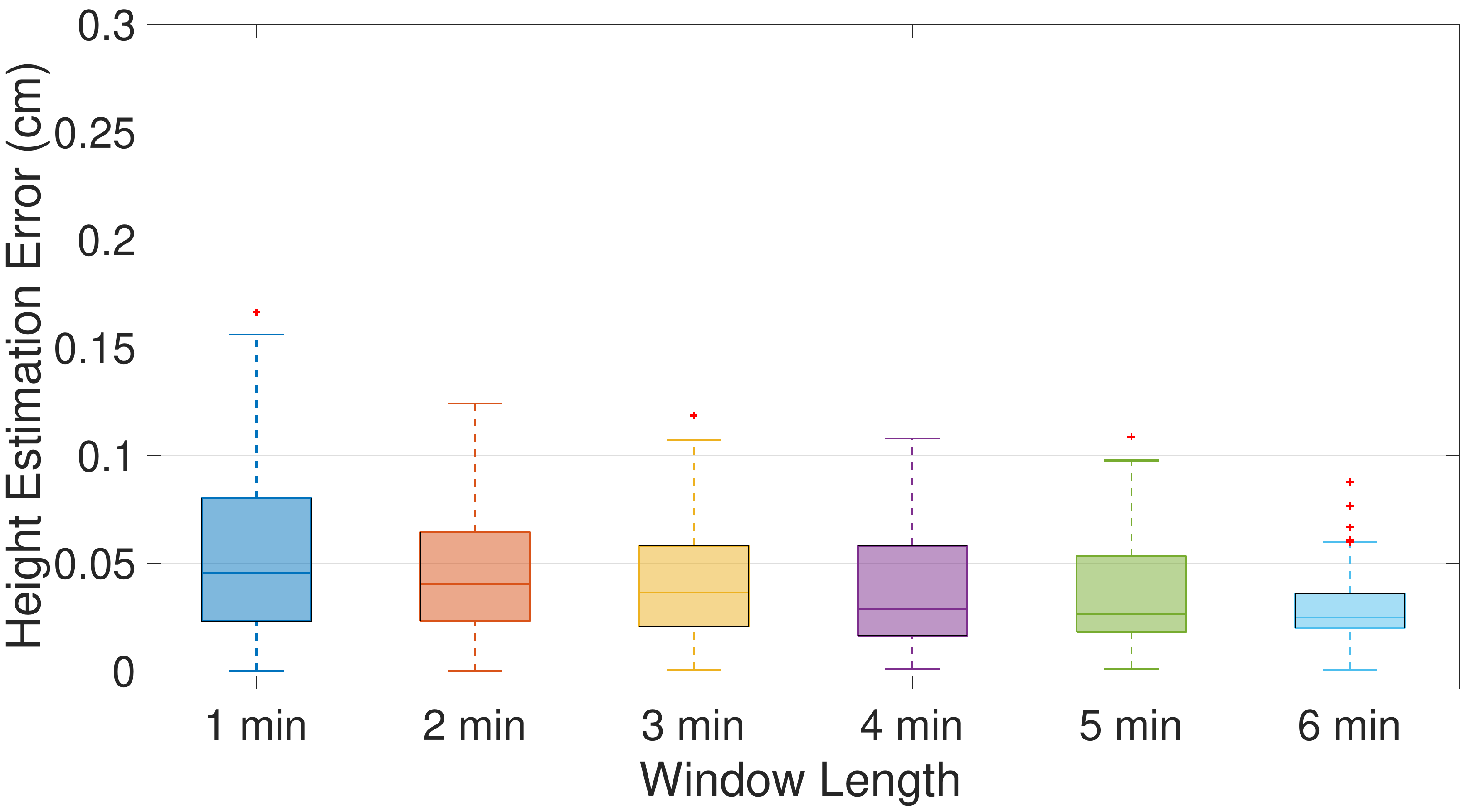}
	\subcaption{Tank$\rightarrow$Pool (mmWave)}
	\label{Fig10a}
\end{subfigure}
\begin{subfigure}[t]{0.49\textwidth}
	\centering
	\includegraphics[width=\textwidth]{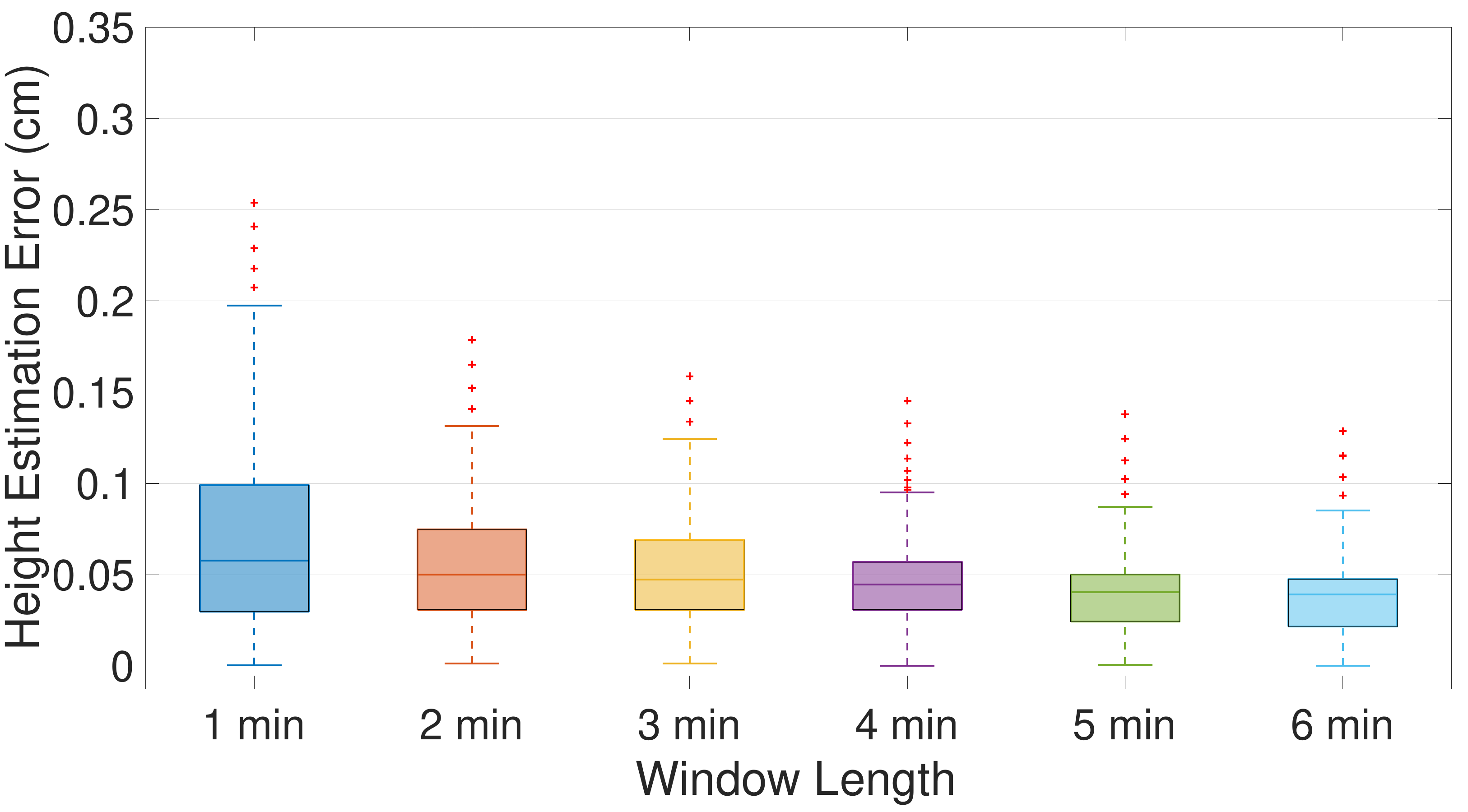}
	\subcaption{Pool$\rightarrow$Tank (mmWave)}
	\label{Fig10b}
\end{subfigure}
\caption{Impact of window length.}
\label{Fig10}
\end{minipage}
\vspace{-1.5em}
\end{figure*}

\subsection{Water Level Sensing Performance}
This section evaluates the proposed water level sensing method using mmWave and LTE signals. The 3.1 GHz LTE system, with three Rx antennas, supports spatial filtering processing, while the 28 GHz mmWave system, utilizing a single RF channel, does not apply the filtering. In Fig. \ref{Fig5}-Fig. \ref{Fig8}, subfigures (a) display the raw CSI amplitude of a subcarrier, uniformly sampled over 8 minutes during \textit{in-flow} (water flowing from the tank to the pool) and \textit{out-flow} (water flowing from the pool to the tank). Subfigures (b) present downsampled CSI, simulating real-world data collection. Subfigures (c) and (d) depict the extracted Doppler and range features. Subfigures (e) and (f) illustrate the normalized amplitude and phase features for water level sensing. Subfigures (g) show the variations in water level height. Key observations are described as follows:

\textit{Impact of Water Level Variation on Signals.} As shown in subfigures (a), the mmWave signal, with a wavelength of 1.07 cm, undergoes multiple cycles of amplitude changes; for the LTE signal, with a longer wavelength of 9.68 cm, the amplitude changes over roughly a quarter of the wavelength. We can see that directly distinguishing water level changes solely based on raw signal amplitude is challenging. For mmWave signals in Fig. \ref{Fig5} and Fig. \ref{Fig6}, amplitude variations remain similar for both rising and falling water levels, making it challenging to distinguish between them. In contrast, LTE signals in Fig. \ref{Fig7} and Fig. \ref{Fig8} exhibit some monotonic trends, but these variations depend on the relative positions of the transmitter, receiver, and water level, leading to inconsistencies in different scenarios. According to our signal model, water level changes may cause phase shifts within different intervals: for example, from $0$ to $\pi/2$ (amplitude increases), $\pi/4$ to $3\pi/4$ (amplitude rises then falls), or $\pi/2$ to $\pi$ (amplitude decreases). Furthermore, environment interference in real-world scenarios further complicates signal differentiation for them.

\textit{Water Level Rise (In-flow, Tank$\rightarrow$Pool).} As water flows into the pool, the NLOS path length shortens, increasing signal amplitude. Figs. \ref{Fig5}e and \ref{Fig7}e demonstrate that the amplitude features can match theoretical expectations, while Figs. \ref{Fig5}f and \ref{Fig7}f illustrate the decreasing phase. For mmWave signals, the phase decreases by 7.96 radians, indicating a path change of about 1.35 cm, while for LTE signals, the phase decreases by approximately 1.1 radians, corresponding to a path change of about 1.7 cm. These findings underscore the higher resolution of phase features than amplitude features, allowing for precise path length tracking. Despite phase periodicity introducing ambiguity, our Kalman filter-based phase unwrapping effectively ensures accurate tracking. Furthermore, the shorter wavelength of mmWave signals enhances sensing precision and provides greater resilience to noise compared to LTE signals.

\textit{Water Level Fall (Out-flow, Pool$\rightarrow$Tank).} When water flows out of the pool, the opposite trends are observed. As the water level decreases, the NLOS path length increases, leading to a decrease in CSI amplitude and an increase in phase. Fig. \ref{Fig6}e and Fig. \ref{Fig8}e show the amplitude features with a gradual decline, while Fig. \ref{Fig6}f and Fig. \ref{Fig8}f illustrate increasing phase features. These observations further confirm the effectiveness of our algorithm in accurately tracking water level variations.

\textit{Water Level Estimation Accuracy.} Our scheme estimates water level height using extracted phase and known transceiver geometry. As discussed in Section IV-B, with equal Tx and Rx heights, the angle is approximated as $45^\circ$. Subfigures (g) in Fig. \ref{Fig5}-\ref{Fig8} show the water level estimation accuracy. The mmWave system achieves a higher accuracy, with an average error of 0.021 cm for in-flow and 0.029 cm for out-flow, while the LTE system has larger average errors of 0.246 cm and 0.149 cm. The higher accuracy of the mmWave system is due to its shorter wavelength, which enables finer resolution in detecting subtle water level variations.

\subsection{Time, Frequency, and Spatial Feature Analysis}
Our algorithm leverages the time (Doppler), frequency (Delay), and spatial (AoA) domains to suppress noise and extract the water surface reflection path for robust water level sensing.  

For the mmWave system, Figs.~\ref{Fig5} and \ref{Fig6} (subfigures c and d) illustrate the extracted Doppler and range features through time- and frequency-domain analyses, respectively. Only Doppler and range features are analyzed, as it operates with a single-channel output. The shorter wavelength enables finer resolution, making these features highly reliable for capturing small water level variations. Despite the absence of spatial filtering, these features can ensure precise tracking. Building on this, the extracted phase information is even more reliable, allowing highly precise water level height estimation.

For the LTE system, due to its limited bandwidth, Figs.~\ref{Fig7} and \ref{Fig8} (subfigures c and d) show range features exhibit some instability, whereas Doppler features offer relatively stable results for capturing the water level changes. However, occasional outliers in Doppler estimates are observed. Purely relying on Doppler for water level tracking may lead to accumulated errors over time. Fig.~\ref{Fig9} illustrates the extracted AoA features, which are also inadequate for reliably indicating water level changes due to limited spatial resolution. In our scheme, the spatial-domain processing is mainly employed to refine  the extraction of the water surface reflection.

\begin{figure*}
\centering
\begin{subfigure}{0.253\textwidth}
	\centering
	\includegraphics[width=\textwidth]{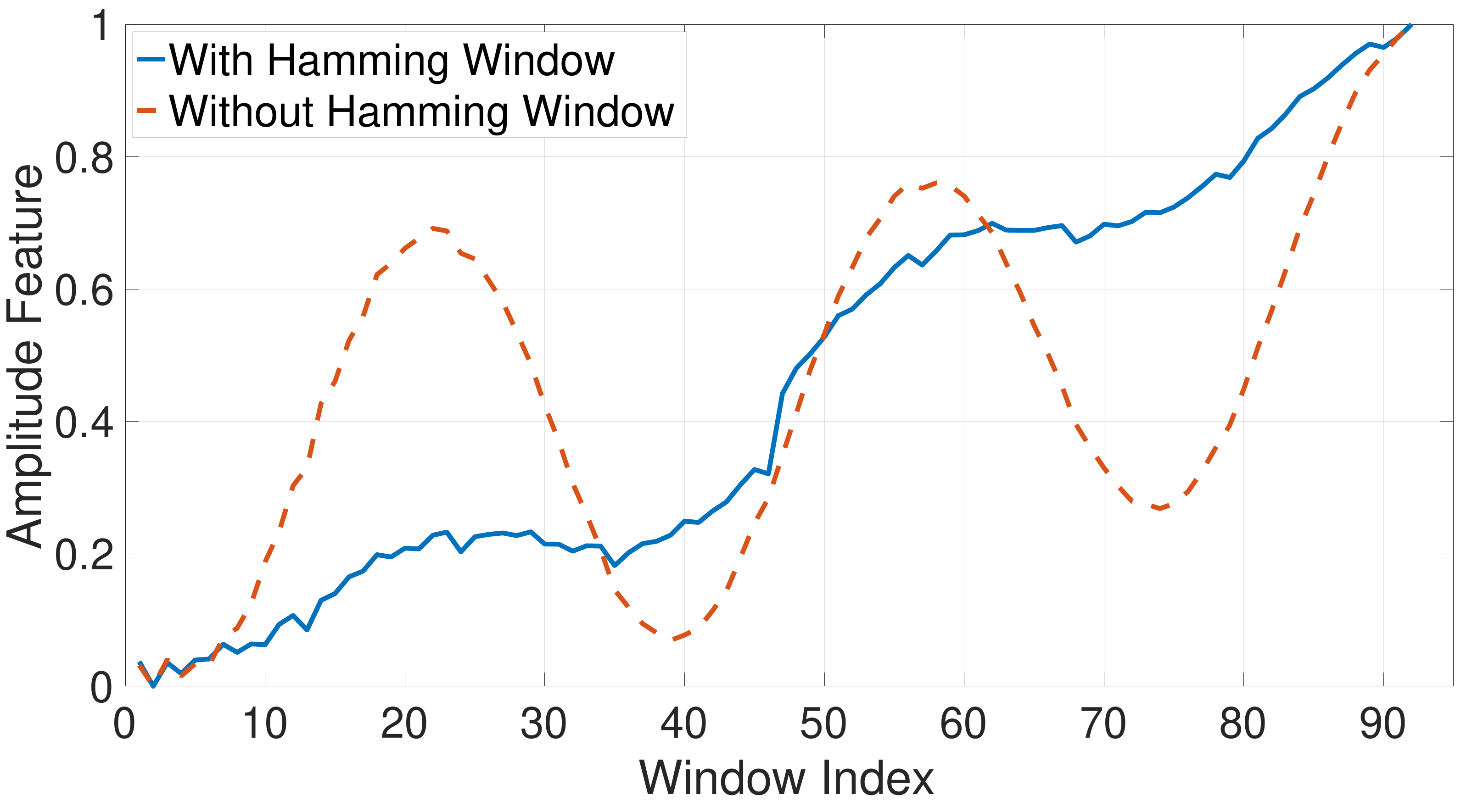}
	\subcaption{Amplitude feature (Tank$\rightarrow$Pool)}
	\label{Fig11a}
\end{subfigure}
\begin{subfigure}{0.253\textwidth}
	\centering
	\includegraphics[width=\textwidth]{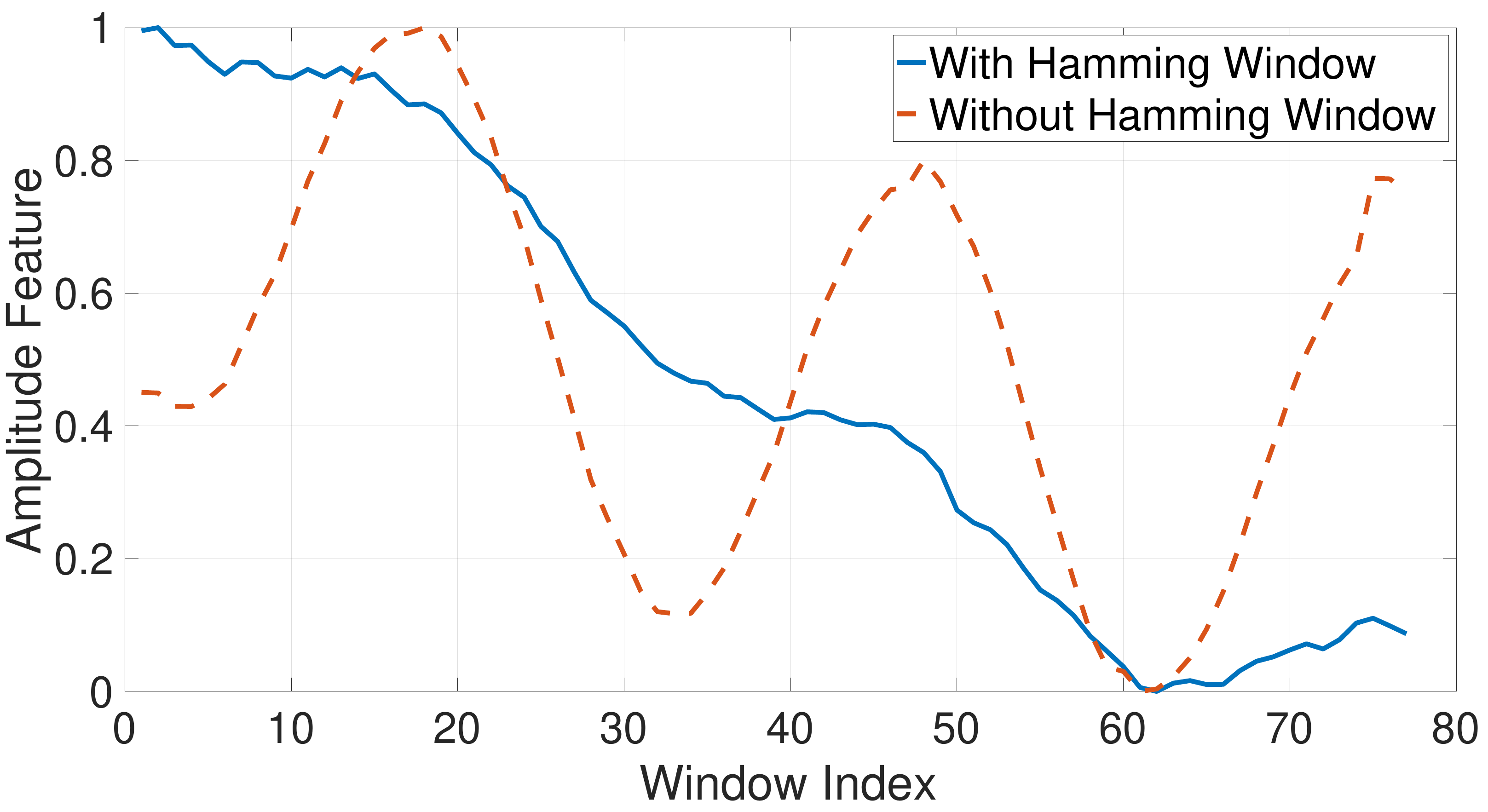}
	\subcaption{Amplitude feature (Pool$\rightarrow$Tank)}
	\label{Fig11b}
\end{subfigure}
\begin{subfigure}{0.238\textwidth}
	\centering
	\includegraphics[width=\textwidth]{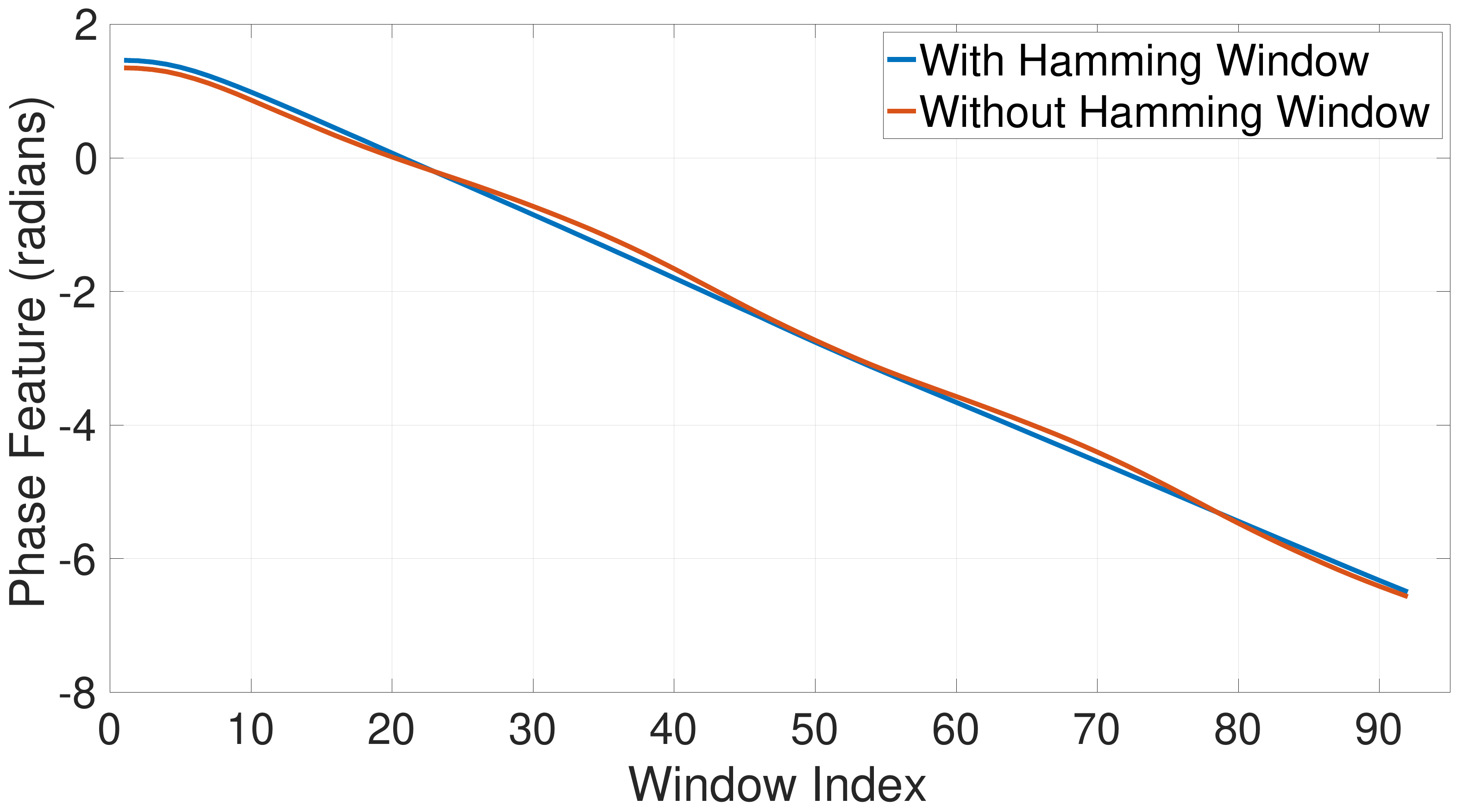}
	\subcaption{Phase feature (Tank$\rightarrow$Pool)}
	\label{Fig11c}
\end{subfigure}
\begin{subfigure}{0.238\textwidth}
	\centering
	\includegraphics[width=\textwidth]{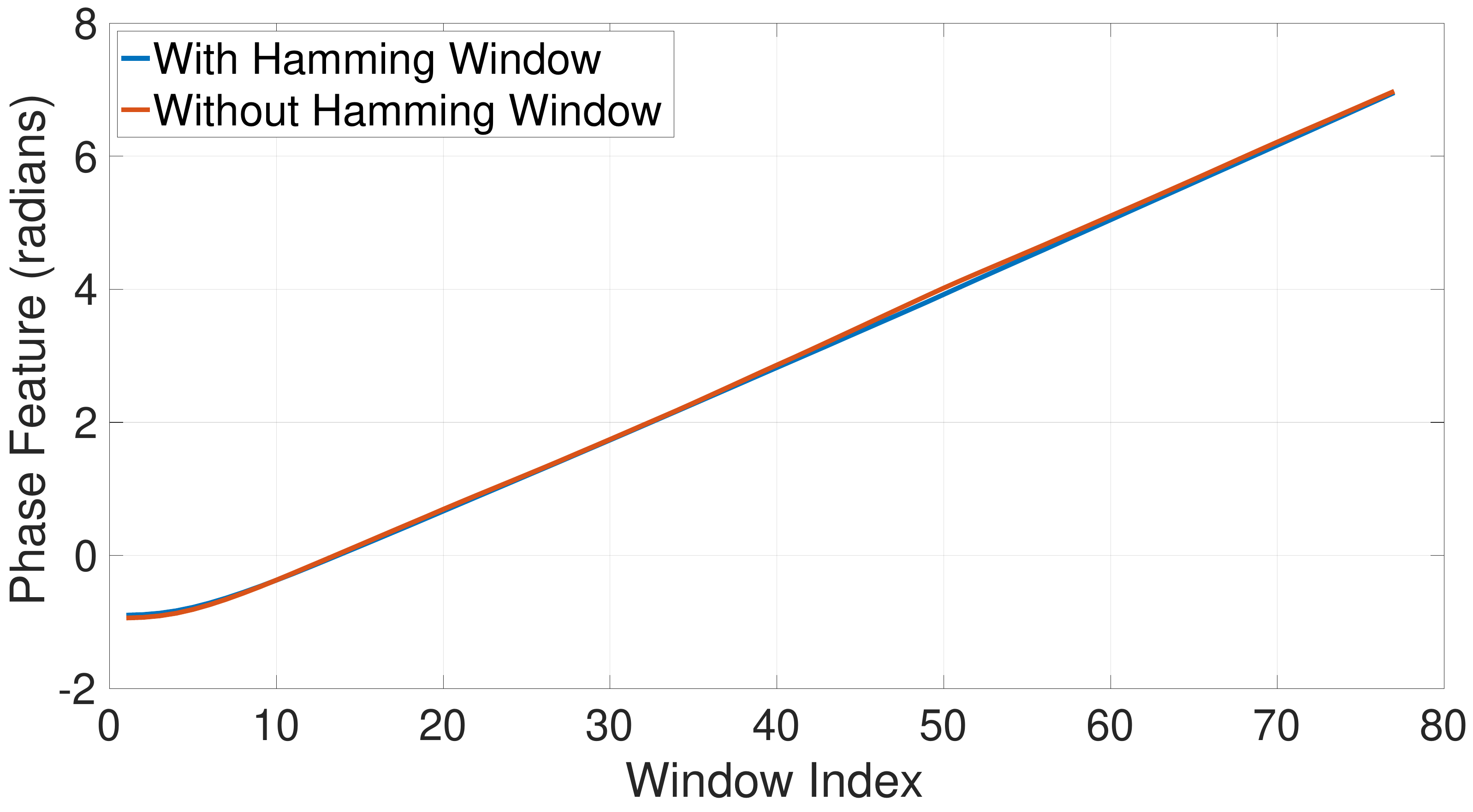}
	\subcaption{Phase feature (Pool$\rightarrow$Tank)}
	\label{Fig11d}
\end{subfigure}
\caption{Impact of Doppler windowing function.}
\label{Fig11}
\vspace{-1em}
\end{figure*}

\begin{table}
\caption{\textbf{Water level variation detection accuracy (\%)}. The table presents the true positive (TP) and false positive (FP) counts, along with the final accuracy of water level variation detection.}
\vspace{-1em}
\label{table1}
\center
\begin{tabular}{c|c|c|c}
\hline
\bfseries{Signals}  & \bfseries{FP (\#)} & \bfseries{TP (\#)} & \bfseries{Accuracy (\%)}\\
\hline
{Level Variation (mmWave)} &  0 &  1686  & 100.0\\
{Level Variation (LTE)}    & 50 &  1442  & 96.64 \\
\hline
{No Level Variation (mmWave)}  & 1 &  916 & 99.89 \\
{No Level Variation (LTE)}     & 12 &  721 & 98.36 \\
\hline
\end{tabular}
\vspace{-1.5em}
\end{table}

\subsection{Water Level Variation Detection Accuracy}
We evaluate the performance of our CFAR-based water level variation detection. In our scheme, we use 8 reference cells, 4 guard cells, and a threshold factor of 0.01. Peaks exceeding the adaptive threshold are classified as water level variations. As shown in Table. \ref{table1}, when the water level changes, the mmWave system achieves 100\% detection accuracy, while the LTE system reaches 96.64\%. In the absence of water level variations, their detection precisions are 99.89\% and 98.36\%, respectively. As expected, the mmWave system benefits from its shorter wavelength and wider bandwidth, enhancing its sensitivity to small water level variations.

\subsection{Real-time Performance} 
\begin{table}
\caption{\textbf{Running time (seconds)}}
\vspace{-1em}
\label{table2}
\center
\resizebox{0.49\textwidth}{!}{
\begin{tabular}{c|c|c|c|c}
\hline
\bfseries{Signals} & \bfseries{Language} & \bfseries{Platform}  & \bfseries {Mean (s)} & \bfseries {Std. (s)} \\
\hline
mmWave & Python & MacBook Pro 2019  &  0.11  &  0.011\\
LTE  & Python &  MacBook Pro 2019   &  1.92  &  0.109 \\ 
\hline
\end{tabular}}
\vspace{-1em}
\end{table}
 
We evaluate the real-time capability on a MacBook Pro (2019) with a 2.6 GHz 6-Core Intel Core i7 processor. CSI data from both mmWave and LTE systems is transmitted via UDP to the laptop, where processing is performed in Python without hardware acceleration. As shown in Table~\ref{table2}, the mmWave system achieves an average computation time of 0.11 s per window, while the LTE system requires 1.92 s. Given the LTE system's 2 kHz sampling rate compared to 100 Hz for mmWave, its CSI data volume is 20 times larger. With a 20-second step size between adjacent windows, the scheme meets real-time processing requirements. Computational efficiency is improved by employing a non-uniform FFT for Doppler estimation across subcarriers and antennas, significantly reducing processing load. Additionally, object detection is applied to the Doppler-Range heatmap to identify the most likely target, enabling efficient spatial filtering. The Kalman filter, operating linearly, further ensures low computational complexity.

\subsection{Impact of Window Length}
We vary the CSI window length from 1 to 6 minutes using the mmWave system to analyze its impact on water level sensing performance. Within each window, CSI data is still collected using non-uniform sampling, where a 2-second session is followed by a 20-second interval. As shown in Fig.~\ref{Fig10}, the estimation error decreases as the window length increases, ranging from approximately 0.05 cm at 1 minute to 0.02 cm at 6 minutes for in-flow and our-flow cases. The mmWave system maintains high precision even with smaller data windows. A larger window improves reflection signal separation but also increases initialization time, leading to larger absolute height differences relative to current water level changes. In practical applications, a trade-off between accuracy and real-time performance should be considered.

\begin{figure}
\centering
\begin{subfigure}{0.24\textwidth}
	\centering
	\includegraphics[width=\textwidth]{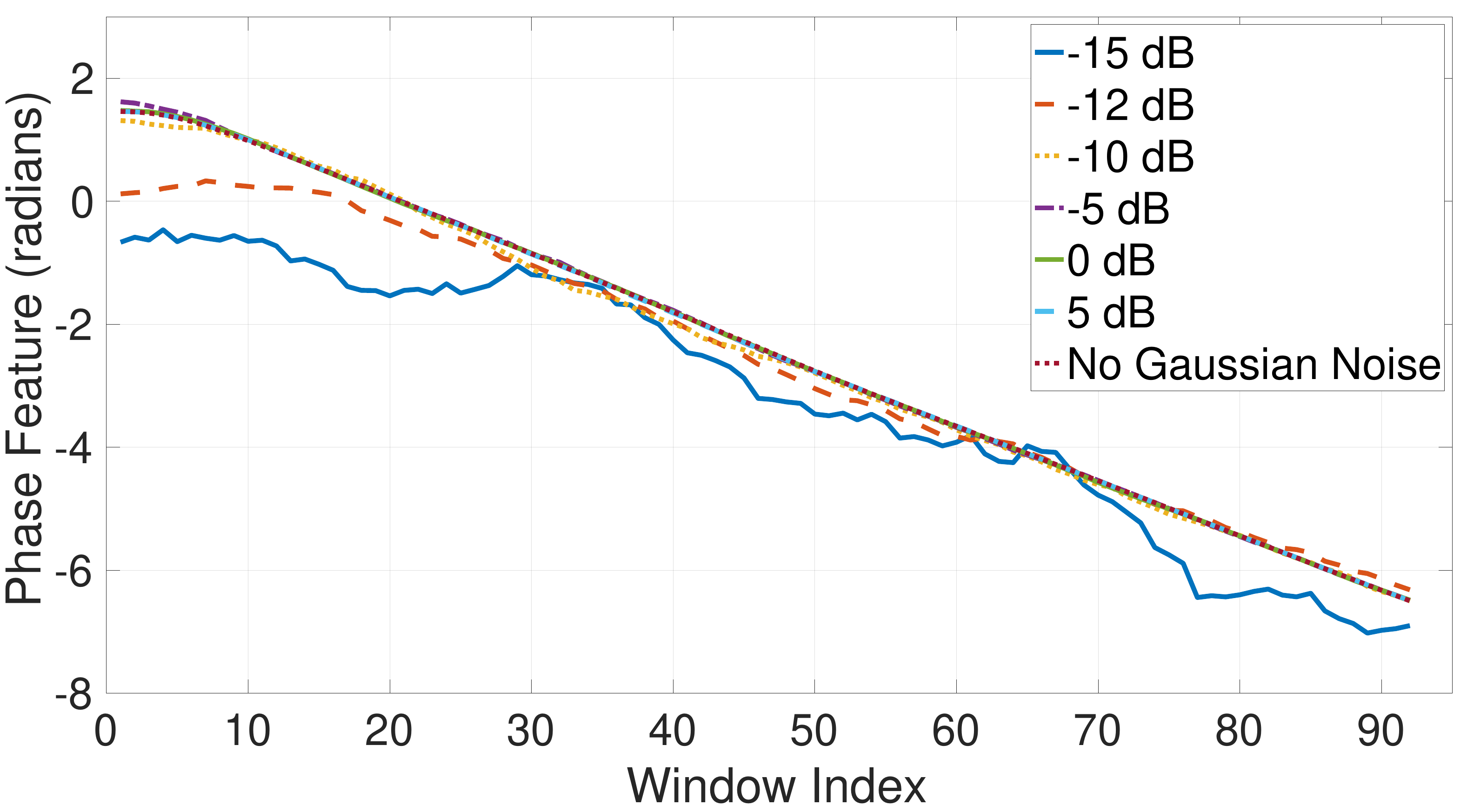}
	\subcaption{Phase feature (Tank$\rightarrow$Pool)}
	\label{Fig12a}
\end{subfigure}
\begin{subfigure}{0.24\textwidth}
	\centering
	\includegraphics[width=\textwidth]{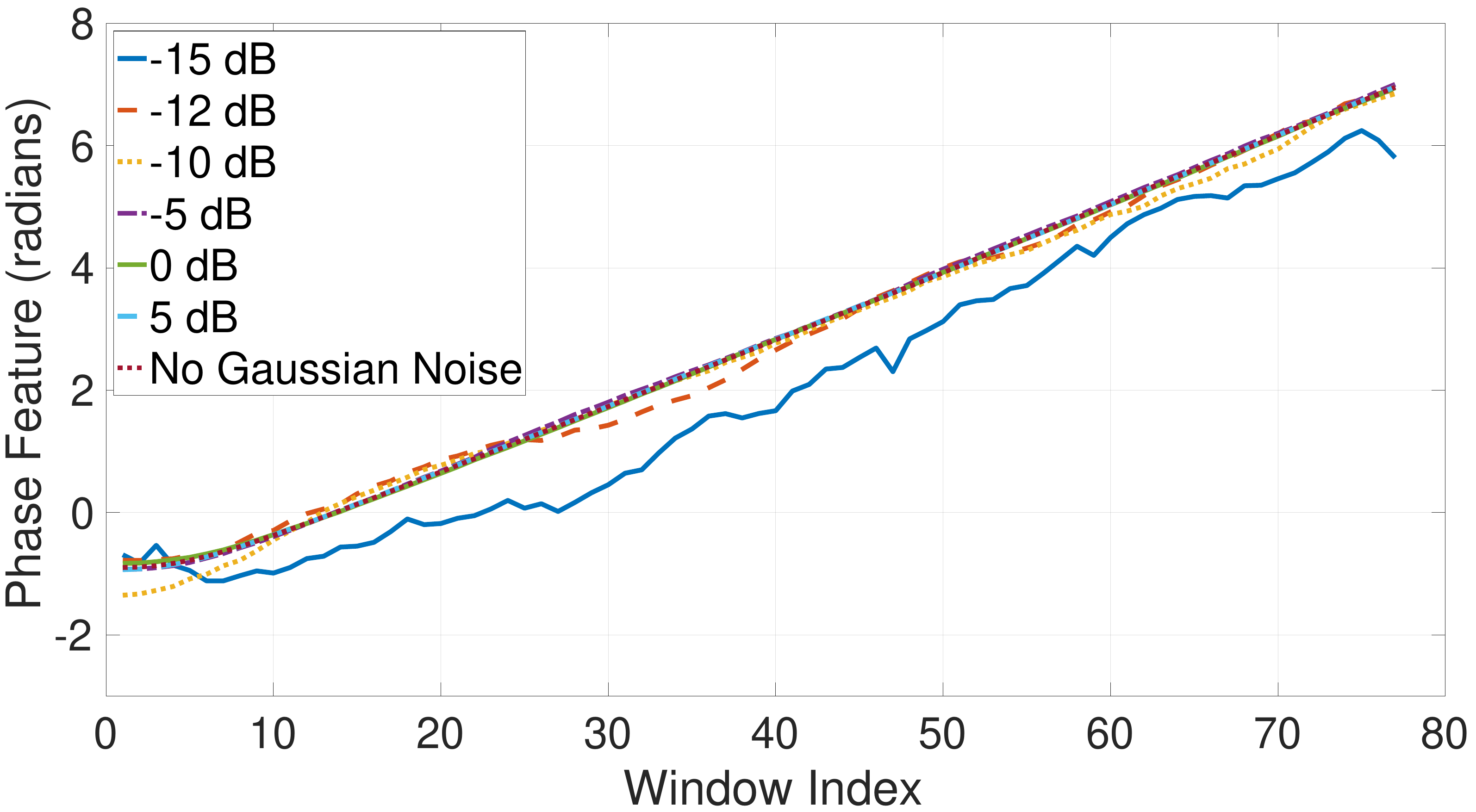}
	\subcaption{Phase feature (Pool$\rightarrow$Tank)}
	\label{Fig12b}
\end{subfigure}
\caption{Impact of environment interference.}
\label{Fig12}
\vspace{-1.5em}
\end{figure}

\begin{figure*}
\centering
\includegraphics[width=0.95\textwidth]{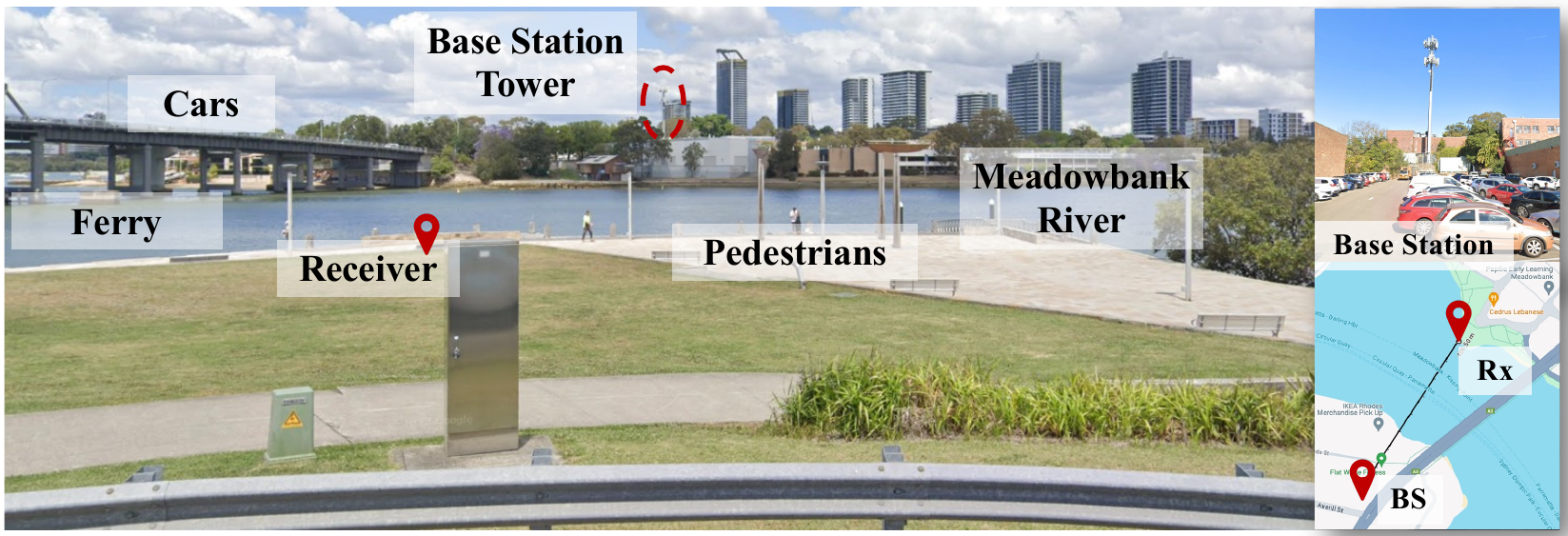}
\caption{Outdoor experiment on water level sensing at the Parramatta River near Meadowbank, Sydney, Australia.}
\label{Fig13}
\vspace{-1em}
\end{figure*}

\begin{figure*}
\centering
	\begin{minipage}[t]{0.329\linewidth}
	\centering
		\includegraphics[width=\textwidth]{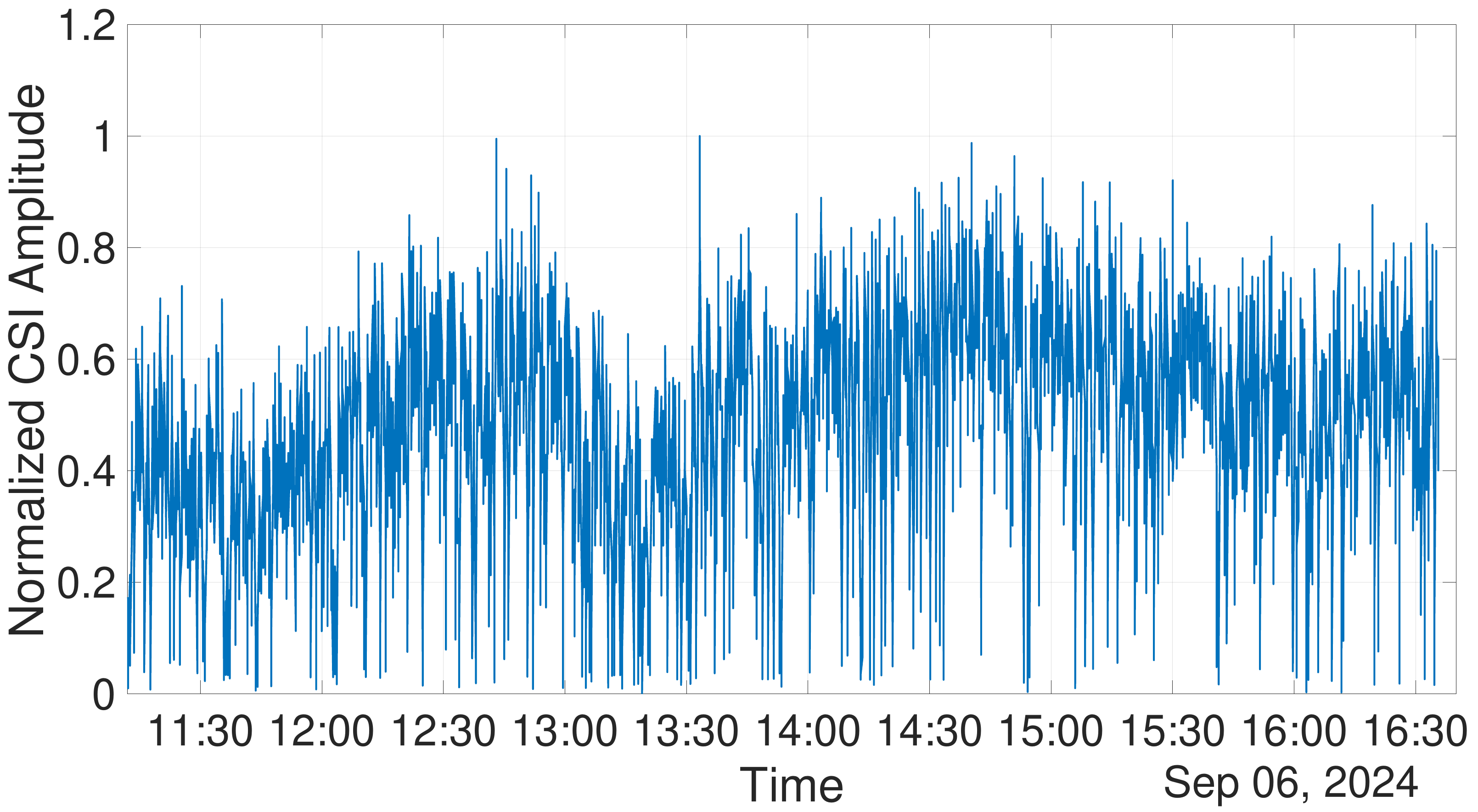}
		\caption{LTE CSI amplitude.}
		\label{Fig14}
	\end{minipage}
	\begin{minipage}[t]{0.329\linewidth}
	\centering
		\includegraphics[width=\textwidth]{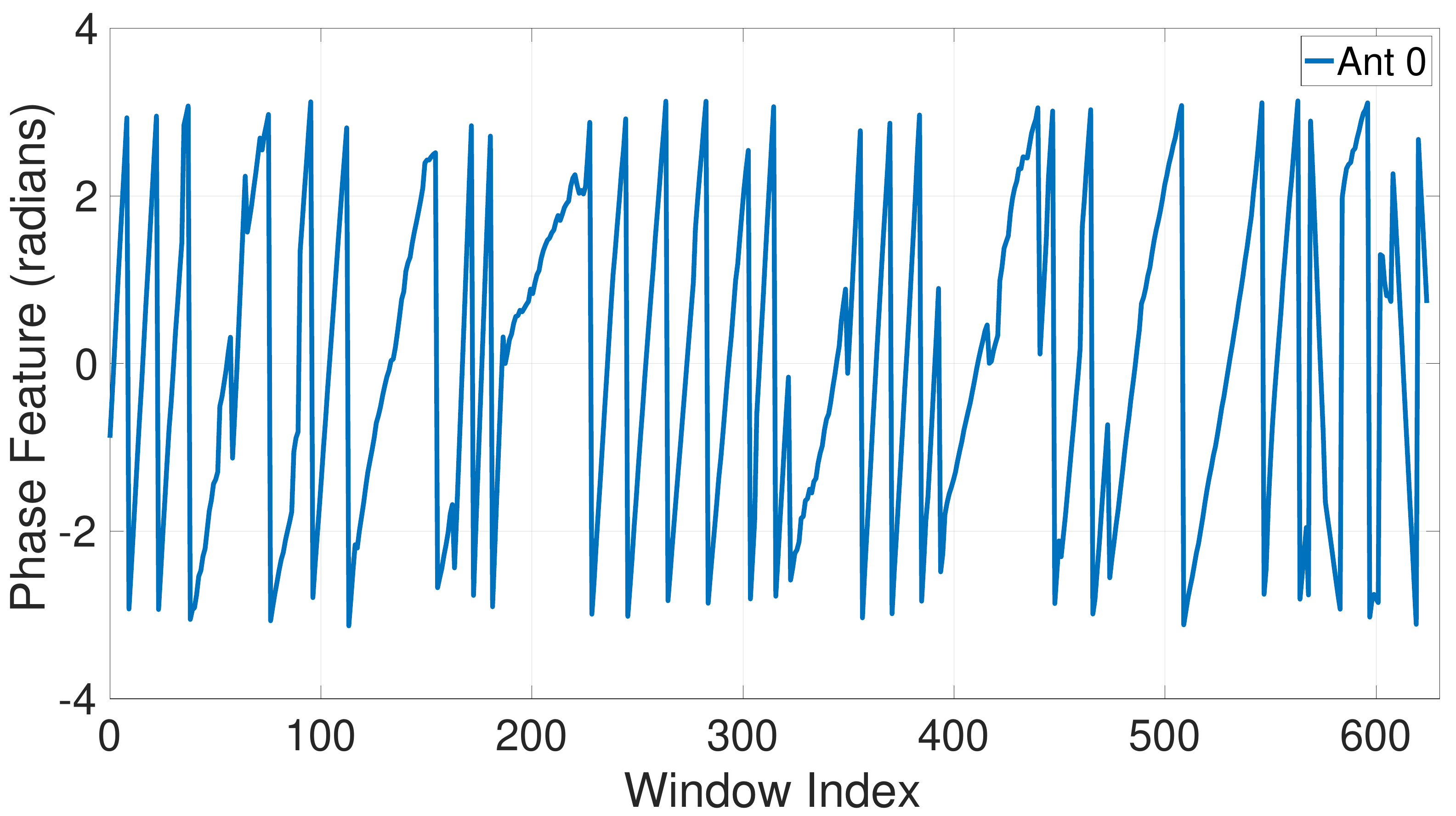}
		\caption{Extracted phase feature.}
		\label{Fig15}
	\end{minipage}
	\begin{minipage}[t]{0.329\linewidth}
	\centering
		\includegraphics[width=\textwidth]{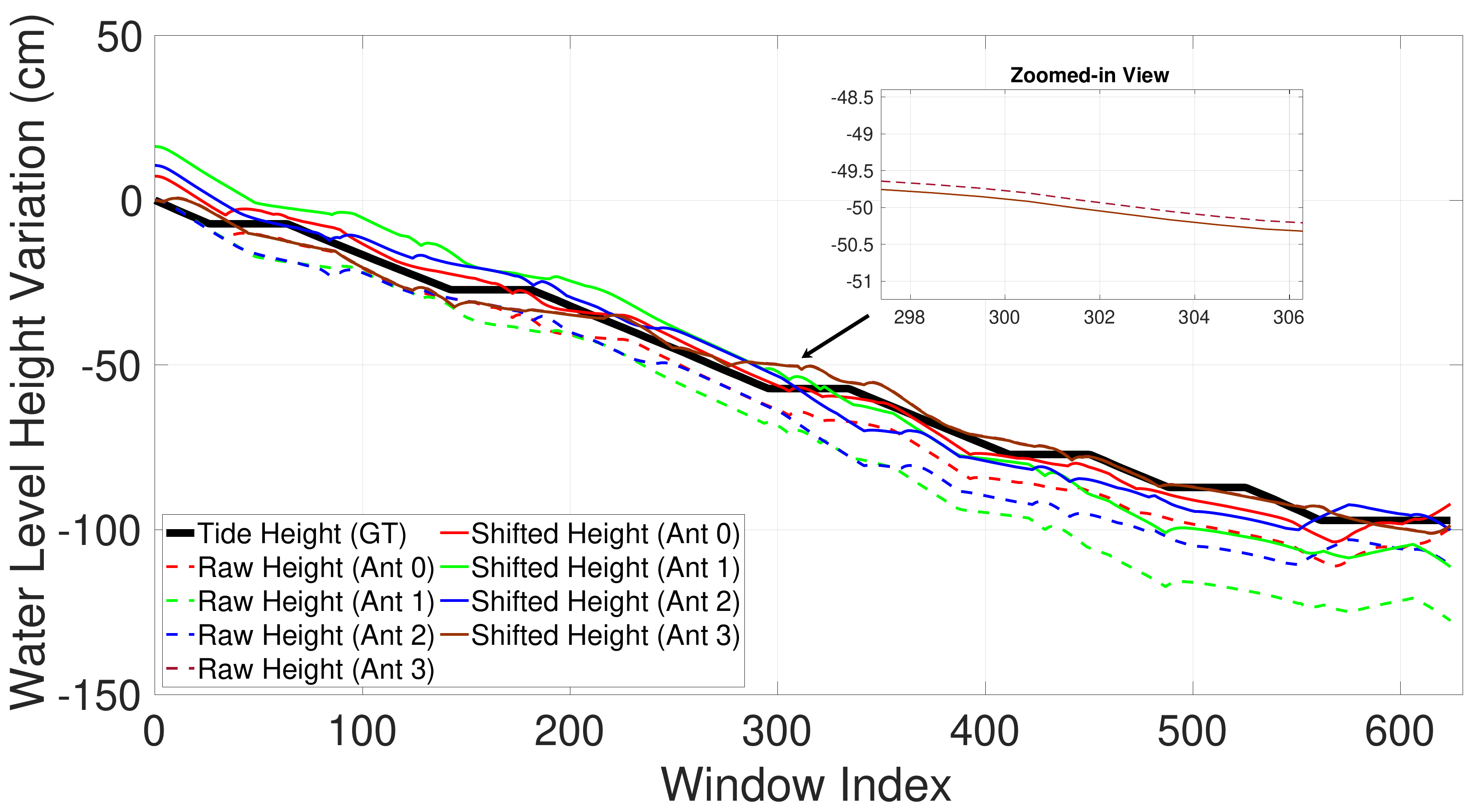}
		\caption{Water level height variation.}
		\label{Fig16}
	\end{minipage}
\vspace{-1.5em}
\end{figure*}

\subsection{Impact of Doppler Windowing Function}
To mitigate low-frequency components caused by power variations, we assess the effect of Doppler windowing on sensing performance. As shown in Fig.~\ref{Fig11}, applying a Hamming window produces smoother amplitude and phase features. Without windowing, spectral leakage and sidelobe effects from abrupt signal truncation introduce periodic fluctuations, further amplified by power variations. The Hamming window reduces these distortions by tapering the signal edges, improving amplitude stability. Given its effective sidelobe suppression, it is recommended for Doppler analysis.

\subsection{Impact of Environment Interference}  
We evaluate the scheme's robustness under different noise conditions. Additive White Gaussian Noise (AWGN) is introduced at varying intensities to the original CSI, with the Signal-to-Noise Ratio (SNR) defined as $P_{\text{noise}} = {P_{\text{CSI}}}/\left({10^{\text{SNR(dB)}/10}} \right)$, where $P_{\text{CSI}}$ represents the average CSI power. The SNR is set to -15 dB, -12 dB, -10 dB, -5 dB, 0 dB, 5 dB, and a no-noise baseline, respectively. Fig.~\ref{Fig12} shows the phase features remain closely aligned with the ground truth across different noise levels. Even under strong interference conditions (SNR = -15 dB), the scheme can effectively extracts phase features with relatively minimal degradation.

In our real-time demo mentioned in Section I, a person moving randomly near the pool introduces multipath reflections and noise. Despite this, the extracted phase features still accurately track water level variations. This robustness is due to the distinct Doppler characteristics of human motion compared to water level changes. The human-induced Doppler shifts are typically higher, irregular, and short-term, whereas water level variations exhibit small and gradually changing Doppler trends. By leveraging the Doppler patterns, the scheme can suppresses human motion interference, ensuring reliable water level sensing even in dynamic environments.

\section{Experiments in a Real Environment}
This section reports the field-trial results of water level sensing in a river using real-world downlink LTE signals. 

\textit{Experiment Scenario.} Fig.~\ref{Fig13} shows the experimental setup to monitors water level variations at the Parramatta River near Meadowbank, Sydney, Australia, which has an approximate width of 260 m. The scenario includes various dynamic interferences such as pedestrians walking along the riverbank, vehicles passing over a nearby bridge, and occasional ferries on the river. The BS tower has a height of 35 m and is located 160 m from the riverbank. The height of the receiver is approximately 1.5 m, placed 0.75 m away from the riverbank. The distance between the Tx and Rx is roughly 424 m.

\textit{Ground Truth.} The ground-truth data is obtained from a reliable tidal reporting website\footnote{\url{https://tides.willyweather.com.au/nsw/sydney/meadowbank.html}} based on tide height measurements collected by physical sensors above the river. The water level height decreases from 1.5 m to 0.5 m during the collection period, with a sampling interval of 18 minutes and a resolution of 0.05 m. The average water level variation is approximately 20 cm per hour, except for the last hour, where the water level remains nearly constant.

\textit{LTE Data Collection.} LTE data with a center frequency of 2.6598 GHz, a bandwidth of 20 MHz, and a 4-antenna MIMO transmitter configuration is collected using Xilinx ZC706 radio hardware as a signal receiver with four Rx antennas. Our data collection is conducted between 11:12:09 and 16:35:35 on 2024/09/06, spanning approximately 5 hours. Each collection captures CSI data over a 0.2-second period, resulting in approximately 1300 CSI samples across 1200 subcarriers. The interval between consecutive collections is 30 seconds.

\textit{System Parameters.} Our experiment uses the single-antenna sensing method without performing spatial filtering. Data from each antenna is processed independently for water level sensing\footnote{This is because the prototyping antennas need extra calibration for coherent processing, which is out of the scope of this work.}. Each estimation is performed using a 30-minute data window. To convert NLOS path length changes into actual water level height variations, the reflection angle is calculated based on the transceiver geometry using Eq.~(\ref{equation25}) and Eq.~(\ref{equation26}). The estimated reflection angle is approximately $85^\circ$. Since the ground truth measurements have a longer sampling interval, we interpolate the data to align with the CSI sampling rate.

Fig.\ref{Fig14} shows the amplitude variations of the CSI at different time (10$\times$ downsampling for clearer visualization). It is evident that water level changes are challenging directly inferred from the raw amplitude variations. Fig.\ref{Fig15} illustrates the phase variations, where periodicity and estimation errors in some instances may lead to erroneous phase unwrapping. Fig.\ref{Fig16} presents the estimated water level height changes from four Rx antennas. Due to the 30-minute data window, the output values have a corresponding time delay of 30 minutes compared to the ground truth. After applying shifting to align the estimated values with the ground truth, the results from all four antennas closely follow the actual water level variations. The estimated water level height shows an average error of approximately 4.8 cm, with a standard deviation of 2.5 cm. These results demonstrate the effectiveness of our scheme, even under real-world conditions with environmental interferences.

\section{Related Work}
Traditional contact-based water level gauges, such as float-based, pressure-based, and reflective systems, typically offer measurement accuracies ranging from 1 to 3 cm \cite{wu2023review}. Despite their reliability, these systems present challenges in high maintenance demands, susceptibility to environmental factors (e.g., sediment buildup and biofouling), and complex deployment requirements in large-scale or remote locations. Additionally, they primarily measure relative water level changes, making them prone to cumulative errors over time. To overcome these limitations, various emerging non-contact technologies have been developed, including ultrasonic systems, mmWave radars, satellite-based remote sensing, vision-driven methods, and communication signal-based approaches. This section reviews these advanced measurement techniques.

\textit{Ultrasonic and mmWave Radar-Based.} They measure a distance by calculating the round-trip time of signals reflected from the water surface. The ultrasonic systems remain popular because of their low cost, ease of installation, and minimal maintenance in controlled environments \cite{sahoo2019novel, dswilan2021flood}. However, they are sensitive to environmental factors, such as temperature fluctuations, wind, humidity, and short effective distances (up to $\sim$10 meters), with a typical accuracy of 1 to 3 cm. Recent mmWave radars, operating at high frequencies of 60-64 GHz and 77 GHz, have gained attention for its high precision, better resilience to environmental changes, and longer range. TI's mmWave radar solution \cite{ti_water_level_sensing} achieves millimeter-level accuracy.  Beamforming techniques \cite{hua2024wide} can suppress surrounding interference in complex environments. However, they face challenges when deployed in large-scale environments.

\textit{Satellite-Based.} Satellites have become a promising tool for large-scale hydrological observation, particularly in remote or inaccessible regions. The method \cite{tanaka2019development} relies on data from optical sensors or satellite radar altimeters combined with digital elevation models to estimate water levels. Optical remote sensing \cite{li2019water} offers high spatial coverage but is limited to decimeter-level accuracy due to resolution constraints and is  affected by cloud cover and lighting conditions. In contrast, satellite radar altimeters improve accuracy by emitting pulse signals toward the water surface and calculating the water level based on the satellite's position and the reflected signal \cite{shu2020evaluation}. However, due to their orbital cycle, they provide infrequent measurements, limiting their suitability for real-time monitoring \cite{narin2023multi}. Multi-satellite approaches \cite{han2024enhanced} are proposed to address this issue by improving temporal resolution and measurement reliability, especially for small and complex inland water bodies. Overall, satellite-based methods are effective for large-scale water monitoring but, due to limited temporal resolution, are less suitable for real-time applications.

\textit{Vision-Based.} The computer vision based methods have gained popularity due to advancements in computer vision and the availability of low-cost cameras. These methods infer water levels from images or videos using image processing or deep learning techniques. Traditional approaches rely on reference targets within the camera's view. Chapman et al. \cite{chapman2022open} develop an open-source tool for measuring water levels using calibrated markers. Zhang et al. \cite{zhang2023robust} propose a robust image processing-based method that performs well under varying conditions. Recent deep learning-based methods can improve performance and robustness. Qiao et al. \cite{qiao2022water} apply YOLOv5 for water level detection, achieving high precision across different environments. Liang et al. \cite{liang2023v} introduce V-FloodNet, a video segmentation system for urban flood detection and water level quantification, which proved effective in dynamic urban scenarios. Liu et al. \cite{liu2024evaluation} evaluate deep learning models for real-time river water level measurement, demonstrating their reliability in natural environments. Furthermore, Zhang et al. \cite{zhang2025cascade} propose a cascade method that combines traditional computer vision and deep learning techniques, offering improved accuracy while reducing computational overhead. Eltner et al. \cite{eltner2023image} also explored real-time image-based estimation, emphasizing its applicability in practical monitoring tasks. Despite their benefits, vision-based methods face challenges such as sensitivity to lighting conditions, weather, and water surface reflections. Achieving high precision depends on using high-resolution cameras, while calibration marker-based methods require accurate calibration to prevent systemic errors.

\textit{Communication Signal-Based.} This emerging water level monitoring approach leverages existing wireless communication infrastructure, such as WiFi or cellular networks. To the best of our knowledge, the related works in this area remains limited but shows significant potential. Key advantages of this method include scalability for large-scale monitoring, low deployment costs, and the capability for real-time sensing. The technique works by analyzing variations in wireless signal propagation characteristics, such as amplitude and phase. Muratore et al. \cite{muratore2024flood} investigate the impact of floods on cellular signal propagation, demonstrating its applicability for large-scale hydrological sensing. Our work PMNs-WaterSense propose a novel framework for water level estimation using CSI from communication signals. By addressing challenges such as environmental noise and multipath interference through advanced signal processing techniques, our method ensures reliable and accurate water level tracking, making it a practical solution for scalable hydrological applications.

\section{Conclusion}
This paper presents a novel framework for real-time water level sensing using CSI from existing communication signals. Key challenges are addressed through innovations such as single-antenna correlation to mitigate clock asynchronization, multi-domain processing for interference suppression, a Kalman filter-based phase unwrapping method for continuous water level tracking, and a geometry-based transformation for height estimation. Experimental results validate the system's efficacy, with average estimation errors of 0.025 cm for 28 GHz mmWave signals and 0.198 cm for 3.1 GHz LTE signals in our lab. In an outdoor scenario, the system maintains an estimation error within 4.8 cm, demonstrating its practicality. Future work will focus on enhancing system stability to further improve sensing performance in diverse environments.

\section*{Acknowledgments}
We thank Ayoob Salari for collecting LTE data in outdoor experiments and his valuable support.

\bibliographystyle{IEEEtran}
\bibliography{Water_Level_Sensing.bbl}

\vspace{11pt}


\vfill

\end{document}